\documentclass[
  prd,
  reprint,
  nofootinbib,
  floatfix,
  aps,
]{revtex4-2}

\usepackage{graphicx}
\usepackage{xcolor}

\usepackage{amsmath,amssymb}
\usepackage{multirow}

\usepackage{gensymb}      

\usepackage[normalem]{ulem} 
\usepackage{orcidlink}

\usepackage{hyperref}
\usepackage{cleveref}

\newif\ifincludesupplement
\includesupplementtrue  

\newcommand{\nuwro}{\text{\sc NuWro}}

\begin{document}

\title{Transfer Learning for Neutrino Scattering: Domain Adaptation with GANs}

\author{Jos\'e L. Bonilla\,\orcidlink{0009-0009-3240-1494}}
\email{joseluis.bonillaramirez@uwr.edu.pl}

\author{Krzysztof M. Graczyk\,\orcidlink{0000-0002-0038-6340}}
\email{krzysztof.graczyk@uwr.edu.pl}

\author{\\Artur M. Ankowski\,\orcidlink{0000-0003-4073-8686}}
\author{Rwik Dharmapal Banerjee\,\orcidlink{0000-0003-3639-7532}}
\author{Beata E. Kowal\,\orcidlink{0000-0003-3646-1653}}
\author{Hemant Prasad\,\orcidlink{0009-0003-1897-9616}}
\author{Jan T. Sobczyk\,\orcidlink{0000-0003-4991-2919}}

\affiliation{Institute of Theoretical Physics, University of Wroc\l aw, plac Maxa Borna 9,
50-204, Wroc\l aw, Poland}

\date{\today}%

\begin{abstract}
Transfer learning (TL) is used to extrapolate the physics information encoded in a Generative Adversarial Network (GAN) trained on synthetic neutrino-carbon inclusive scattering data to related processes such as neutrino-argon and antineutrino-carbon interactions. We investigate how much of the underlying lepton-nucleus dynamics is shared across different targets and processes. We also assess the effectiveness of TL when  training data is obtained from a different neutrino–nucleus interaction model. Our results show that TL not only reproduces key features of lepton kinematics, including the quasielastic and $\Delta$-resonance peaks, but also significantly outperforms generative models trained from scratch. Using data sets of 10,000 and 100,000 events, we find that TL maintains high accuracy even with limited statistics. Our findings demonstrate that TL provides a well-motivated and efficient framework for modeling (anti)neutrino-nucleus interactions and for constructing next-generation neutrino-scattering event generators, particularly valuable when experimental data are sparse.

\end{abstract}

\maketitle

\normalem

\section{Introduction}

Significant experimental efforts have been devoted to studying (anti)neutrino–nucleus interactions~\cite{Avanzini:2021qlx,NuSTEC:2017hzk} in the energy range relevant for next-generation neutrino oscillation experiments, such as Hyper-Kamiokande~\cite{Hyper-KamiokandeProto-:2015xww} and DUNE~\cite{DUNE:2020lwj}. In parallel, theoretical models describing these interactions have been developed~\cite{NuSTEC:2017hzk}. The outcomes of both experimental and theoretical advances are incorporated into Monte Carlo (MC) event generators, which simulate (anti)neutrino–nucleus collisions under realistic conditions~\cite{campbell2024eventgeneratorshighenergyphysics,Hayato:2021heg,Andreopoulos:2009rq,Mosel:2018qmv,Golan:2012wx,Isaacson:2022cwh}.

MC generators are often tuned to reproduce experimental observations, relying on adjustable parameters fitted to available data~\cite{GENIE:2021zuu}. However, this tuning process cannot fully compensate for the fundamental limitations of the underlying models, especially those relying on complex approximations, such as nuclear modeling. Consequently, there is a growing interest in alternative approaches to traditional MC event generation—methods that can learn directly from experimental or synthetic data and dynamically refine their predictions. Among the most promising directions is the development of artificial intelligence (AI)-driven tools for simulating particle collisions~\cite{Radovic:2018dip,Alanazi_2021,Butter2022}.

In this paper, we present results from our development of generative deep learning models to simulate neutrino and antineutrino interactions with nuclei. Specifically, we focus on generative adversarial networks (GANs)~\cite{goodfellow2014generativeadversarialnetworks}, which have recently attracted much interest in high-energy and nuclear physics~\cite{deOliveira:2017pjk,Monk:2018zsb,Alanazi:2020jod,Butter:2022rso,Ghosh:2022zdz,Ilten:2022jfm,Chan:2023ume}.

Our previous paper~\cite{bonilla2025generativeadversarialneuralnetworks} presented one of the first applications of GAN models for simulating neutrino collisions with nuclei. Additionally, Refs.~\cite{ElBaz:2023ijr, ElBaz:2025qjp} provide further examples of using generative deep-learning techniques in the simulation of neutrino–nucleus interactions, although in those studies, a method of normalizing flows was employed.

The main objective of this study is to examine the generalization ability of deep generative models, following the idea introduced in our earlier work~\cite{graczyk2024electronnucleuscrosssectionstransfer}. There, we showed that a model trained on inclusive electron–carbon scattering data could be adapted via transfer learning (TL) to predict cross sections for electron scattering on helium-3, lithium, oxygen, aluminum, calcium, and iron. This suggested that the network had captured universal features of lepton–nucleus dynamics, such as kinematic constraints and the relative position and strengths of quasielastic and $\Delta(1232)$ resonance peaks.

Here, we extend that concept to neutrino interactions. We employ a pre-trained GAN model~\cite{bonilla2025generativeadversarialneuralnetworks}, trained on synthetic inclusive charged-current (CC) muon-neutrino scattering data obtained from the \nuwro{} MC generator, as a baseline model encoding nuclear relationships. The model generates lepton kinematics in the quasielastic, $\Delta(1232)$ resonance and inelastic regions for a given input neutrino energy.

We hypothesize that this model retains generic relationships common to CC neutrino scattering processes, independent of the target nucleus or probe type. To test this, we apply domain adaptation, a form of TL in which a model trained in one domain (the source) is fine-tuned to operate effectively in another, related domain (the target). This allows us to probe how much of the encoded physics can be reused when data are limited or originate from a different generator configuration.
The baseline model is adapted to generate predictions for three processes: (A) muon-neutrino scattering off an argon nucleus; (B) muon-antineutrino scattering off a carbon nucleus; and (C) muon-neutrino scattering off a carbon nucleus, using data from an alternative version and configuration of \nuwro{}.

By comparing the domain-adapted and from-scratch models, we demonstrate that TL preserves the essential quasielastic and $\Delta(1232)$-resonance structure of the interactions and achieves high accuracy even with reduced training statistics. 
The GAN primarily encodes universal kinematic constraints and the overall quasielastic (QE) and $\Delta$-resonance pattern, while target-dependent spectral-function details are corrected by fine-tuning.
These findings suggest that TL is not only a computationally efficient\footnote{A computational efficiency of TL in classification of neutrino interaction processes was studied in Ref.~\cite{Chappell:2022yxd}.} method but also a physically interpretable tool for constructing (anti)neutrino–nucleus scattering models, enabling insight into which aspects of lepton–hadron dynamics are universal and which require re-optimization.

This paper is structured as follows.  In Sec.~\ref {sec:TL}, we discuss the idea of transfer learning and its applications to building (anti)neutrino-scattering models. In Sec.~\ref{sec:method}, key features of the GAN models and training procedures are explained. In Sec.~\ref{sec:results}, we present the numerical results. Finally, in Sec.~\ref{sec:summary} we summarize the paper. The paper is accompanied by supplemental material~Sections \ref{Supplement:S1}, \ref{Supplement:S2}, and \ref{Appendix:loss}, which includes figures showing the performance of the obtained models and a detailed description of the loss functions used in the numerical analysis.

\section{Neutrino interactions from transfer learning}
\label{sec:TL}

\subsection{Transfer learning}

The TL or knowledge transfer originally emerged from pedagogical and psychological research.~\cite{STEINER200115845}. It is founded on the idea that skills acquired in one task, such as playing the piano, can be applied to solve different tasks, like mathematical problems. Both types of challenges are addressed using the same set of neurons, which can either be used directly for both tasks or be slightly modified to tackle different kinds of problems.

TL has emerged as a powerful technique in deep learning~\cite{transfer_learning_survey,zhuang2020comprehensive,tan2018survey,Weiss2016}. Models that are initially trained for one type of problem can later be adapted, typically through fine-tuning, to address a different problem. The fundamental idea is that the initial layers of a neural network, once adapted to a specific task, extract universal features from the data that can be applied almost directly to related problems.

Our goal is to showcase the effectiveness of TL in simulating (anti)neutrino-nucleus interactions. Specifically, we aim to begin with a GAN trained on synthetic data generated by the theoretical framework, particularly the \nuwro{} MC generator. The predictions made by this generator align with reality to some extent, like any other theoretically based model. Therefore, the GAN model trained on \nuwro{} generated data will serve as a reference for creating more realistic simulators.

In our previous studies~\cite{graczyk2024electronnucleuscrosssectionstransfer}, using the TL technique, we showed that the inclusive electron-carbon interaction model can be adapted with a small amount of information to predict electron scattering off targets ranging from helium-3 to iron. In this paper, we demonstrate that, from a GAN simulation of neutrino-carbon scattering, one can obtain models for neutrino-argon, antineutrino-carbon, and neutrino-carbon (with a different nuclear model than the reference) scattering.

Similarly to Ref.~\cite{graczyk2024electronnucleuscrosssectionstransfer}, we assume that GAN's generator and discriminator for neutrino-carbon interactions, obtained in our previous work~\cite{bonilla2025generativeadversarialneuralnetworks}, have already learned the fundamental patterns underlying neutrino–nucleus interactions, and we will fine-tune this pre-trained model to obtain models for related processes.

One of the key advantages of the TL technique is that it reduces the need for large amounts of new data when developing models for related processes. Since the pre-trained model already encodes the fundamental relationships underlying neutrino scattering, only a relatively small amount of additional data is needed to fine-tune it. This property makes TL particularly attractive for neutrino–nucleus interactions, where experimental data are often sparse.

\begin{figure*}\centering
    \includegraphics[width=0.46\textwidth]{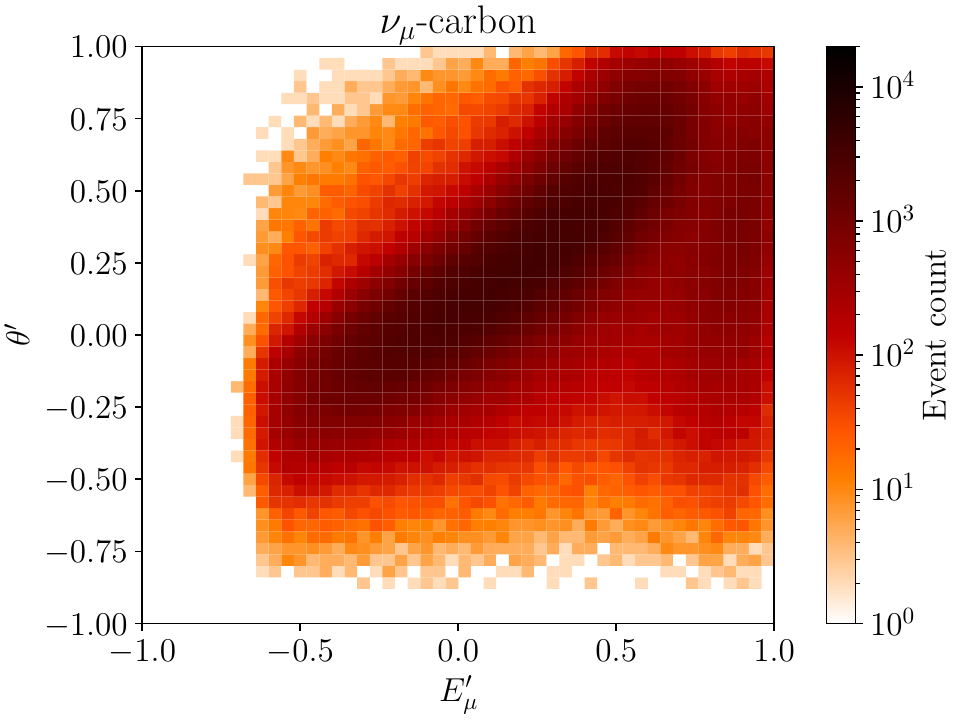} \ \
    \includegraphics[width=0.46\textwidth]{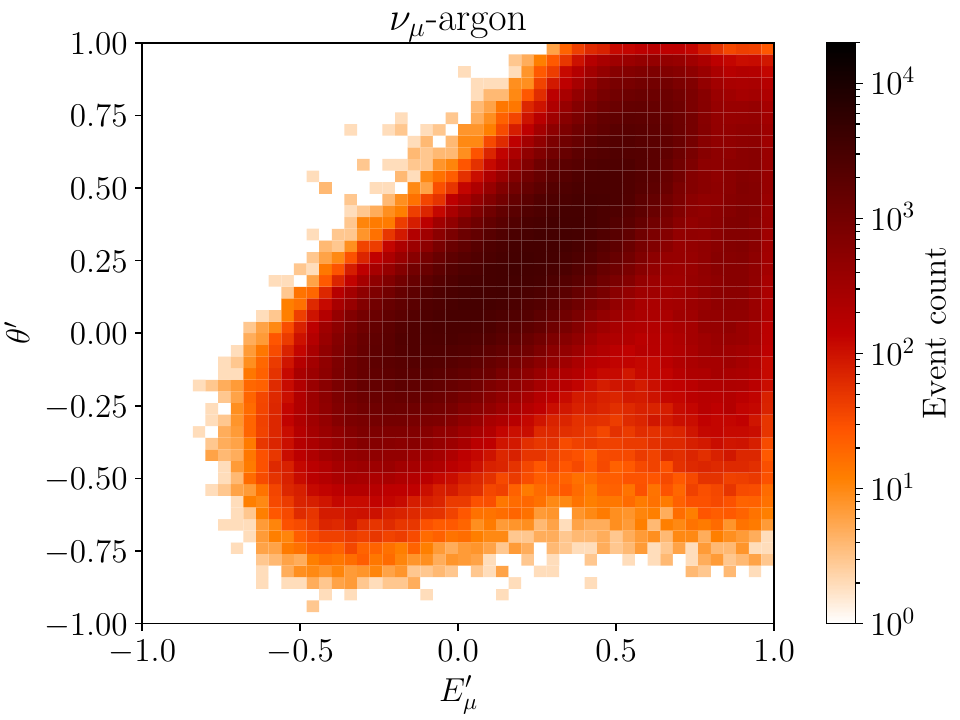}
    \includegraphics[width=0.46\textwidth]{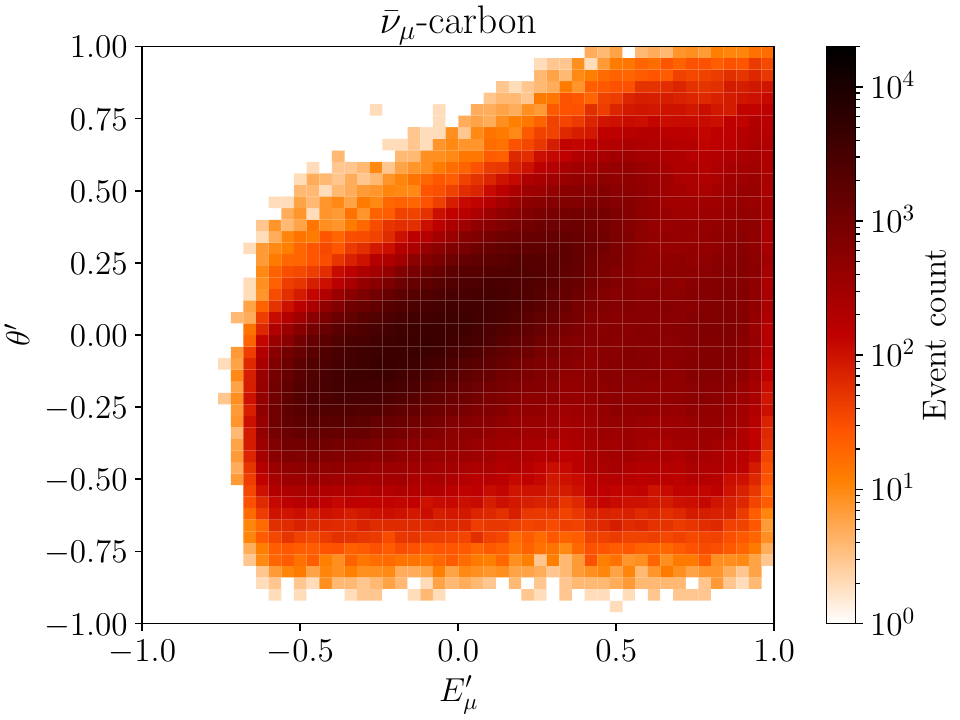} \ \
    \includegraphics[width=0.46\textwidth]{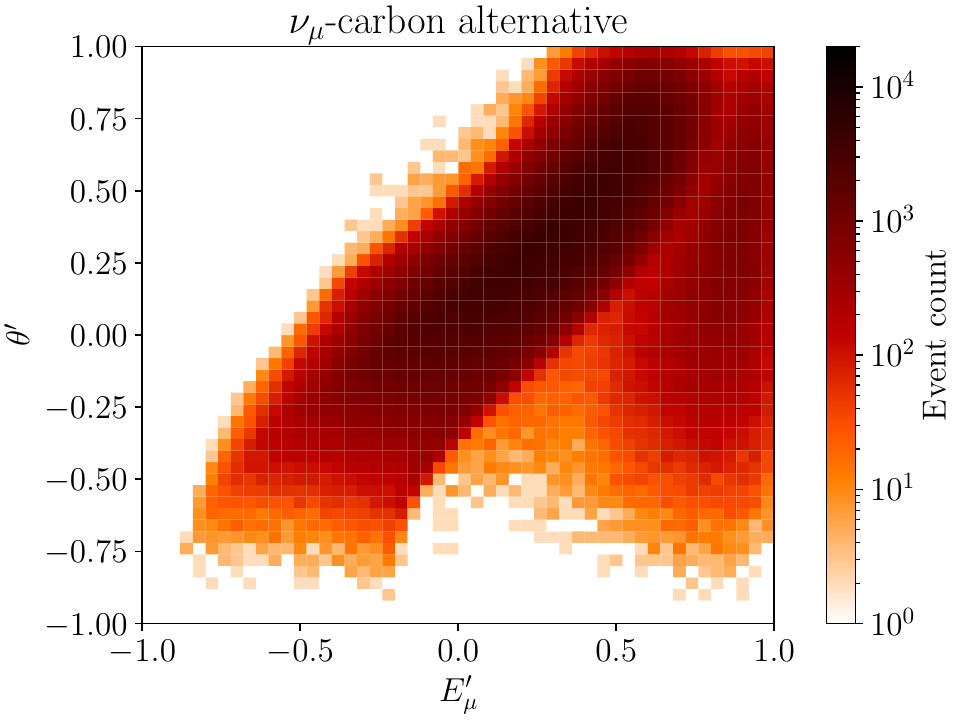}
    \caption{Test data samples $(E'_\mu,\theta')$ distributions for $E_\nu = 500$ MeV. On the top left, we show the original MC $\nu_\mu$-carbon scattering; on the top right, the $\nu_\mu$-argon scattering; on the bottom left, the $\bar\nu_\mu$-carbon scattering; and on the bottom right, the distributions of the alternative $\nu_\mu$-carbon samples.
    \label{fig:test_samples}}
\end{figure*}

\subsection{Domain adaptation}

As mentioned above, we consider the GAN model generating muon kinematical variables for the inclusive CC muon neutrino-carbon scattering. This model will be fine-tuned to obtain a description of three types of inclusive CC scattering processes:
\begin{enumerate}
    \item[(A)] $\nu_\mu$-argon  scattering;
    \item[(B)] $\bar\nu_\mu$-carbon scattering;
    \item[(C)] $\nu_\mu$-carbon scattering  with a different model of interaction than one used in the pre-training.
\end{enumerate}
All models assume a similar (uniform) initial spectrum of energy for (anti)neutrinos, and the target space encompasses configurations of kinematic variables that characterize the charged muon. Because the pre-trained model is trained on data from a single process (source domain), the resulting models are expected to generalize to related processes defined in a distinct domain. Consequently, the TL approach employed in this paper can be regarded as an instance of domain adaptation.

The pre-trained model generates events as predicted by the \nuwro{} ver.\texttt{21.09}, which employs the spectral function approach to model the carbon ($^{12}$C) nucleus in the QE channel~\cite{Benhar:1994hw}, and the local Fermi gas (LFG) model for other interaction channels. 
The generated events may include a charged muon and one or more nucleons, as well as occasionally pions, resulting from final-state interaction (FSI) effects. Pion production is modeled mostly through the $\Delta(1232)$ resonance excitation~\cite{Sobczyk:2004va,Juszczak:2005zs,Graczyk:2009qm}, while contributions from more inelastic processes are modeled using parton distribution functions (PDFs). 

In scenario (A), we consider the argon ($^{40}$Ar) nucleus, described by a spectral function (SF)~\cite{Banerjee:2023hub} derived from the analysis of Jefferson Lab coincidence electron–argon scattering data~\cite{PhysRevD.105.112002,PhysRevD.107.012005}. This setup allows us to test the effectiveness of TL when the nuclear target is changed from carbon to a structurally different nucleus. Note that the \nuwro{} SF model for carbon includes an optical potential, which is not present in the argon model\footnote{The latest version of \nuwro{} keeps optical potential for both targets.}.

In Scenario (B), we use the same model for nuclear and hadronic interactions; however, instead of neutrino–carbon interactions, we consider muon antineutrino–carbon interactions. From a theoretical point of view, on the three-level, the main difference between neutrino and antineutrino interactions is the sign between vector and axial contributions in the hadronic transition current. It results in the redistribution of the differential cross sections and the suppression of the total cross section with respect to the neutrino scattering cross section. 

Scenario (C) addresses a case where the default model used during pre-training encodes interaction dynamics that differ from those in the fine-tuning data. Specifically, the fine-tuning dataset is generated using the version of \nuwro{} \texttt{25.03}, which includes an improved treatment of resonant pion production extended to invariant masses up to $1.9$~GeV. In addition, we deliberately modify key model theoretical parameters: the axial mass in the QE channel is increased from $1.03$~GeV to $1.350$~GeV, and the axial mass in the pion production model is changed from $0.94$~GeV to $1.034$~GeV. Finally, the nuclear model is altered by replacing the spectral function with the LFG model.

The three scenarios differ in their kinematic domains, displaying notable variations in coverage of the relevant phase space, as shown in Fig.~\ref{fig:test_samples}. This figure presents the $(E^\prime_\mu, \theta^\prime)$ distributions, see Sec.~\ref{subsec:kiematics}, for 1,000,000 test events, for $E_\nu = 500$ MeV, in each case: the original test dataset ($\nu_\mu$–carbon), and the three test scenarios: $\nu_\mu$–argon, $\bar\nu_\mu$–carbon, and $\nu_\mu$–carbon with a modified interaction model\footnote{Note that in the supplement we show also $(E_\mu, \theta)$ distributions for $E_\nu = 500$ MeV for neutrino scattering with carbon and argon.}. All the distributions are strongly dominated by QE events. The four datasets exhibit significant differences in their distributions. The most pronounced divergence is observed between the (anti)neutrino–carbon cases and the two involving $\nu_\mu$–argon and the altered $\nu_\mu$–carbon interaction model. In the latter, a distinct band associated with the QE peak is visible. The differences observed between neutrino–carbon and neutrino–argon event distributions arise from the distinct spectral functions of the two nuclei. Argon, being much heavier and more complex than carbon, with more occupied shells, more nucleons, and a stronger Coulomb potential, has a spectral function characterized by larger and more widely distributed removal energies, as well as a different (and potentially stronger) high-momentum component. There is also an important effect originating from the optical potential included in the carbon spectral function.
Consequently, the QE peak in argon is broader than that observed for carbon targets (see Fig.~\ref{fig:test_samples_appendix} in the supplement). In the antineutrino–carbon event distribution, the QE peak is less pronounced than in the neutrino case. Finally, in the alternative carbon model using the LFG, the QE peak becomes noticeably sharper.

\section{GAN framework}
\label{sec:method}

\subsection{Kinematics}
\label{subsec:kiematics}
\begin{figure}\centering
    \includegraphics[scale=1.5]{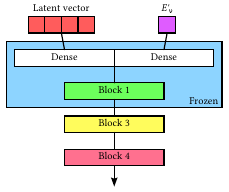}
    \caption{Generator architecture. The blue block encloses the portion of the network that remains frozen when training with TL. The output of the network is given by the proxy vector Eq.~\ref{Eq:proxy}. Similar to Ref.~\cite{bonilla2025generativeadversarialneuralnetworks}, the skip connections are applied but not shown in the graph above.  \label{fig:generator}}. 
\end{figure}
\begin{figure}\centering
    \includegraphics[scale=1.5]{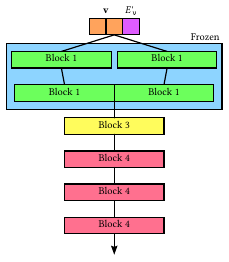}
    \caption{Discriminator architecture. The blue block encloses the portion of the network that remains frozen when training with TL. The network takes for the input $\mathbf{v}$ (the proxy variable vector defined by Eq.~\ref{Eq:proxy}) and proxy neutrino energy $E'_\nu$. Similar to Ref.~\cite{bonilla2025generativeadversarialneuralnetworks}, the skip connections are applied but not shown in the graph above.\label{fig:discriminator}}
\end{figure}

Our goal is to obtain a model that generates the kinematic characteristics of the charged muon produced in the inclusive CC $\nu_\mu$-nucleus and $\overline{\nu}_\mu$-nucleus scattering. We consider the same generator and discriminator models as in our previous paper~\cite{bonilla2025generativeadversarialneuralnetworks}.

Note that our research plan involves generating comprehensive event information, including the hadronic components. However, creating all final-state particles is complex due to the variability in the number of final
particles and the need to ensure charge, energy, and momentum conservation. For this reason, we have chosen
first to investigate inclusive contributions to test models and various solutions.

The generator for given normalized (anti)neutrino energy $E_\nu^\prime$ gives as an output two kinematic variables characterizing the muon, namely,  $E_\mu^\prime$ and $\theta^\prime$, where
\begin{equation}
\label{Eq:Eprime}
	E_{\nu}^\prime = 2 \frac{E_\nu - E_{\nu, min}}{E_{\nu, max} - E_{\nu, min}} - 1,
\end{equation}
$E_\nu$ is the (anti)neutrino energy, $E_{\nu, min} = 0.3$~GeV, $E_{\nu, max} = 10$~GeV, 
\begin{eqnarray}
\label{Eq:Emuprime}
  E_\mu^\prime & = & 2\sqrt{1-\frac{E_\mu-m_\mu}{\Delta E}} - 1, \quad \Delta E = E^-_\nu - m_\mu \\
 \theta^\prime & = & 2\sqrt{\frac{\theta}{\pi}} - 1, 
 \label{Eq:thetaprime}
\end{eqnarray}
and $m_\mu$ is the muon mass, $\theta$ is the muon scattering angle relative to  the neutrino beam direction.

The choice of the proxy variables  is discussed in the Ref.~\cite{bonilla2025generativeadversarialneuralnetworks}. The values $E_{\nu, min}$ and $E_{\nu, max}$ correspond to minimal and maximal neutrino energy generated by the model.
In this paper, we consider a slightly different value of $E^-_\nu$ than in the previous study~\cite{bonilla2025generativeadversarialneuralnetworks}. While the previously introduced boundaries prevented the production of non-physical events, they turned out to be too strong for the applications discussed in the present study. Since in experimental data we expect not to follow the same physics as the \nuwro{} simulation, the constraint has been relaxed to $E^-_\nu = E_\nu-4$~MeV.

In our previous studies, we optimized the model using $\nu_\mu$-carbon ($^{12}$C) scattering data produced with \nuwro{} ver.~\texttt{21.09} generator. In the present analysis, we generate  data using the same version of \nuwro{}. However, for one discussed case, we consider the \nuwro{} ver.~\texttt{25.03}. As we changed the boundary conditions, we repeated the previous analysis and obtained a new model for neutrino-carbon scattering. This model serves as the baseline for subsequent studies.

\subsection{Networks}

We consider the same network architecture as in our previous paper~\cite{bonilla2025generativeadversarialneuralnetworks} (model for inclusive CC neutrino-carbon scattering). The generator takes as input a random vector of length $50$ (drawn from a Gaussian multivariate distribution) and neutrino energy $E'_\nu$. Both inputs are processed independently by two distinct dense layers. Then, the output is concatenated and processed by three blocks, as shown in Fig.~\ref{fig:generator}. The output of the generator is given by a last dense layer that produces the proxy vector
\begin{equation}
\label{Eq:proxy}
\mathbf{v}=(\theta',{E'}_\mu).
\end{equation}

To obtain the GAN model, one needs to train the discriminator model, which takes as input the proxy variables  
$\mathbf{v}$ and neutrino energy $E'_\nu$, which are concatenated and processed by two independent blocks. The output is then concatenated and transformed. Eventually, the output is transformed by four blocks, as shown in Fig.~\ref{fig:discriminator}. Finally, the discriminator uses dense layer with single output value.

The fundamental building block of our network consists of a fully connected (dense) layer, followed by layer normalization~\cite{ba2016layernormalization}, an activation function—\texttt{ReLU}\cite{pmlr-v15-glorot11a} for the generator and \texttt{PReLU}\cite{2015arXiv150201852H} for the discriminator, and a dropout layer. We refer to this configuration as \texttt{Block1}. Introducing a skip connection from the input to the output of \texttt{Block1} yields \texttt{Block2} (the latter is not used in the models of the present study). When the $E'_\nu$ information is concatenated to the input of \texttt{Block1} and subsequently \texttt{Block2}, we obtain \texttt{Block3} and \texttt{Block4}, respectively. Furthermore, Gaussian noise is injected into both the generator and the discriminator after each layer containing learnable parameters. A comprehensive description of the GAN architecture is provided in Ref.\cite{bonilla2025generativeadversarialneuralnetworks}.

Although the architectural scheme remains unchanged, we introduce several modifications compared to the previous analysis. Specifically, Gaussian noise is now injected into both the generator and the discriminator, whereas previously it was applied only for the discriminator. In addition, \texttt{Blocks 1 and 2} in the main body of the generator have been replaced by \texttt{Blocks 3 and 4}, respectively, as illustrated in Fig.~\ref{fig:generator}. It is important to note that the discriminator architecture remains unchanged (see Fig.~\ref{fig:discriminator}). Notably, the discriminator is approximately twice as deep as the generator, and the number of neurons per layer is configured to yield roughly twice the number of trainable parameters per layer compared to the generator.

To adapt the pre-trained model to new data, we freeze the first block of both the generator and discriminator, as illustrated in Figs.~\ref{fig:generator} and \ref{fig:discriminator}. The remaining blocks are fine-tuned during the optimization process.

\subsection{Loss and metrics}

The loss function has two components: one for the discriminator and one for the generator. As in our previous study, the discriminator loss is binary cross-entropy. In the case of generator loss, we tested three cases: the same choices as in our previous work, the ``classical'' one initially proposed in Ref.~\cite{goodfellow2014generativeadversarialnetworks}, the heuristic (non-saturating) loss also proposed in  Ref.~\cite{goodfellow2014generativeadversarialnetworks}, and the sum of both. These preliminary studies showed that the last proposal is the most efficient. As it is not a common choice of the loss, we discuss its properties in the supplement section \ref{Appendix:loss}.

The network models are optimized using the \textit{AdamW}~\cite{loshchilov2018decoupled} algorithm in the initial stage of the training and then with \textit{Amsgrad}~\cite{j.2018on}.

\begin{table}
    \caption{Performance metrics of the pre-trained $\nu_\mu$-carbon model. \label{tab:Inlusivemetrics}}
    \begin{ruledtabular}
        \begin{tabular}{lccccc}
            $E_\nu$ & EMD  & MAP  & (3D) & MAP w/o tails & (3D) \\ \hline
            500 MeV & 0.15 & 1.09 & ---  & 1.09         & ---  \\ 
            1 GeV   & 0.15 & 1.05 & ---  & 1.05         & ---  \\ 
            3 GeV   & 0.15 & 1.00 & ---  & 1.00         & ---  \\ 
            5 GeV   & 0.15 & 1.00 & ---  & 0.98         & ---  \\ 
            9 GeV   & 0.14 & 0.97 & ---  & 0.95         & ---  \\ 
            All     & 0.12 & 0.88 & 0.89 & 0.86         & 0.88
        \end{tabular}
    \end{ruledtabular}
\end{table}

To evaluate the quality of the model, we consider two metrics: the Mean Averaged Pull (MAP)~\cite{DemortierEverythingYA,Alanazi:2020jod,bonilla2025generativeadversarialneuralnetworks} and the Earth Mover's Distance (EMD)~\cite{710701}. In our earlier studies, we used MAP in its two-dimensional form; however, in this analysis, we adapt the MAP to three dimensions (MAP-3D). This adaptation involves creating a histogram of the variables ($E^\prime_\nu$, $E^\prime_\mu$, $\theta^\prime$) using a uniformly distributed sample of $E_\nu$ with a binning of $14\times14\times14$\footnote{The number of bins in the MAP-3D calculation is approximately similar to the number of $50\times50$ bins used for the 2-dimensional version in the previous version.}.

For the MAP-3D metric, we consider two versions: one that includes all tails, where the bin content of both the \nuwro{} histogram ($n_{nuw,i}\neq0$) and the GAN histogram ($n_{gan,i}\neq0$) is used, and another that excludes low-content bins, where we only consider bins with $n_{nuw,i}>5$ and $n_{gan,i}>5$. Additionally, we compute the MAP metrics using histograms of the two-dimensional distribution of $E_{\mu}^\prime$ and $\theta^\prime$, with a binning of $50\times50$. The metric values during optimization are calculated on a validation dataset consisting of one million events.

Typically, the best model is the one that minimizes both metrics, but, in the present analysis, to choose the  optimal model, 
we evaluate an average of both MAP-3D metrics 
$\langle\text{MAP}\rangle=\frac{1}{2}(\text{MAP-3D}+\text{MAP-3D w/o tails})$.

\begin{figure}\centering
    \includegraphics[width=0.46\textwidth]{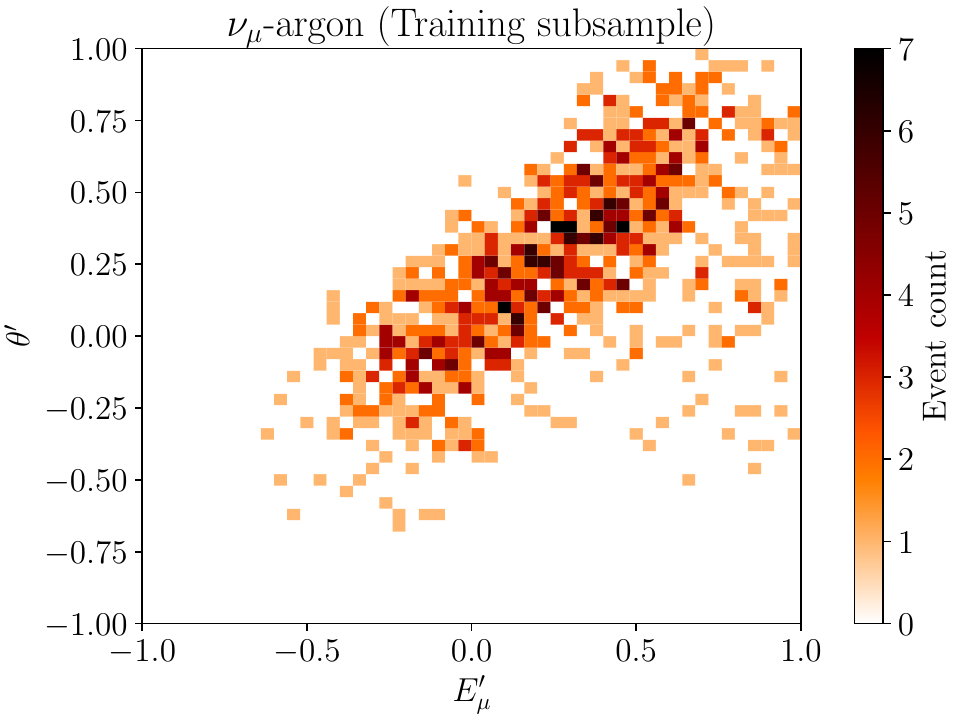} \ \
    \includegraphics[width=0.46\textwidth]{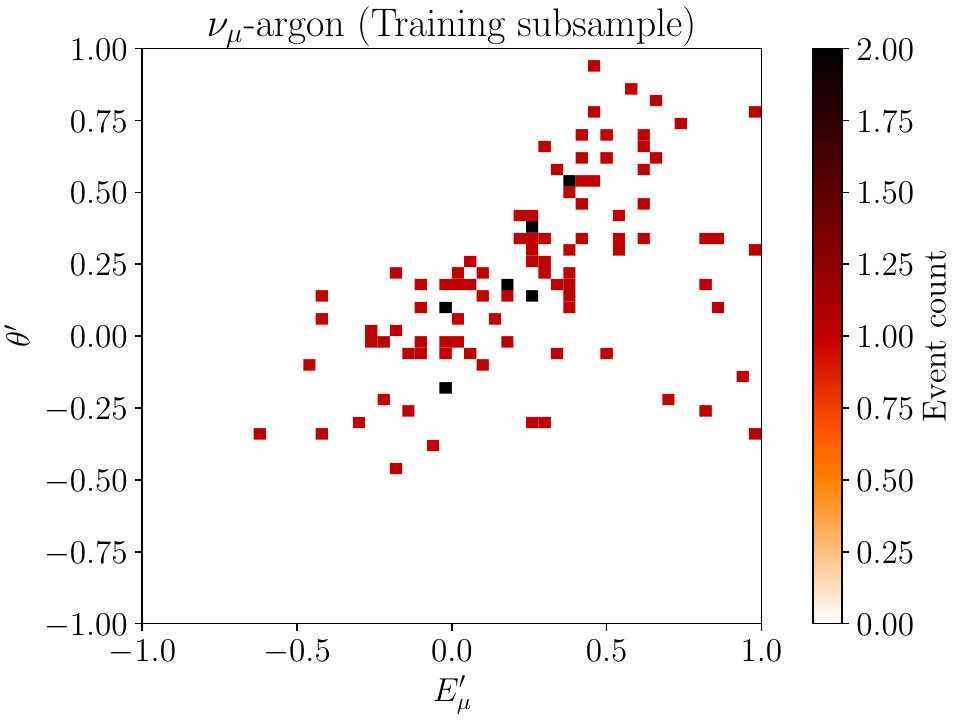}
    \caption{Training data $450 < E_\nu < 550$ MeV sub-samples $(E'_\mu,\theta')$ distributions. On the top, we have the sub-sample from a training sample of $100$k events, while on the bottom is the sub-sample from a sample of $10$k events.
    \label{fig:train_subsample}}
\end{figure}
\begin{figure*}\centering
    \includegraphics[width=0.4\textwidth]{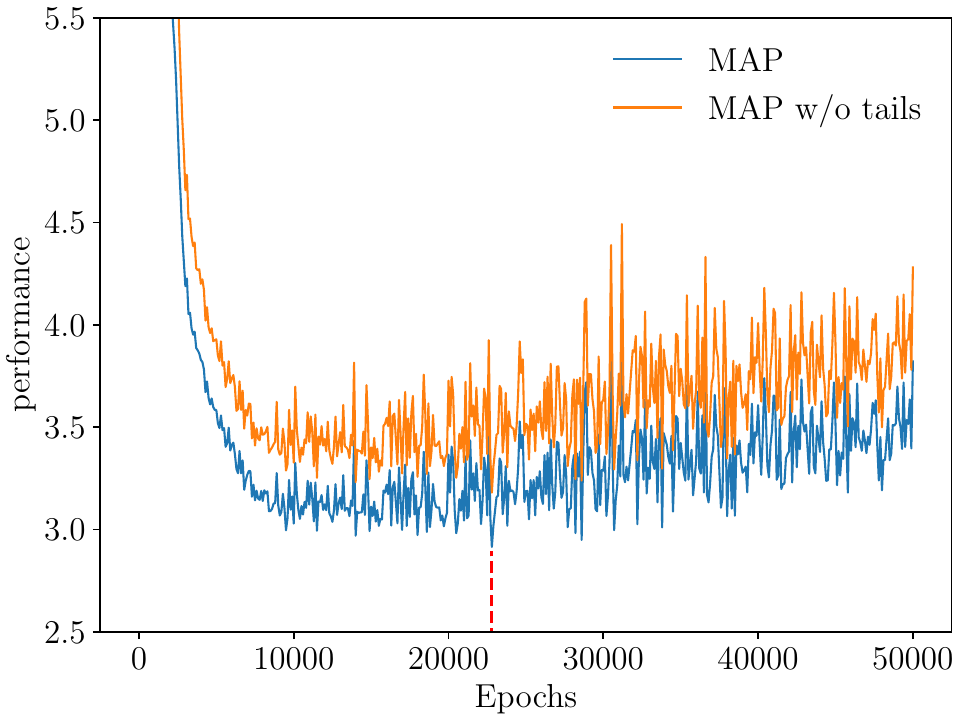} 
    \includegraphics[width=0.4\textwidth]{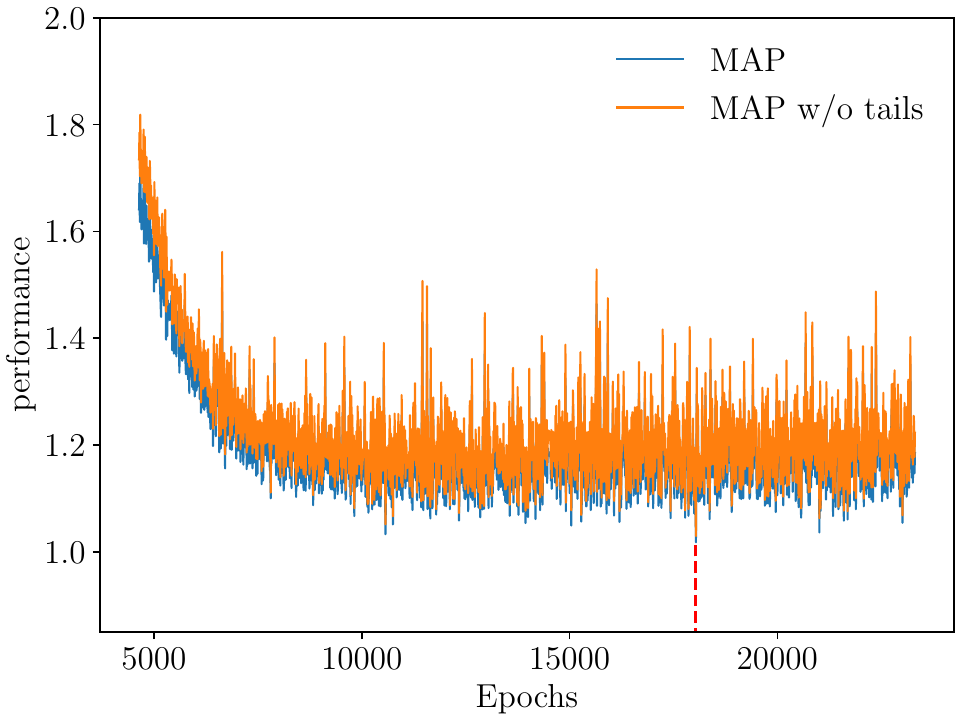} \\
    \includegraphics[width=0.4\textwidth]{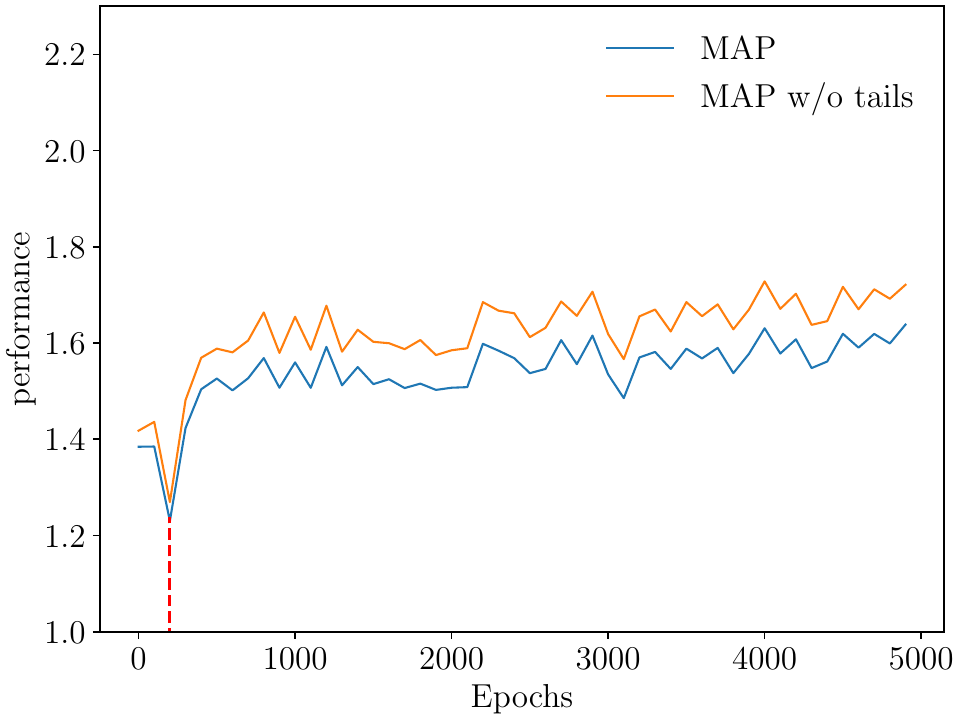}  
    \includegraphics[width=0.4\textwidth]{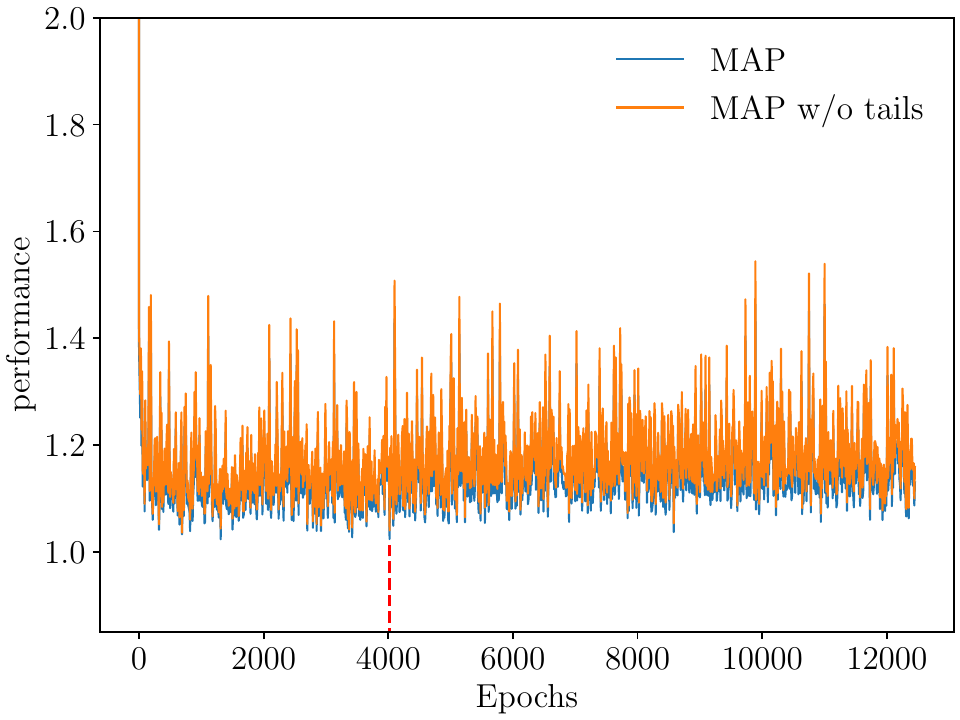}
    \caption{Dependence of the MAP-3D metric on the number of optimization epochs for the $\nu_\mu$–argon models. The left and right columns correspond to optimizations using 10,000 and 100,000 events, respectively. The top and bottom rows show results for models trained from scratch and with transfer learning (TL), respectively. Red dashed vertical lines mark the epochs of the selected models.}
    \label{fig:nuAr40performance}
\end{figure*}

\begin{figure}[btp]\centering
    \includegraphics[width=0.46\textwidth]{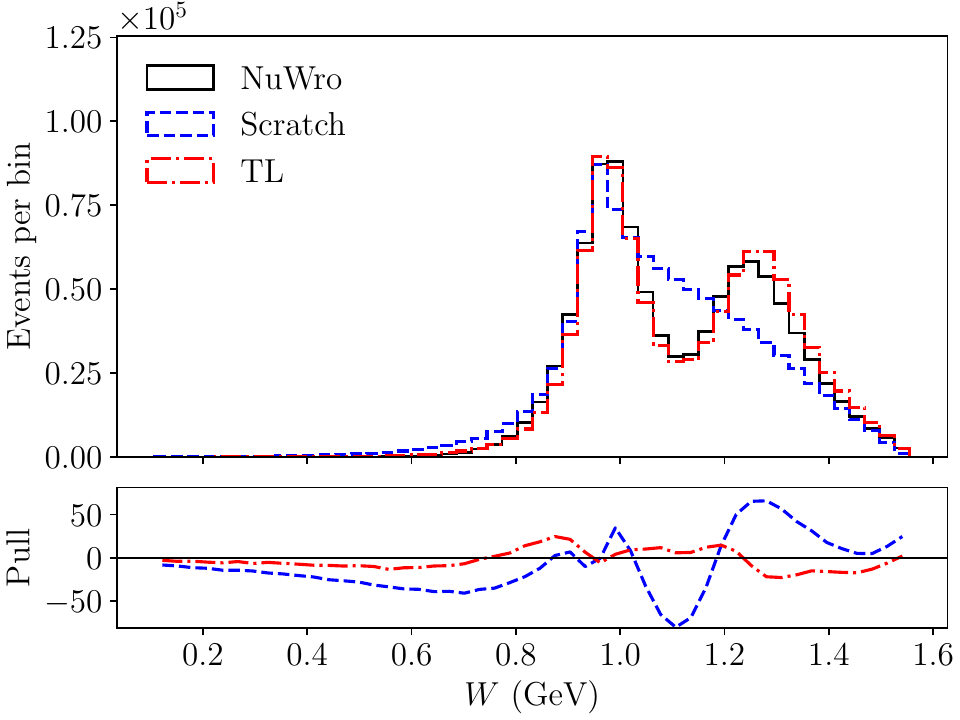}\\
    \includegraphics[width=0.46\textwidth]{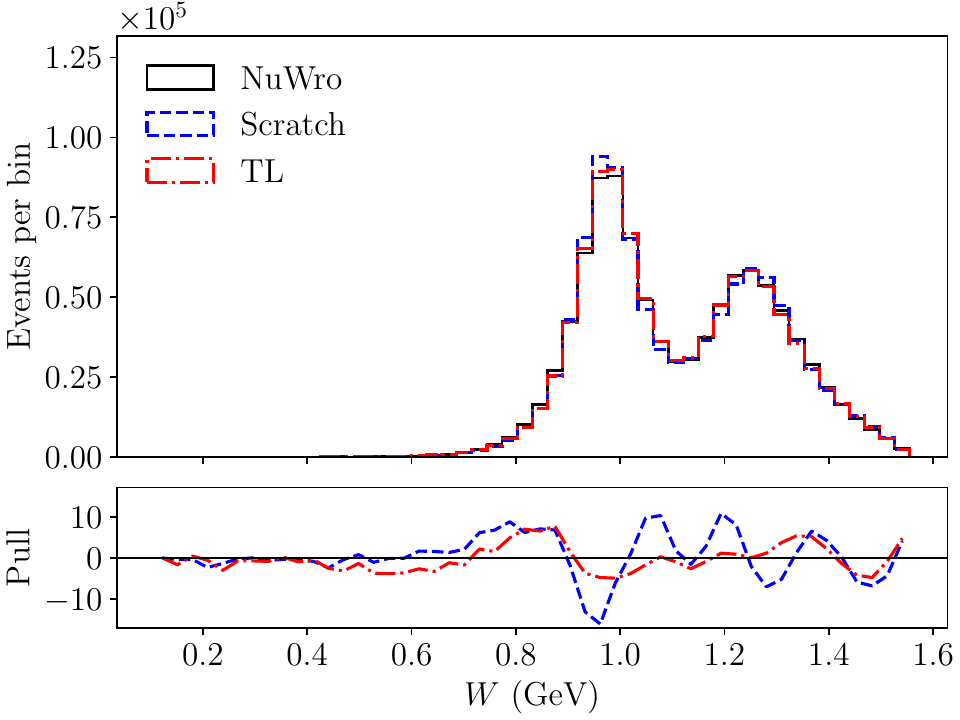}
    \caption{Reconstructed hadronic invariant mass $W$-distributions, for $E_\nu = 1$ GeV and for inclusive CC $\nu_\mu$-argon scattering, generated by \nuwro{} and GAN model trained from scratch and using TL, for the optimization  with 10,000 (top) and 100,000 (bottom) events (top). 
    \label{fig:nuAr40-10k}}
\end{figure}

\begin{table*}[htbp]
    \caption{Performance metrics for generators of CC inclusive $\nu_\mu$–argon scattering events. The left and right panels display results for the models trained from scratch and using transfer learning, respectively. The training dataset comprises 10,000 (top table) and 100,000 (bottom table) events.
    \label{tab:Ar40-10k}}
\begin{ruledtabular}
        \begin{tabular}{lcccccccccc}
             \multicolumn{11}{c}{Trained on 10,000 events}  \\
                    & \multicolumn{5}{c}{Scratch} & \multicolumn{5}{c}{TL} \\
                    \cline{2-6}\cline{7-11}
            $E_\nu$ & EMD  & MAP  & (3D) & MAP w/o tails & (3D)  & EMD  & MAP  & (3D) & MAP w/o tails & (3D) \\ \hline
            500 MeV & 0.96 & 6.46 & ---  & 7.63          & ---  & 0.36 & 2.51 & ---  & 2.71          & ---  \\ 
            1 GeV   & 0.84 & 5.42 & ---  & 6.17          & ---  & 0.37 & 2.42 & ---  & 2.67          & ---  \\ 
            3 GeV   & 0.73 & 3.92 & ---  & 4.30          & ---  & 0.25 & 1.40 & ---  & 1.44          & ---  \\ 
            5 GeV   & 0.65 & 3.50 & ---  & 3.81          & ---  & 0.23 & 1.32 & ---  & 1.35          & ---  \\ 
            9 GeV   & 0.65 & 3.58 & ---  & 3.88          & ---  & 0.28 & 1.60 & ---  & 1.65          & ---  \\ 
            All     & 0.36 & 2.01 & 3.01 & 2.05          & 3.22 & 0.17 & 1.07 & 1.27 & 1.08          & 1.29
            \\
            \hline
            \\
             \multicolumn{11}{c}{Trained on 100,000 events}  \\
                    \cline{2-6}\cline{7-11}
            500 MeV & 0.32 & 2.42 & ---  & 2.67          & ---  & 0.21 & 1.63 & ---  & 1.69          & ---  \\ 
            1 GeV   & 0.26 & 1.58 & ---  & 1.64          & ---  & 0.19 & 1.38 & ---  & 1.41          & ---  \\ 
            3 GeV   & 0.20 & 1.30 & ---  & 1.33          & ---  & 0.16 & 1.10 & ---  & 1.10          & ---  \\ 
            5 GeV   & 0.18 & 1.21 & ---  & 1.21          & ---  & 0.18 & 1.14 & ---  & 1.15          & ---  \\ 
            9 GeV   & 0.20 & 1.31 & ---  & 1.33          & ---  & 0.20 & 1.41 & ---  & 1.45          & ---  \\ 
            All     & 0.14 & 0.98 & 1.06 & 0.97          & 1.05 & 0.14 & 0.94 & 1.04 & 0.93          & 1.03
        \end{tabular}
    \end{ruledtabular}
\end{table*}

\begin{figure}{p}\centering
    \includegraphics[width=0.46\textwidth]{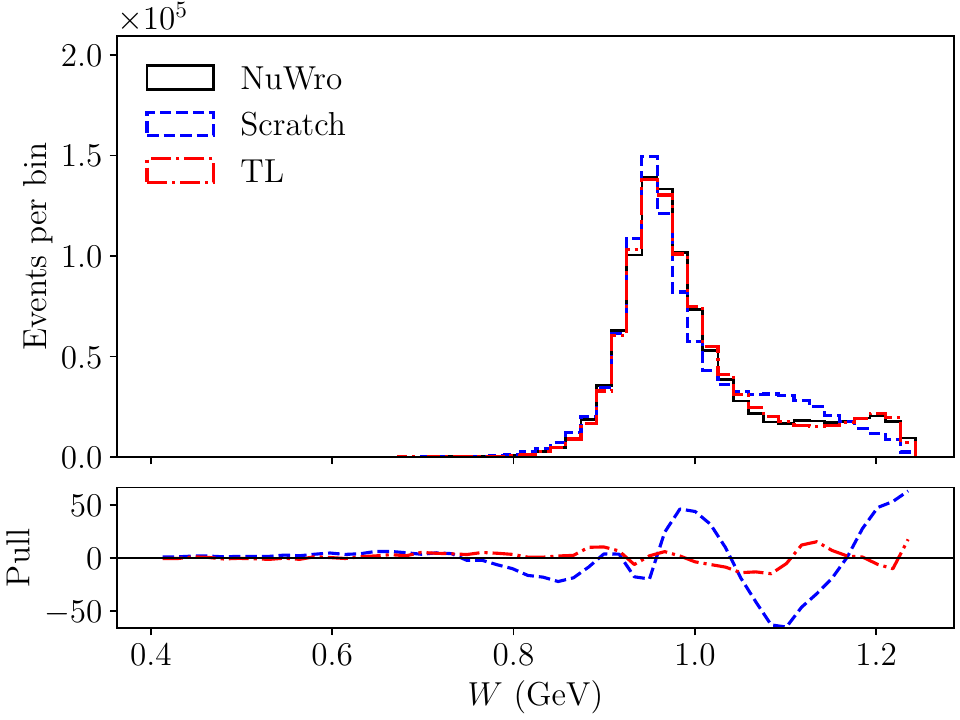}\\
    \includegraphics[width=0.46\textwidth]{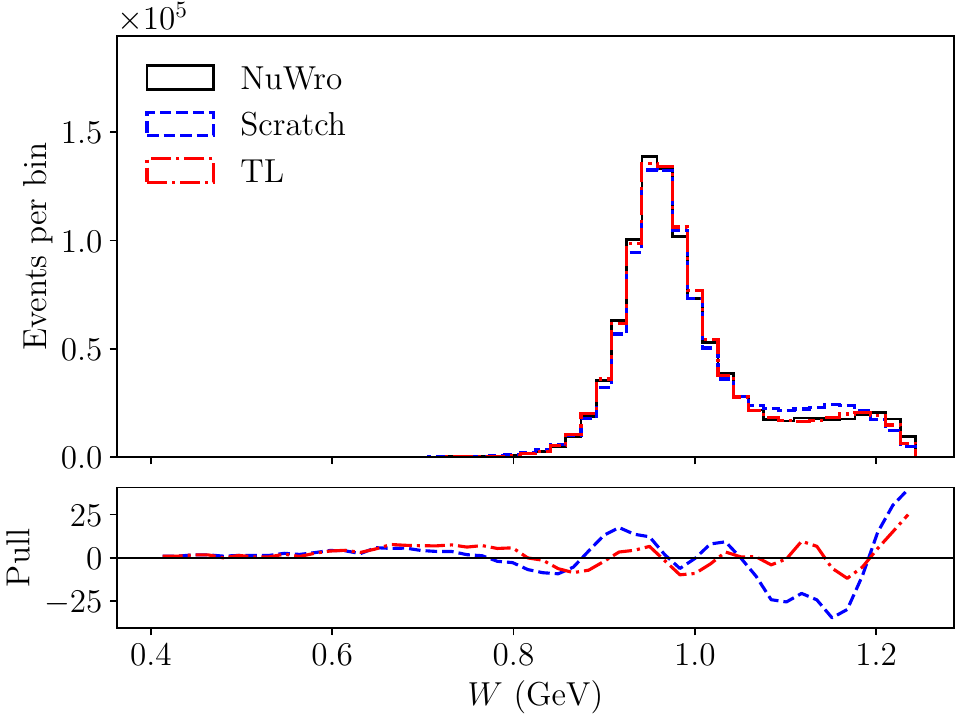}
    \caption{Same as in Fig.~\ref{fig:nuAr40-10k} but for CC $\bar\nu_\mu$-carbon scattering and $E_\nu=500$ MeV.
    \label{fig:antinuC12-10k}}
\end{figure}

\begin{table*}
    \caption{Same as in Table~\ref{tab:Ar40-10k} but for 
    charged-current inclusive $\bar{\nu}_\mu$–carbon scattering. \label{tab:antinu-10k}}
    \begin{ruledtabular}
        \begin{tabular}{lcccccccccc}
            \multicolumn{11}{c}{Trained on 10,000 of events}  \\
             & \multicolumn{5}{c}{Scratch} & \multicolumn{5}{c}{TL} \\
                    \cline{2-6}\cline{7-11}
            $E_\nu$ & EMD  & MAP  & (3D) & MAP w/o tails & (3D)  & EMD  & MAP  & (3D) & MAP w/o tails & (3D) \\ \hline
            500 MeV & 0.64 & 4.65 & ---  & 5.21          & ---  & 0.36 & 2.13 & ---  & 2.32          & ---  \\ 
            1 GeV   & 0.88 & 4.45 & ---  & 5.00          & ---  & 0.38 & 2.57 & ---  & 2.79          & ---  \\ 
            3 GeV   & 0.84 & 4.00 & ---  & 4.46          & ---  & 0.39 & 2.07 & ---  & 2.21          & ---  \\ 
            5 GeV   & 0.67 & 3.57 & ---  & 3.87          & ---  & 0.33 & 1.67 & ---  & 1.75          & ---  \\ 
            9 GeV   & 0.72 & 3.77 & ---  & 3.98          & ---  & 0.32 & 1.85 & ---  & 1.94          & ---  \\ 
            All     & 0.51 & 2.15 & 2.88 & 2.37          & 3.10 & 0.21 & 1.26 & 1.56 & 1.28          & 1.61
            \\
            \hline
            \\
            \multicolumn{11}{c}{Trained on 100,000 of events}  \\
                    \cline{2-6}\cline{7-11}
            500 MeV & 0.31 & 2.27 & ---  & 2.41          & ---  & 0.22 & 1.63 & ---  & 1.66          & ---  \\ 
            1 GeV   & 0.29 & 1.77 & ---  & 1.79          & ---  & 0.25 & 1.52 & ---  & 1.55          & ---  \\ 
            3 GeV   & 0.20 & 1.39 & ---  & 1.40          & ---  & 0.18 & 1.24 & ---  & 1.25          & ---  \\ 
            5 GeV   & 0.21 & 1.24 & ---  & 1.24          & ---  & 0.20 & 1.20 & ---  & 1.21          & ---  \\ 
            9 GeV   & 0.25 & 1.33 & ---  & 1.34          & ---  & 0.22 & 1.24 & ---  & 1.26          & ---  \\ 
            All     & 0.13 & 0.93 & 1.10 & 0.91          & 1.09 & 0.14 & 0.94 & 1.08 & 0.92          & 1.08
        
        \end{tabular}
    \end{ruledtabular}
\end{table*}

\begin{table*}
\caption{Performance metrics for generators of CC $\nu_\mu$–carbon (alternative interaction model) scattering events. The left and right panels display results for the models trained from scratch and using transfer learning, respectively. The training dataset comprises 10,000. \label{tab:nuC12alt-10k}}
    \begin{ruledtabular}
        \begin{tabular}{lcccccccccc}
            & \multicolumn{5}{c}{Scratch} & \multicolumn{5}{c}{TL} \\
                    \cline{2-6}\cline{7-11}
            $E_\nu$ & EMD  & MAP  & (3D) & MAP w/o tails & (3D)  & EMD  & MAP  & (3D) & MAP w/o tails & (3D) \\ \hline
            500 MeV & 1.18 & 8.47 & ---  & 10.20        & ---   & 0.51 & 3.48 & ---  & 3.99        & ---  \\ 
            1 GeV   & 1.05 & 6.27 & ---  & 7.32         & ---   & 0.44 & 2.95 & ---  & 3.25        & ---  \\ 
            3 GeV   & 1.03 & 5.14 & ---  & 5.83         & ---   & 0.35 & 2.16 & ---  & 2.28        & ---  \\ 
            5 GeV   & 0.95 & 4.40 & ---  & 4.86         & ---   & 0.35 & 2.13 & ---  & 2.23        & ---  \\ 
            9 GeV   & 0.86 & 4.46 & ---  & 4.86         & ---   & 0.35 & 2.12 & ---  & 2.22        & ---  \\ 
            All     & 0.47 & 2.58 & 3.54 & 2.64         & 3.86  & 0.20 & 1.23 & 1.63 & 1.23        & 1.68 
        \end{tabular}
    \end{ruledtabular}
\end{table*}

\begin{figure}\centering
    \includegraphics[width=0.46\textwidth]{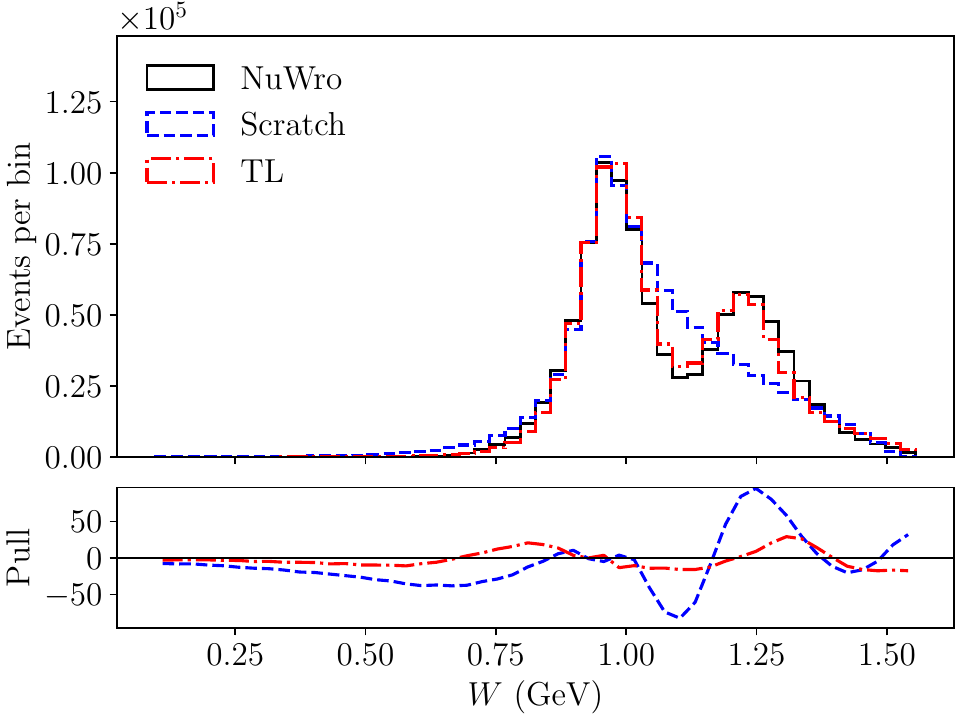}
    \caption{Same as in Fig.~\ref{fig:nuAr40-10k} but for CC $\nu_\mu$-carbon, $E_\nu=1$~GeV (alternative interaction) scattering and the training dataset with 10,000 events.\label{fig:nuC12alt-10k}}
\end{figure}

\section{Numerical results}
\label{sec:results}

To study the  TL approach, we began by obtaining a baseline (pre-trained) generator and discriminator for inclusive neutrino-carbon scattering events. We trained the model on samples uniformly distributed across  neutrino energies from $300$~MeV to $10$~Gev, using a training dataset with four million events. To test the quality of this model, we generated new data at fixed neutrino energies: $500$~MeV, $1$~GeV, $3$~GeV, $5$~GeV, and $10$~GeV, as well as for uniformly distributed samples across the entire energy spectrum. The values of performance metrics of the baseline GAN model are low and comparable to those reported in our previous paper~\cite{bonilla2025generativeadversarialneuralnetworks}. Table~\ref{tab:Inlusivemetrics} summarizes the quality of the baseline model.

To evaluate the effectiveness of the TL technique -- specifically, to determine how well the baseline model captures universal patterns shared across all processes discussed in this paper -- we consider two training scenarios. In particular, we train models using both a large dataset of 100,000 events and a smaller dataset of 10,000 events.

Fig.~\ref{fig:train_subsample} illustrates the differences in event distributions between the two training datasets for $\nu_\mu$–argon scattering events within the neutrino energy range of $450$ to $550$~MeV. The figure clearly shows that training the GAN and discriminator with only 10,000 events poses a significant challenge. 

Using a smaller dataset to fine-tune models mimics experimental reality, where data sets are often limited and sparse. Testing the TL approach in this context allows us to evaluate its potential for tuning generative models with real measurements. It's important to note that experimental data is typically produced in a more complex form, as flux-averaged. In this paper, we present a toy model approach to address this issue, while a more realistic solution is the focus of our current investigations.

We begin our discussion with case (A): neutrino scattering on argon target. 
To highlight the role of training statistics, we compare models trained on 10,000 and 100,000 events. Fig.~\ref{fig:nuAr40performance} shows the evolution of performance metrics during training. For the training with 100,000 events, the model's performance remains stable. It is noteworthy that the best-performing models in the transfer learning framework are reached significantly faster than those trained from scratch.
Furthermore, as summarized in Table~\ref{tab:Ar40-10k}, the TL models consistently achieve lower metric values, indicating better performance. This suggests that when trained from scratch, the network struggles to learn the features of $\nu_\mu$–argon scattering from limited data, whereas the transfer learning approach, which builds on prior knowledge of $\nu_\mu$–carbon interactions, offers a substantial advantage. Eventually, when the number of samples in the training dataset increases, the quality of the obtained GAN models improves. However, even with a large number of samples, the transfer learning model still outperforms the model trained from scratch.

Fig.~\ref{fig:nuAr40-10k} and Figs. \ref{fig:nuAr40-10k_supplement} and \ref{fig:nuAr40-100k_supplement} in the supplement  show the event distributions in reconstructed hadronic invariant mass $W$ as generated by our GAN models and \nuwro{} for training datasets containing 10,000 and 100,000 events. This quantity is defined by  
\begin{equation}
W = \sqrt{M_N^2+2 \omega M - Q^2}, \quad \omega = E_\nu - E_\mu,
\end{equation}
$M_N$ is the average nucleon mass, and the four-momentum transfer squared
\begin{equation} 
Q^2 = 2 E_\nu (E_{\mu} - p_{\mu,z}) - m_\mu^2,
\end{equation}
where $p_{\mu,z}$ is the muon momentum component along the neutrino beam direction. The $W$ distributions clearly highlight the QE and $\Delta(1232)$ resonance peaks. In Fig.~\ref{fig:nuAr40-10k}, we show the results from simulations for neutrino energy $500$ MeV, while for neutrino energies: $1$, $3$, and $5$ GeV, the results are presented in the supplement. As illustrated, the GAN model trained from scratch struggles to capture the correct structure of the distributions across all energy values. In particular, Fig.~\ref{fig:nuAr40-10k} highlights the model's failure to reproduce the peak in the $W$ distribution, whereas the model pre-trained on carbon data achieves a remarkably accurate fit. While using a larger, more informative training dataset improves the performance of the model trained from scratch, it still does not reach the accuracy achieved through TL.

Similarly to the case (A), we perform an analogous analysis for the case (B), involving $\bar\nu_\mu$–carbon interactions. The results are summarized in Table~\ref{tab:antinu-10k}. Models trained using transfer learning consistently outperform those trained from scratch, see Table \ref{tab:antinu-10k}. When trained on a larger and more informative dataset, model performance improves significantly in both scenarios.

Fig.~\ref{fig:antinuC12-10k} and corresponding figures (Figs. \ref{fig:antinuC12-10k_supplemet} and \ref{fig:antinuC12-100k_supplemet}) in the supplement show histograms generated by our models and by \nuwro{} for energies of $500$ MeV, and $1$, $3$, $5$ GeV, respectively. In contrast to the neutrino–argon case, even the models trained from scratch can capture the resonance structures present in the distributions. Nevertheless, models trained with TL  demonstrate superior efficiency and performance.

In the final case (C), we optimize two models, one trained from scratch and the other using TL, using only a small training dataset. Table~\ref{tab:nuC12alt-10k} and Fig.~\ref{fig:nuC12alt-10k} (and Fig. \ref{fig:nuC12alt-10k_supplement}) summarize our results. The model trained from scratch struggles to accurately capture the QE peak and performs poorly in the resonance region. In contrast, the model trained via transfer learning performs significantly better, despite being initially trained on data generated by a completely different \nuwro{} model.

\begin{figure}\centering
    \includegraphics[width=0.46\textwidth]{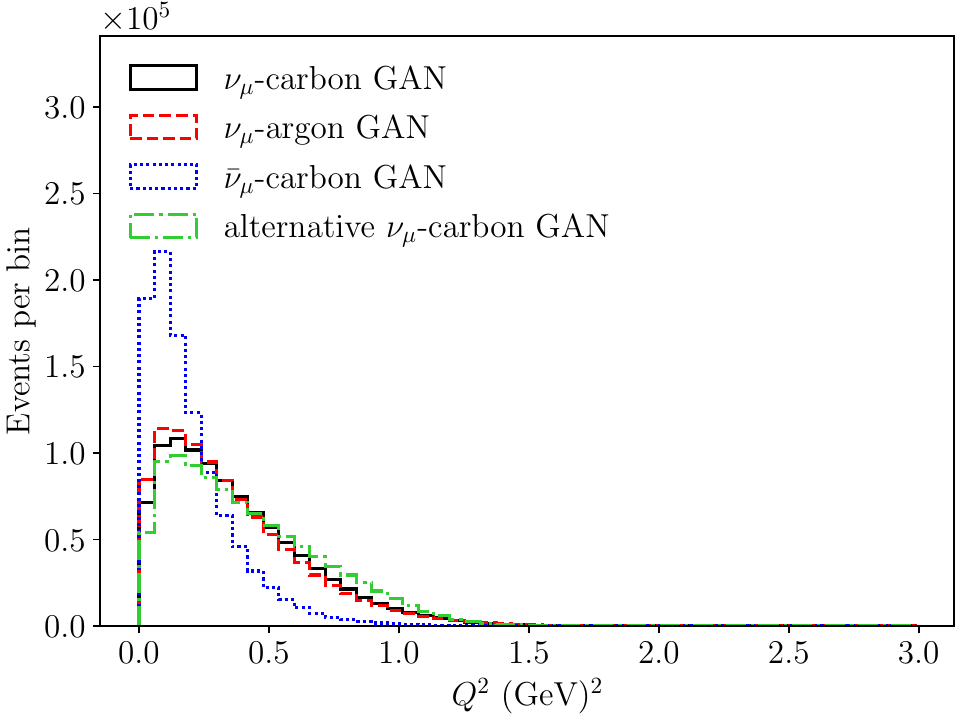}
    \includegraphics[width=0.46\textwidth]{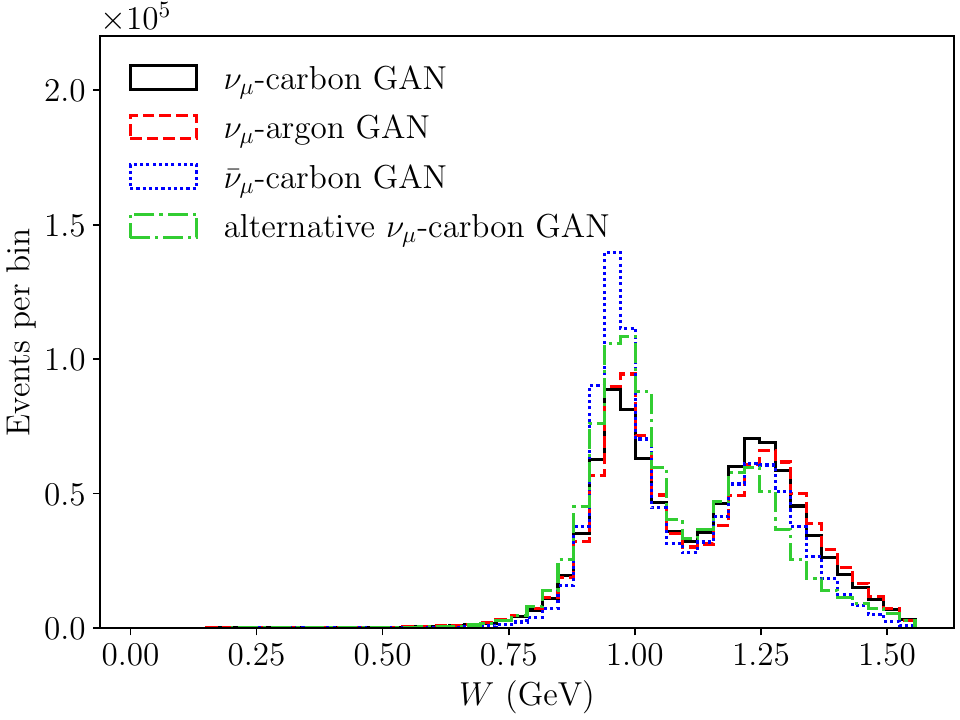} \caption{Event distributions in $Q^2$ and $W$ for $E_\nu=1$~GeV were obtained from our GAN models for neutrino-carbon, neutrino-argon, antineutrino-carbon, and an alternative model of neutrino-carbon scattering processes. The models were trained using 10,000 events and transfer learning.
    \label{fig:gan_comparison_models}}
\end{figure}

As discussed in Sec.~\ref{sec:method}, the event distributions for the four considered processes  exhibit significant differences arising from the use of spectral functions versus the LFG model, as well as from the different ways in which neutrinos and antineutrinos interact with nucleons. This is illustrated in Fig.~\ref{fig:gan_comparison_models}, where we show the event distributions as functions of $Q^2$ (top panel) and $W$ (bottom panel).

In particular, the antineutrino QE peak is more pronounced, though narrower, than in the case of neutrino–nucleus interactions. The neutrino–carbon and neutrino–argon distributions look similar; however, the QE peak is slightly higher for the scattering off the argon nucleus. In contrast, the alternative nuclear model based on the LFG approximation, combined with a different resonance description, produces event distributions with visibly different shapes and notably different strengths of both the QE and the $\Delta(1232)$ resonance peaks.

\section{Summary}
\label{sec:summary}

We have trained GAN models using transfer learning from a pre-trained GAN originally developed on the inclusive CC $\nu_\mu$-carbon scattering MC dataset, adapting them to various other scattering scenarios, including inclusive CC $\bar{\nu}_\mu$–carbon, $\nu_\mu$-argon, and $\nu_\mu$–carbon with alternative interaction models. These models were compared against their counterparts trained from scratch. We observed that initializing from the pre-trained model consistently outperforms training from scratch, particularly when datasets are small or poorly sample the kinematic distributions of the underlying physical processes. This suggests that the pre-trained model captures essential features of neutrino scattering dynamics.

To more thoroughly evaluate the effectiveness of transfer learning, we also trained the same network architecture on datasets with a significantly larger number of events. This demonstrated that the inferior performance of models trained from scratch on smaller datasets is not due to architectural limitations. In this larger-statistics scenario, the model trained from scratch begins to catch up with the transfer-learned model. Nevertheless, the latter continues to show some advantages, particularly in regions where the dataset remains a poor representation of the relevant kinematic phase space.

From a technical standpoint, the frozen initial layers of the generator primarily reshape the latent Gaussian distribution into a generic lepton-kinematic manifold that should be rather universal for various neutrino-induced processes. The subsequent unfrozen layers adapt this manifold to process-specific features. Therefore, the first block is effectively universal for all neutrino and antineutrino scattering tasks, while fine-tuning later layers successfully transfers the model to the target domain. In the case of the discriminator, which takes lepton kinematics as input, the first frozen block is also rather universal and extracts/processes basic features that subsequent blocks transform. These blocks are adapted to a particular type of process. The fact that TL from neutrino-carbon to neutrino–argon and antineutrino-carbon succeeds demonstrates that the GAN primarily encodes universal kinematic constraints and the overall QE and $\Delta$-resonance pattern, while target-dependent spectral-function details are corrected by fine-tuning in subsequent layers.

There are several physical arguments supporting our approach. First: in all the discussed processes, the underlying nuclear physics is the same. The carbon and argon targets differ in nuclear structure,  but, as we showed in our previous paper, the TL approach can be less effective when the differences are radical, as in the case of $^3$He and $^{12}$C targets. However, even in this case, the TL scenario in full-tuning mode still works efficiently. Indeed, when one looks at the QE and $\Delta$ peaks for carbon and argon, they differ but not dramatically. Hence, the carbon-scattering model can be considered a good \textit{a posteriori} model for argon.  A similar argument works for neutrino and antineutrino models. In this case, the target is the same, but the neutrino and antineutrino probe nuclear structure, which is reflected in some differences between the resulting distributions. However, minor tuning  allows us to obtain one model from the other.

Our findings highlight the promising potential of transfer learning in developing  accurate models for (anti)neutrino–nucleus scattering. In particular, one can start from a model trained on MC simulation data and adapt it to more realistic scenarios using comparatively limited data. This approach offers a practical path forward, especially when experimental statistics are sparse. Further investigation in this direction is planned for future work, as well as extending the method to full final-state hadrons, which is part of the ongoing work.

\begin{acknowledgments}
This research is partly (K.M.G., A.M.A., J.T.S.) or fully (B.E.K., J.L.B, H.P., R.D.B.) supported by the Na{\-}tional Science Centre under grant UMO-2021/41/B/ST2/ 02778. J.L.B. and J.T.S. are also supported by the Polish Ministry of Science Grant No. 2022/WK/15. K.M.G is partly supported by the ``Excellence Initiative – Research University" for the years 2020-2026 at the University of Wroc\l aw.
\end{acknowledgments}

\clearpage
\onecolumngrid

\section*{Supplemental Material}

\setcounter{section}{0}
\setcounter{subsection}{0}
\setcounter{equation}{0}
\setcounter{figure}{0}
\setcounter{table}{0}

\renewcommand{\thesection}{S\arabic{section}}
\renewcommand{\thesubsection}{\thesection.\arabic{subsection}}
\renewcommand{\theequation}{S\arabic{equation}}
\renewcommand{\thefigure}{S\arabic{figure}}
\renewcommand{\thetable}{S\arabic{table}}

\renewcommand{\thesection}{S\arabic{section}}

\renewcommand{\thefigure}{S\arabic{figure}}

\section{Event distributions}
\label{Supplement:S1}

\begin{figure*}[hbtp]\centering
    \includegraphics[width=0.46\textwidth]{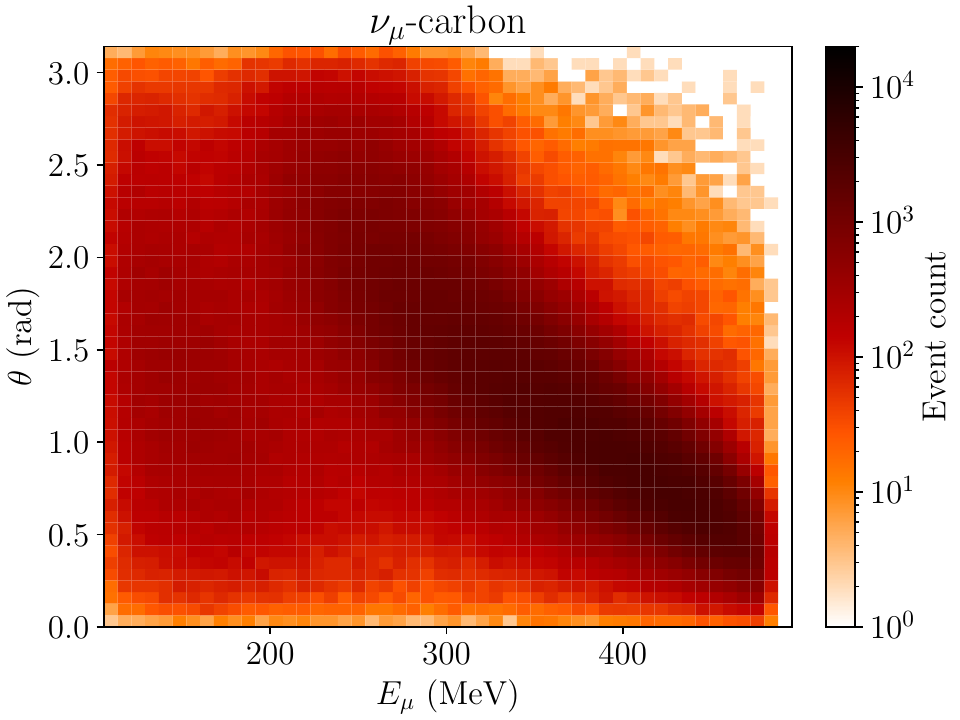}
    \includegraphics[width=0.46\textwidth]{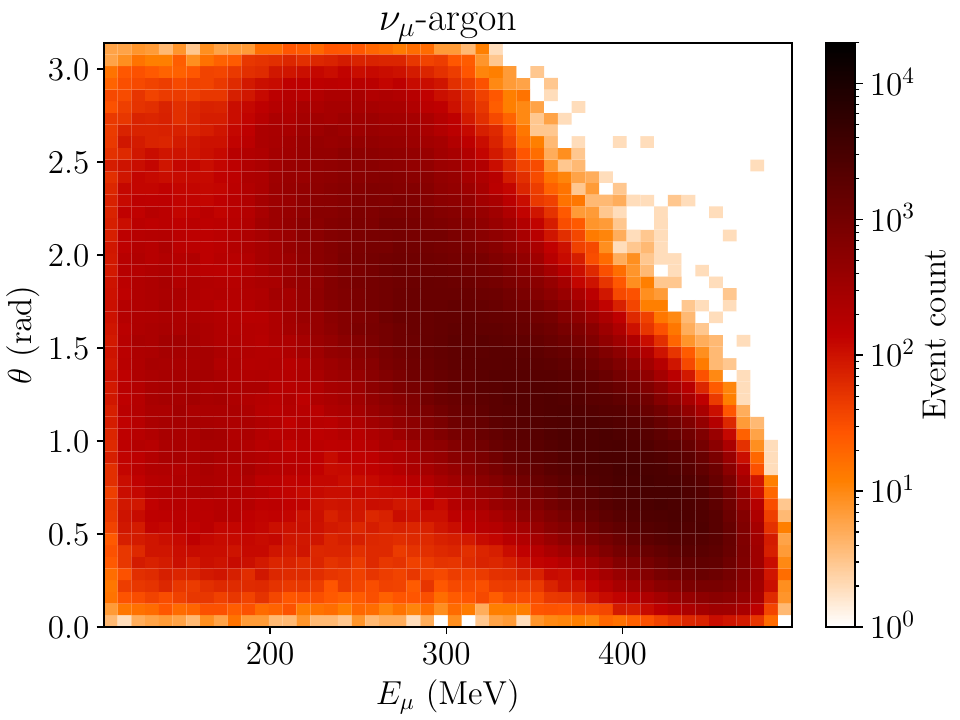}
    \caption{Test data samples $(E_\mu,\theta)$ distributions for $E_\nu = 500$~MeV. On the left, we have the original $\nu_\mu$-carbon scattering; on the  right, the $\nu_\mu$-argon scattering.
    \label{fig:test_samples_appendix}
    }
\end{figure*}

In the main paper, we show the distribution of events, in proxy variables, for all the processes we study. Here, in Fig.~\ref{fig:test_samples_appendix}, we additionally present the analogical distributions but in physical variables $(E_\mu, \theta)$ for neutrino-carbon and neutrino-argon scattering. Spectral functions describe the carbon and argon nuclei from Refs. \cite{Benhar:1994hw} and \cite{PhysRevD.105.112002,PhysRevD.107.012005}, respectively.

\newpage

\section{Reconstructed hadronic invariant mass $W$-distributions}
\label{Supplement:S2}

\begin{figure*}[btp]\centering
    \includegraphics[width=0.46\textwidth]{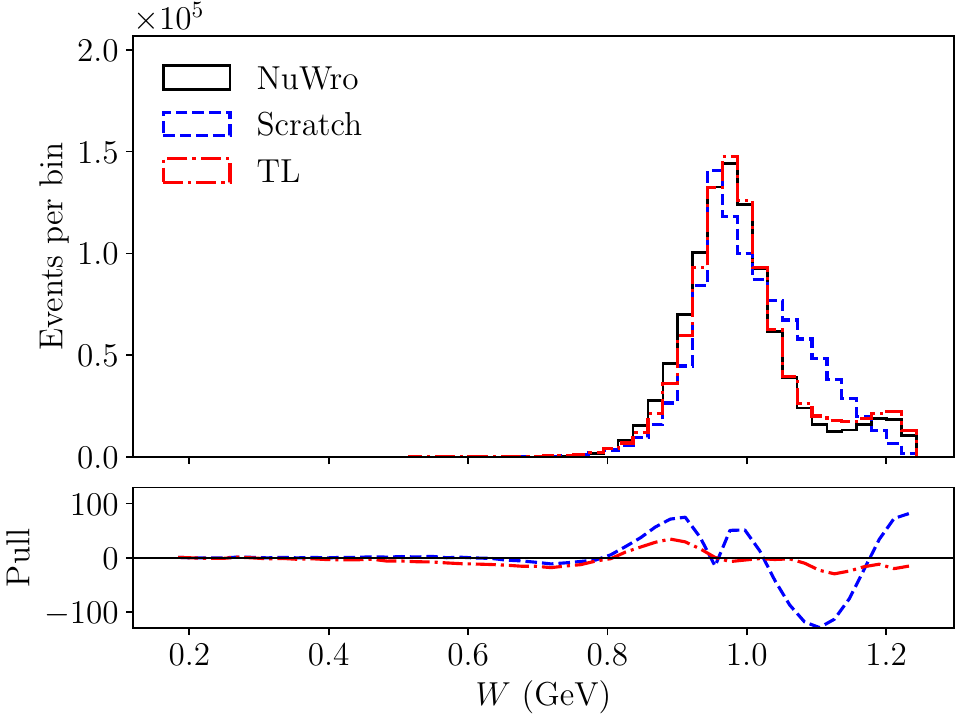}
    \includegraphics[width=0.46\textwidth]{nu_ar40_10kevents_wexp_1000mev}
    \includegraphics[width=0.46\textwidth]{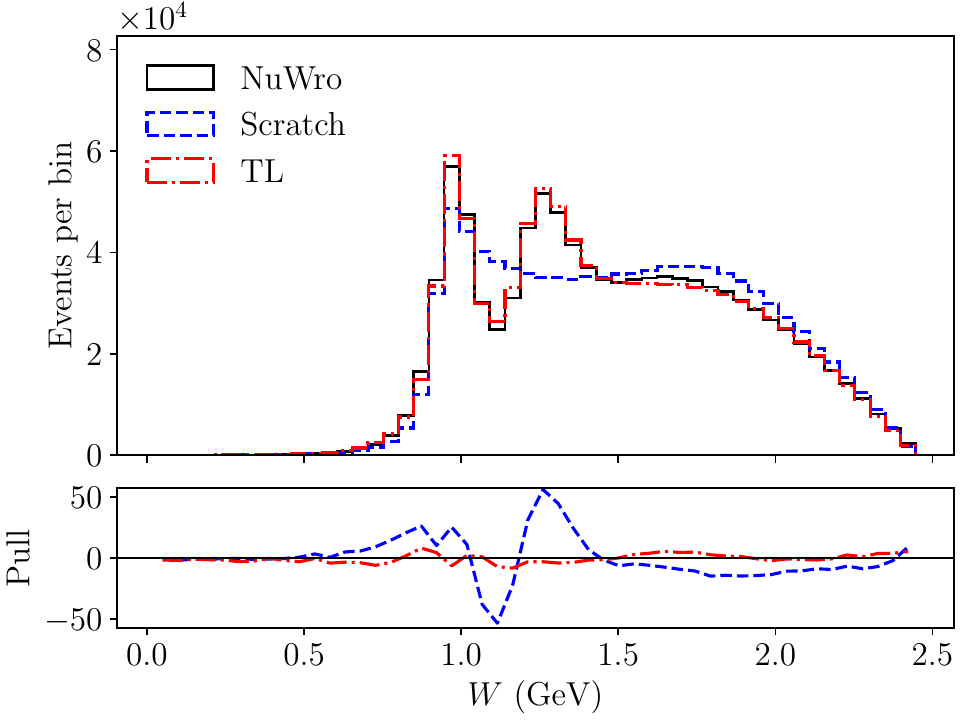}
    \includegraphics[width=0.46\textwidth]{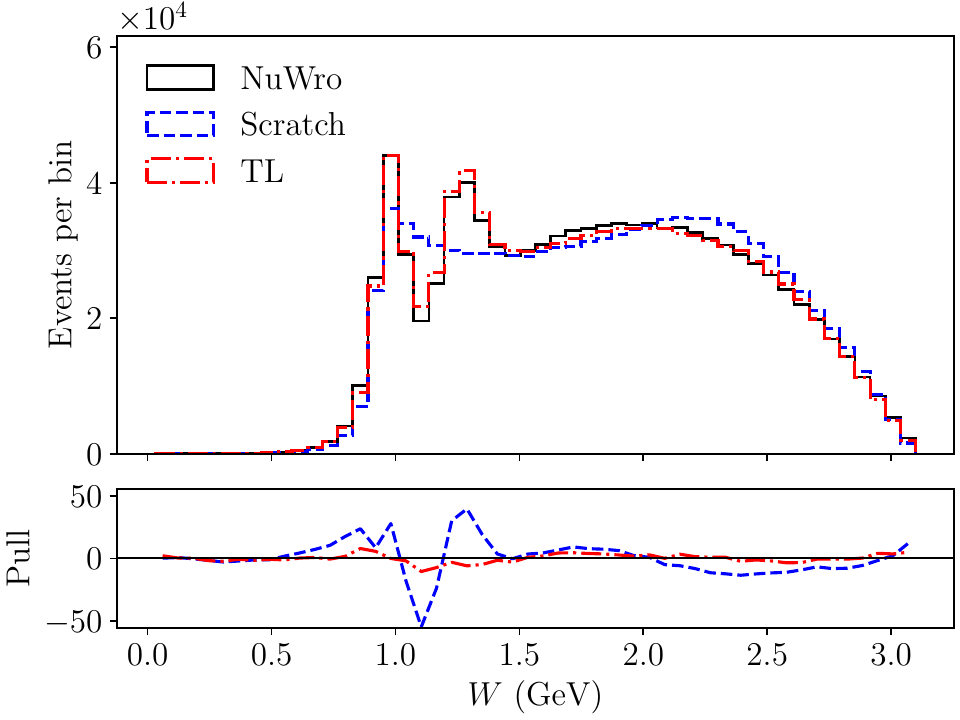}
    \caption{Reconstructed hadronic invariant mass $W$-distributions, for inclusive CC $\nu_\mu$-argon scattering, generated by \nuwro{} and GAN model trained from scratch and using TL, for the optimization  with 10k events. We present the results for $E_\nu = 500$ MeV (top left), $1$ GeV (top right), $3$ GeV (bottom left), and $5$ GeV (bottom right).
    \label{fig:nuAr40-10k_supplement}}
\end{figure*}

\begin{figure*}\centering
    \includegraphics[width=0.46\textwidth]{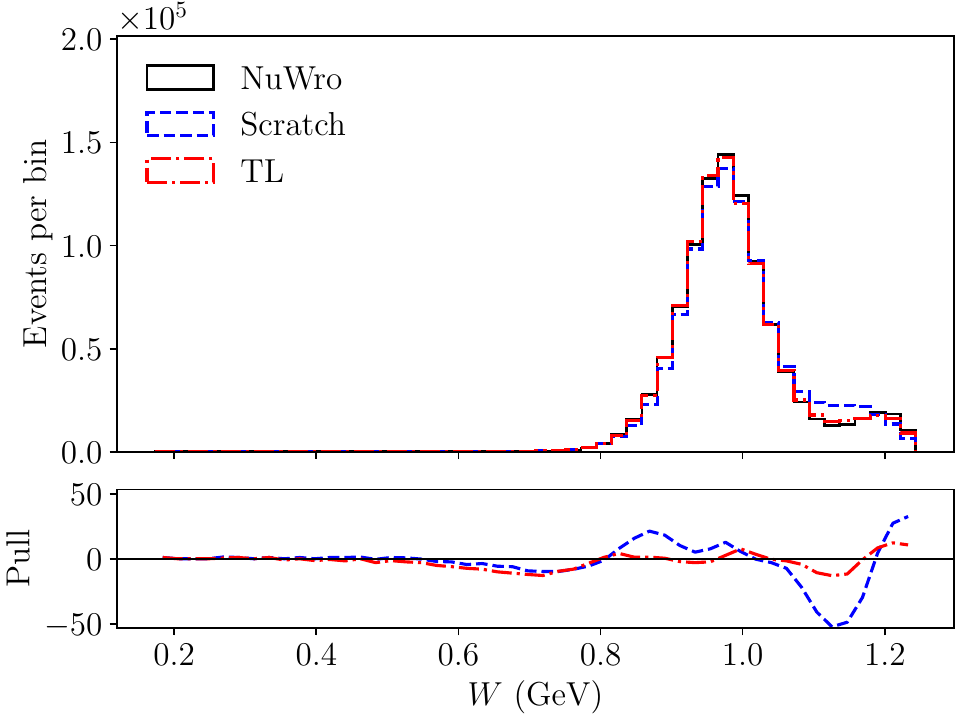}
    \includegraphics[width=0.46\textwidth]{nu_ar40_100kevents_wexp_1000mev}
    \includegraphics[width=0.46\textwidth]{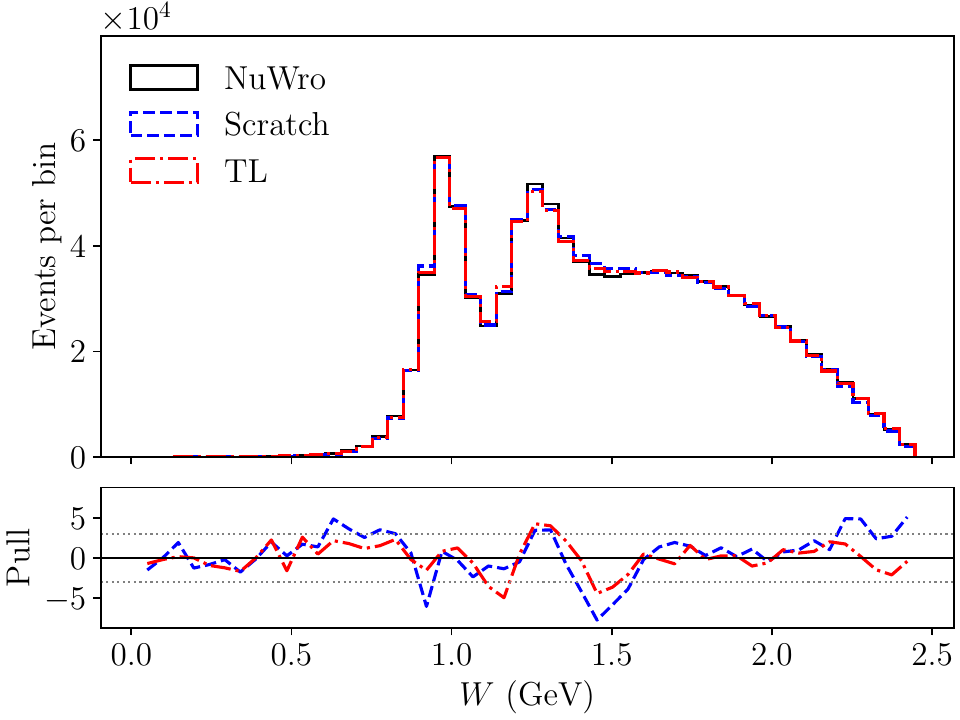}
    \includegraphics[width=0.46\textwidth]{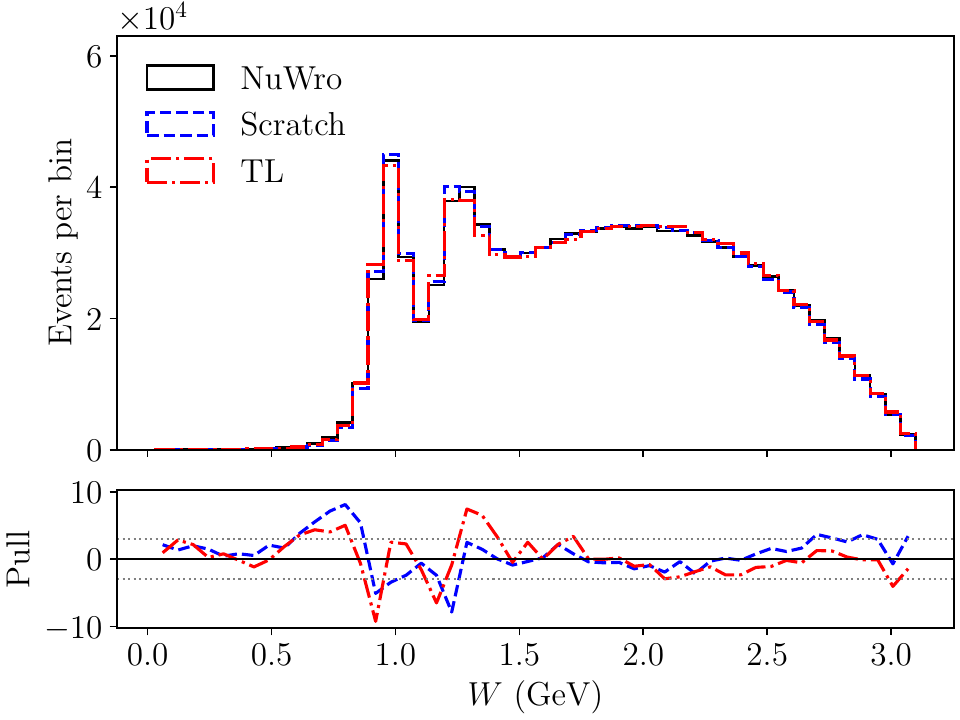}
    \caption{Same as in Fig.~\ref{fig:nuAr40-10k_supplement} but for CC $\nu_\mu$-argon scattering and the training dataset with 100,000 events.
    \label{fig:nuAr40-100k_supplement}}
\end{figure*}

\begin{figure*}\centering
    \includegraphics[width=0.46\textwidth]{antinu_c12_10kevents_wexp_500mev}
    \includegraphics[width=0.46\textwidth]{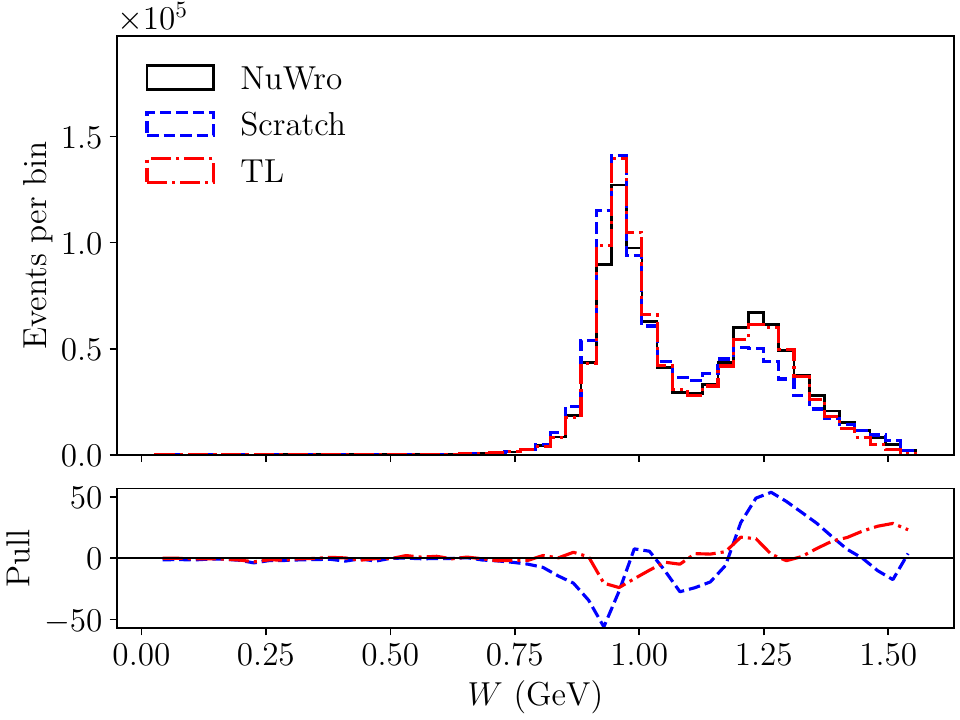}
    \includegraphics[width=0.46\textwidth]{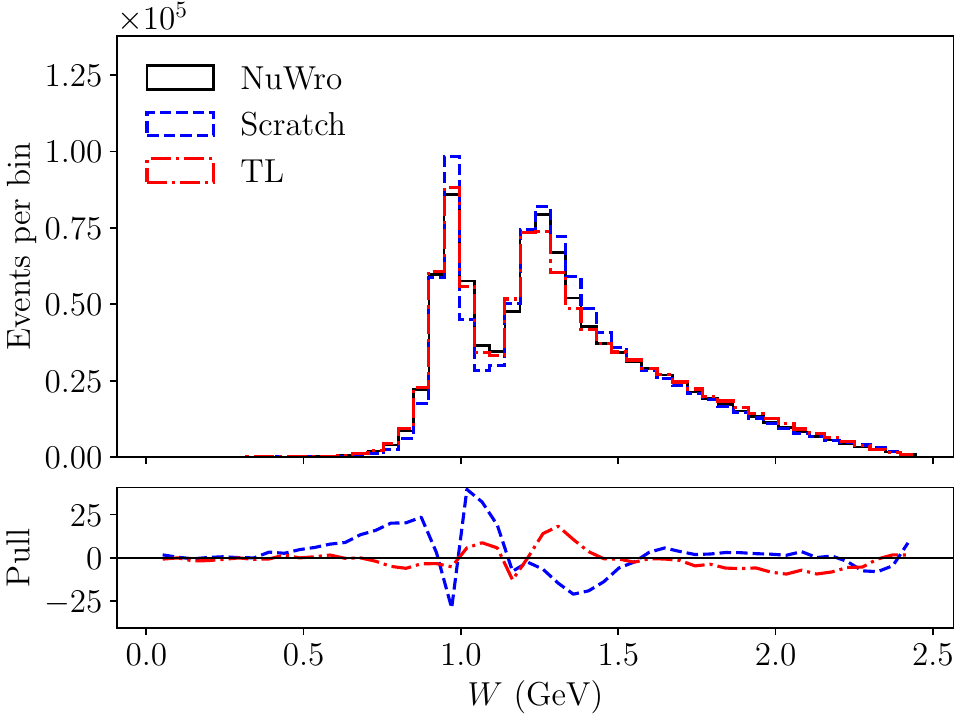}
    \includegraphics[width=0.46\textwidth]{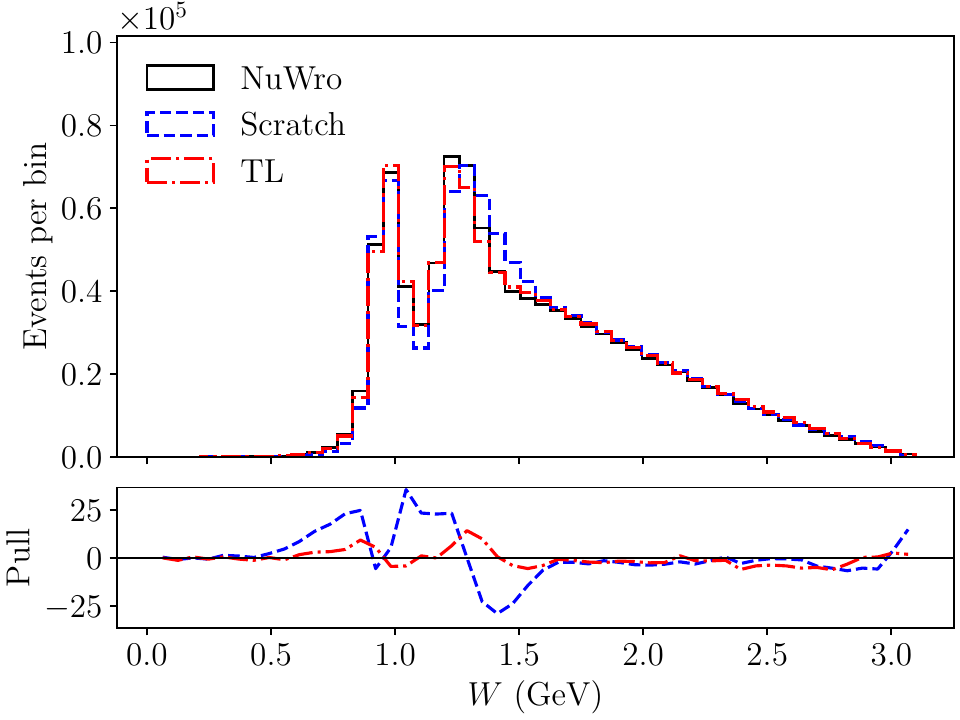}
    \caption{Same as in Fig.~\ref{fig:nuAr40-10k_supplement} but for CC $\bar\nu_\mu$-carbon scattering and the training dataset with 10,000 events.
    \label{fig:antinuC12-10k_supplemet}}
\end{figure*}

\begin{figure*}\centering
    \includegraphics[width=0.46\textwidth]{antinu_c12_100kevents_wexp_500mev}
    \includegraphics[width=0.46\textwidth]{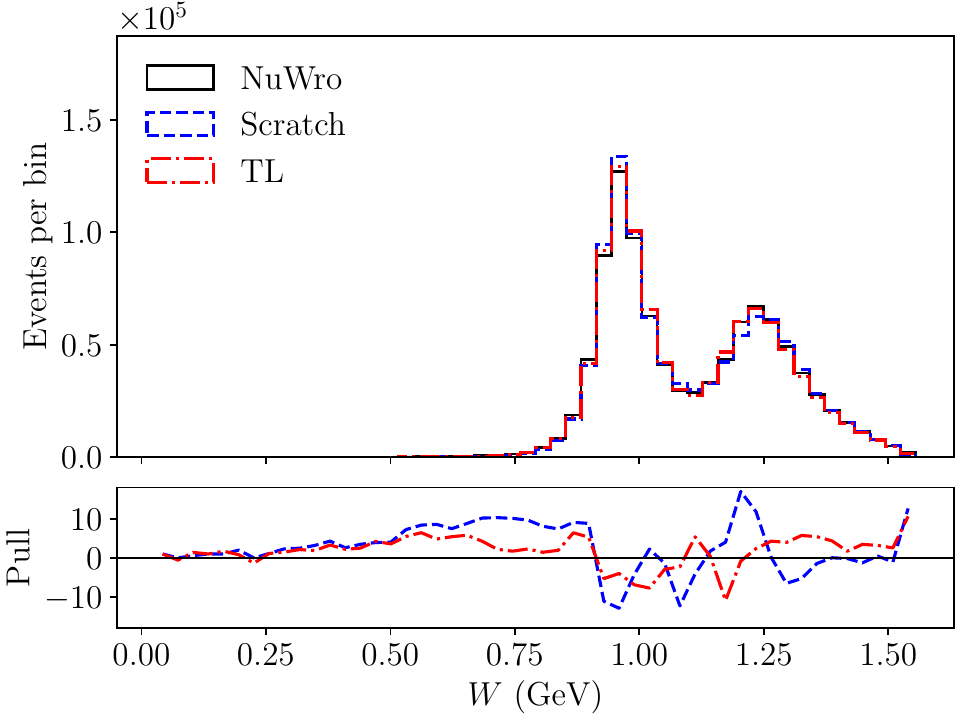}
    \includegraphics[width=0.46\textwidth]{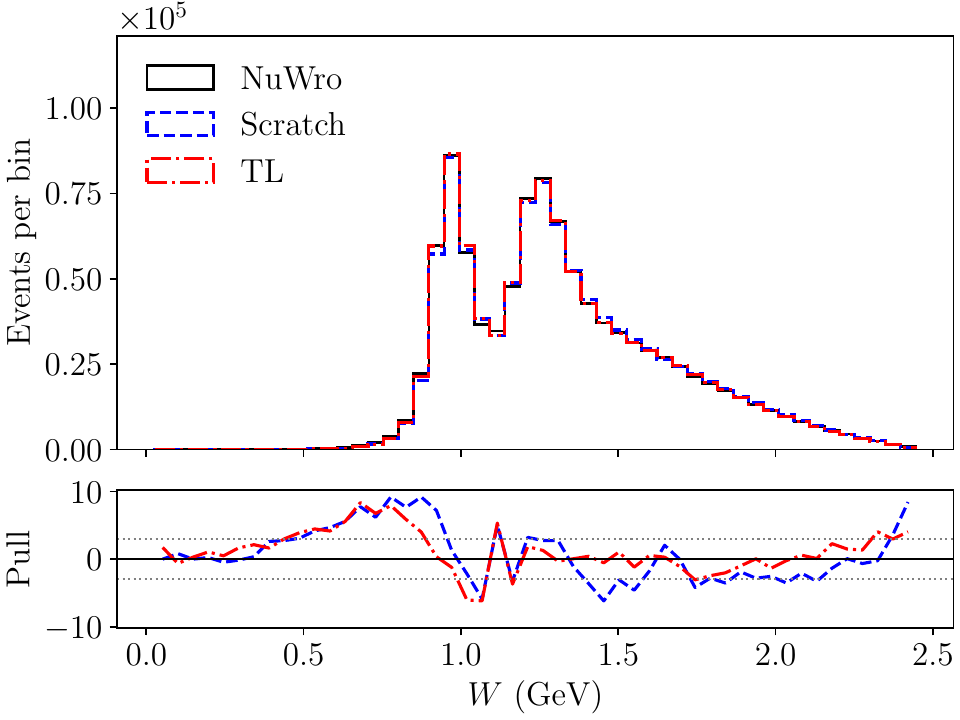}
    \includegraphics[width=0.46\textwidth]{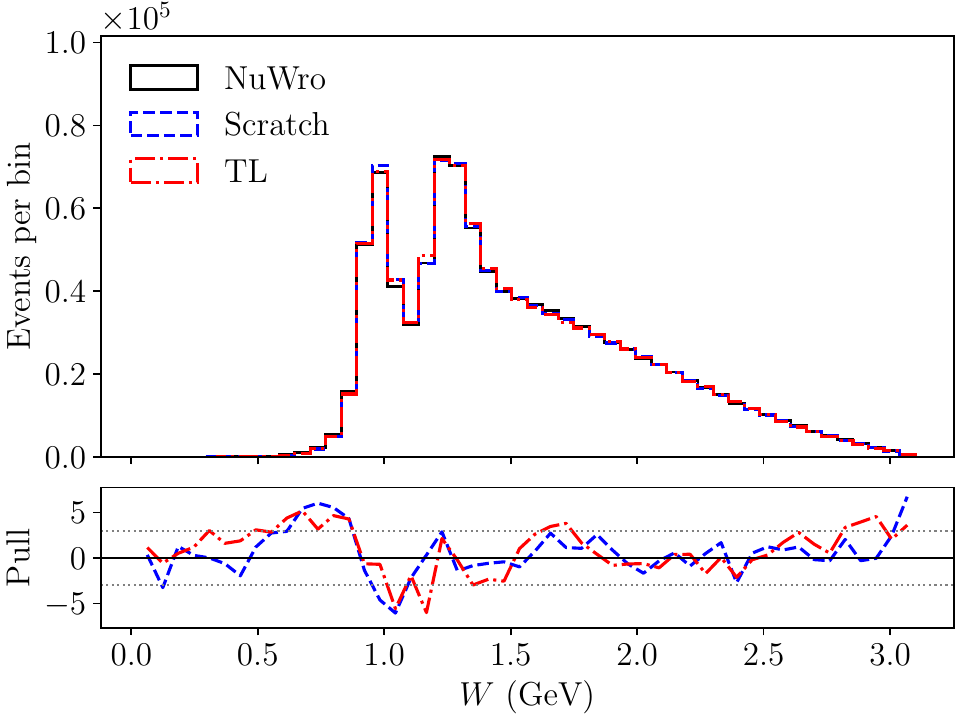}
    \caption{Same as in Fig.~\ref{fig:nuAr40-10k_supplement} but for CC $\bar\nu_\mu$-carbon scattering and the training dataset with 100,000 events.
    \label{fig:antinuC12-100k_supplemet}}
\end{figure*}

\begin{figure*}\centering
    \includegraphics[width=0.46\textwidth]{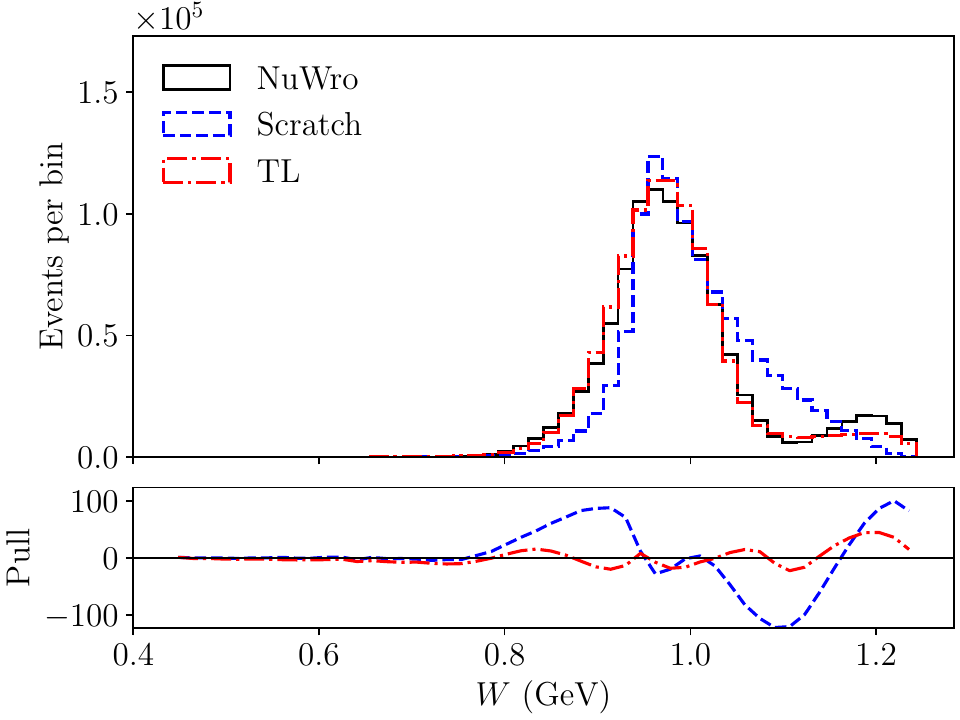}
    \includegraphics[width=0.46\textwidth]{nu_c12_alt_10kevents_wexp_1000mev}
    \includegraphics[width=0.46\textwidth]{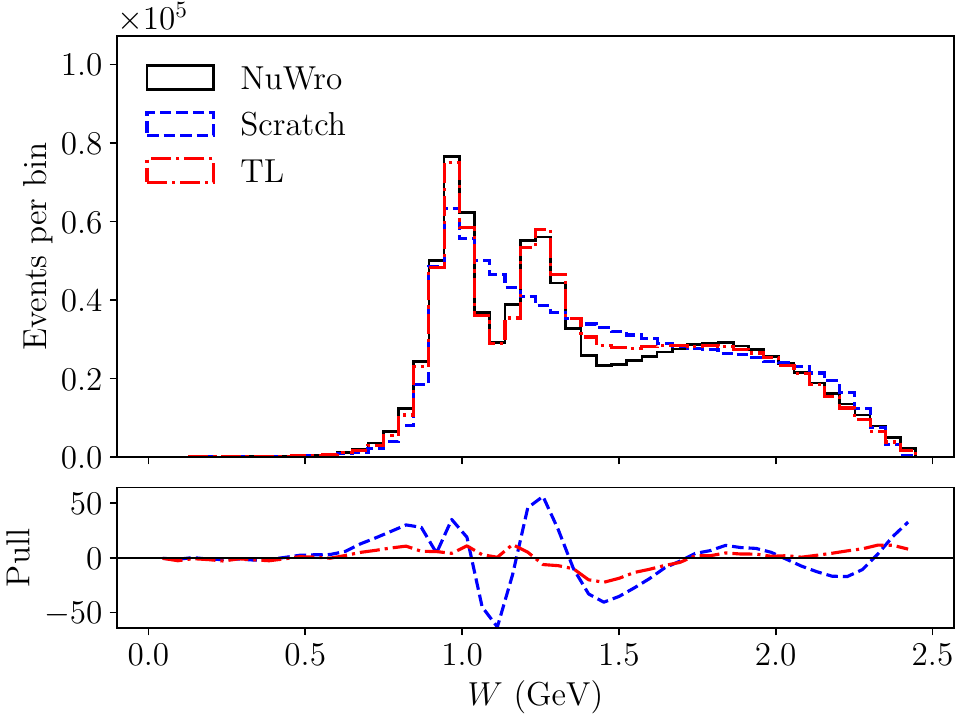}
    \includegraphics[width=0.46\textwidth]{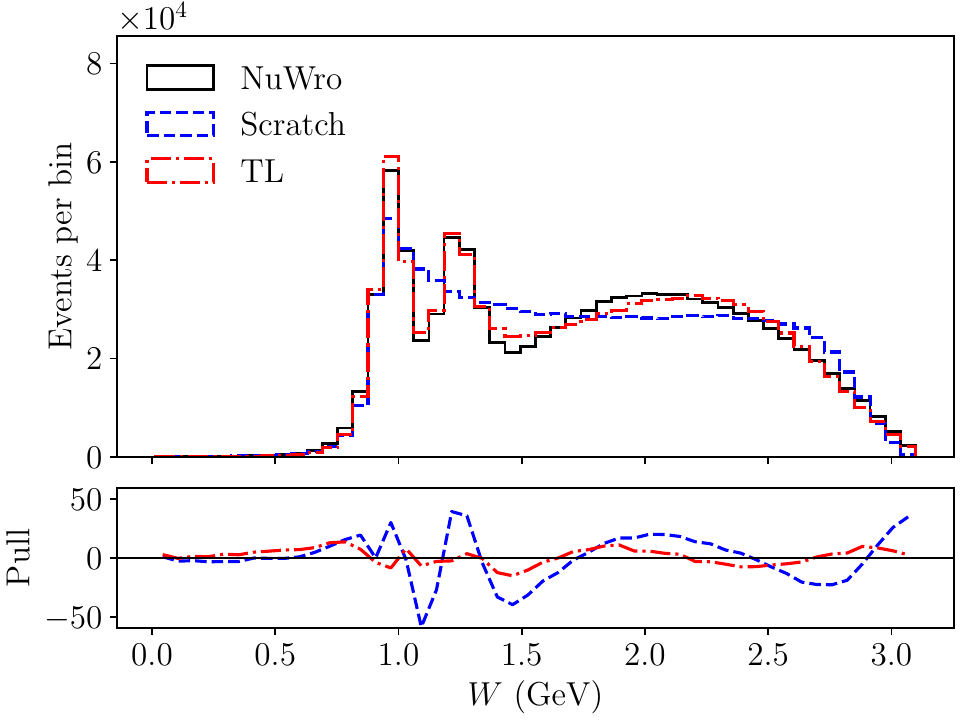}
    \caption{Same as in Fig.~\ref{fig:nuAr40-10k_supplement} but for CC $\nu_\mu$-carbon (alternative interaction) scattering and the training dataset with 10,000 events. 
    \label{fig:nuC12alt-10k_supplement}}
\end{figure*}

This section contains the distribution of events generated by GANs for neutrino-argon scattering for models trained on 10,000 (Fig.~\ref{fig:nuAr40-10k_supplement}) and 100,000 (Fig.~\ref{fig:nuAr40-100k_supplement}) events, respectively. Similarly in 
Figs.~\ref{fig:antinuC12-10k_supplemet} and \ref{fig:antinuC12-100k_supplemet} show the distribution of events generated by the GANs trained on 10,000 and 100,000 events for antineutrino-carbon scattering. Eventually, the Fig.~\ref{fig:nuC12alt-10k_supplement} shows histograms for events generated by the model of neutrino-carbon trajectory on 10,000 events obtained from \nuwro{} using an alternative interaction model.



\clearpage

\section{Loss function}
\label{Appendix:loss}

\begin{figure*}[htbp]\centering
    \includegraphics[width=0.45\textwidth]{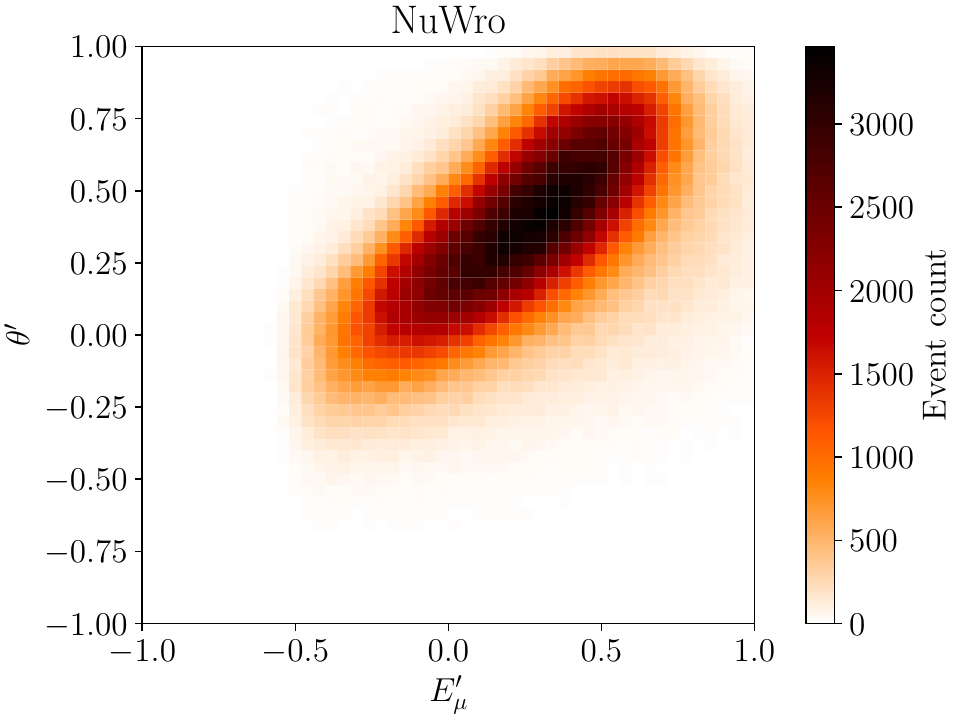}
    \includegraphics[width=0.45\textwidth]{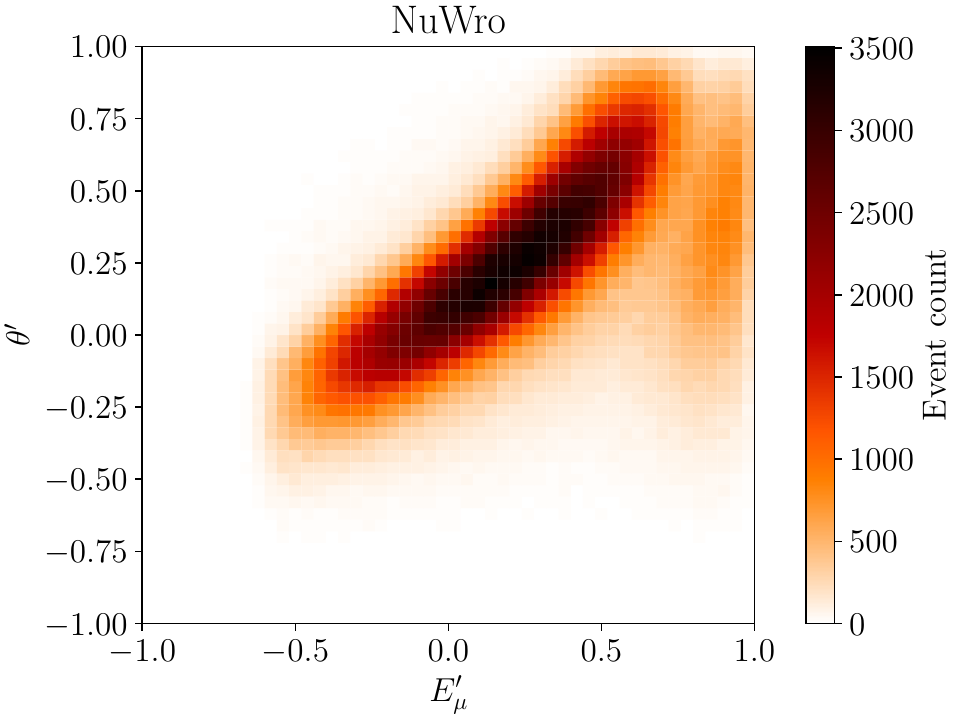}\\
    \includegraphics[width=0.45\textwidth]{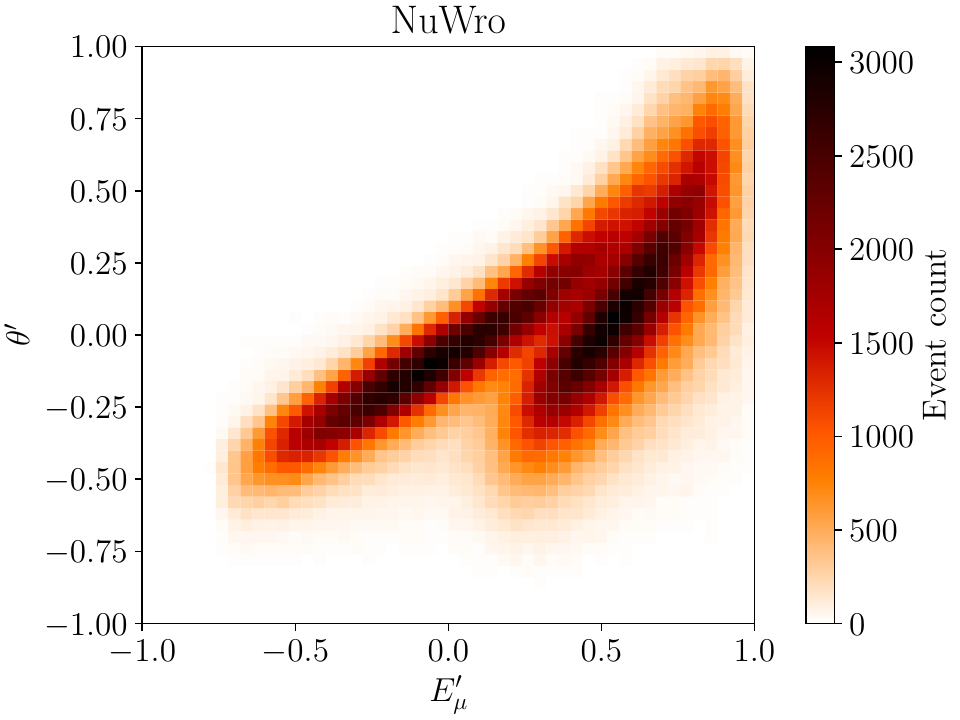}
    \includegraphics[width=0.45\textwidth]{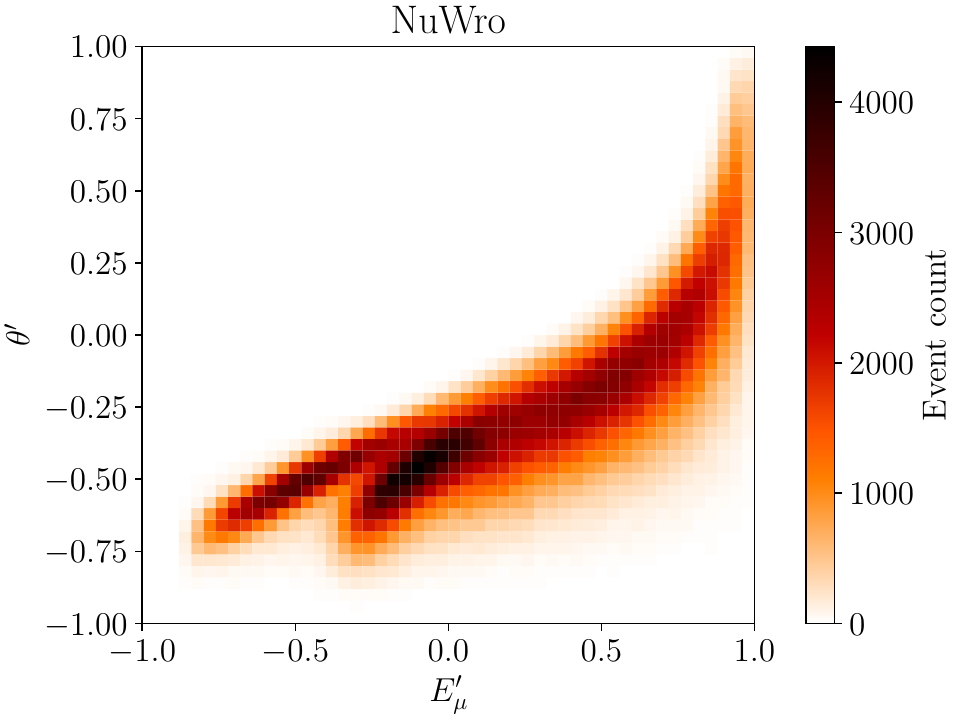}\\
    \includegraphics[width=0.45\textwidth]{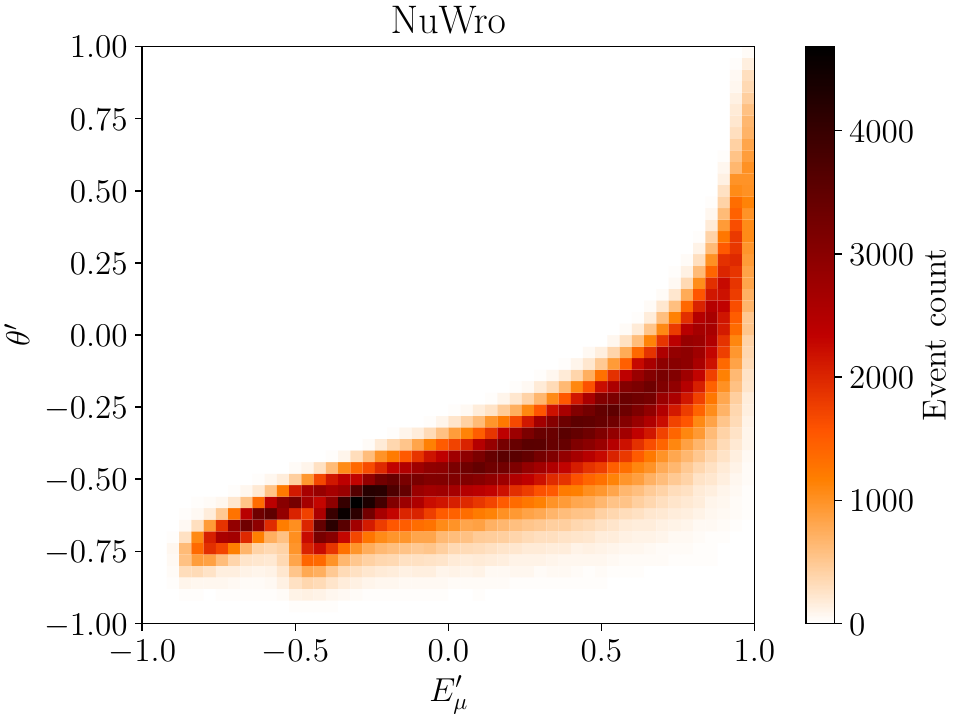}
    \includegraphics[width=0.45\textwidth]{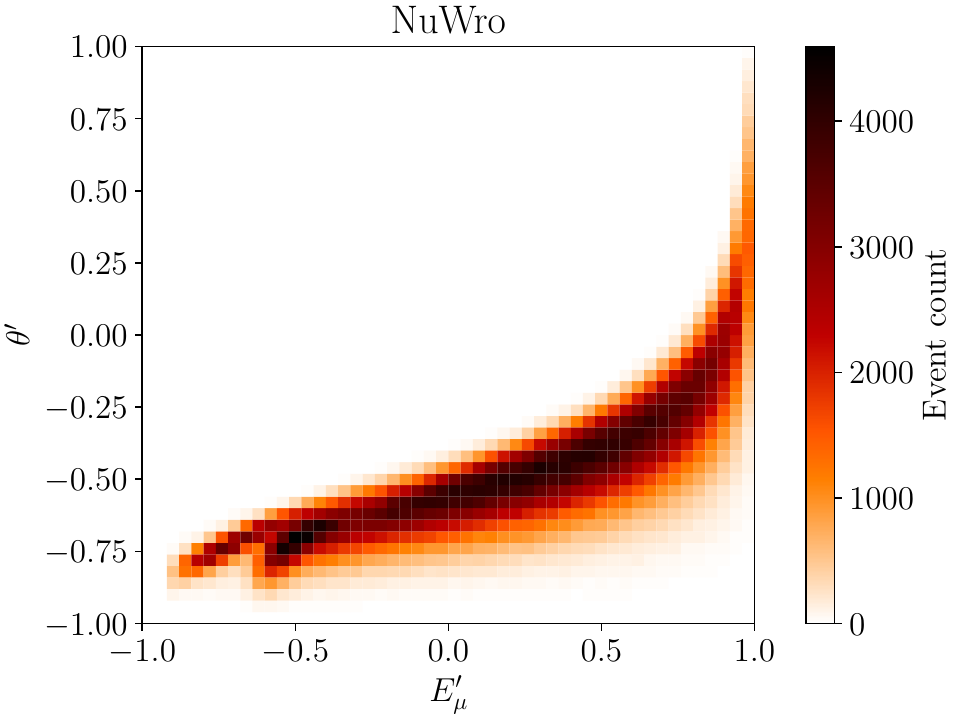}
    \caption{Kinematic distributions for inclusive CC $\nu_\mu$-carbon scattering for $E'_\mu$ and $\theta'$ variables computed for 1M events at different neutrino energies, according to \nuwro{}. From left to right and top to bottom distributions for energies $E_\nu =$ $0.35$, $0.5$, $1$, $3$, $5$, and $9$ GeV.\label{fig:nuwro}}
\end{figure*}

\begin{figure}[b]\centering
    \includegraphics[width=0.49\textwidth]{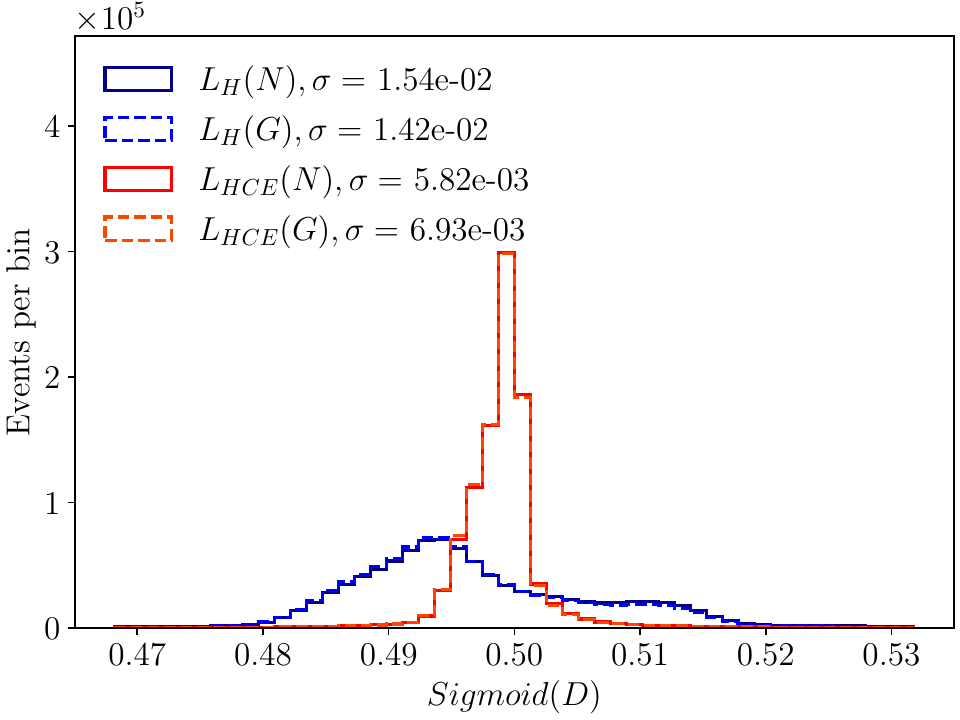}
    \includegraphics[width=0.49\textwidth]{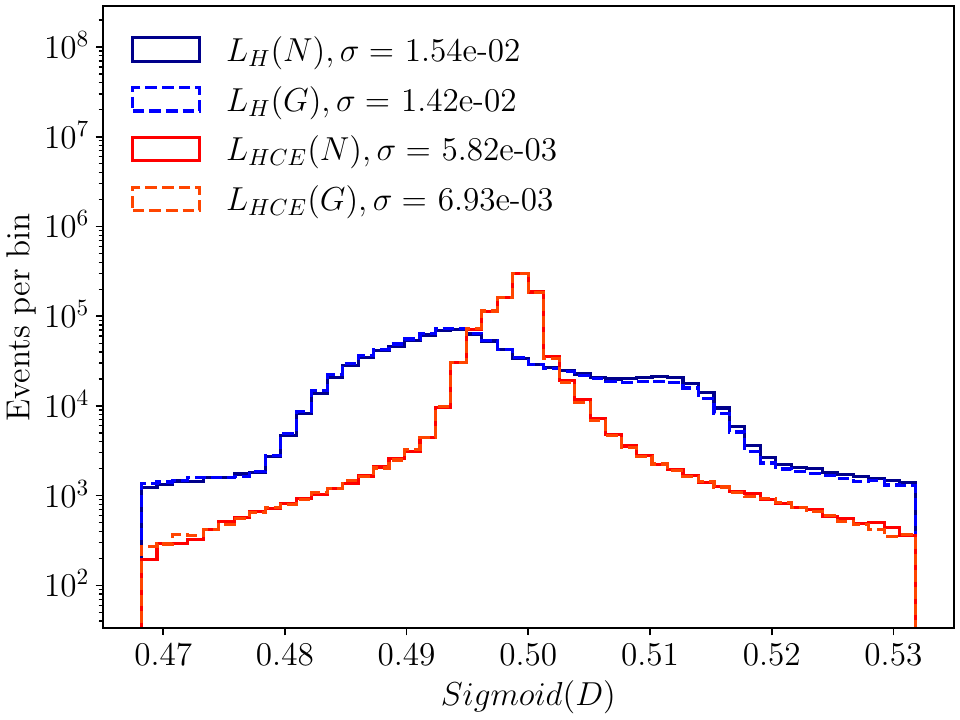}
    \caption{Probability distributions from the discriminator for samples generated by \nuwro{} and by the GAN trained with the heuristic loss $L_H$ [Eq. \ref{Eq:Glossheuristic}] and with our loss $L_{HCE}$ [Eq. \ref{Eq:Glosshce}], for $E_\nu = 350$ MeV. Here, $N$ and $G$ denote samples from \nuwro{} and the GAN, respectively.  \label{fig:350probs}}
\end{figure}
Typically, in most applications, the heuristic non-saturating loss~\cite{goodfellow2014generativeadversarialnetworks}
\begin{equation}
\label{Eq:Glossheuristic}
L_{H}(G) = -\frac{1}{B}\sum_{k=1}^B \log[D(G(\mathbf{x}k, E{\nu,k}), E_{\nu,k})]
\end{equation}
is considered, where $G$ and $D$ denote the generator and discriminator, respectively; $B$ is the mini-batch size; $\mathbf{x}_k$ is the latent space vector; and $E_{\nu,k}$ represents the energy of the interacting neutrino.

The rationale for using the loss (\ref{Eq:Glossheuristic}) stems from the fact that the maximization of the conventional cross-entropy loss,
\begin{equation}
\label{Eq:Glossce}
L_{CE}(G) = \frac{1}{B}\sum_{k=1}^B \log[1 - D(G(\mathbf{x}k, E{\nu,k}), E_{\nu,k})],
\end{equation}
tends to saturate early in the training process, as the discriminator quickly becomes adept at distinguishing generated events from real ones. Nonetheless, $L_H$ has a saturation region as well, just in the ``opposite side'' compared to $L_{CE}$.

However, we observe that employing the non-saturating loss in Eq.~\ref{Eq:Glossheuristic} can lead to uneven performance of the discriminator across different $E_\nu$ regions. Specifically, the GAN appears to favor optimization at higher neutrino energies, where the model benefits from smoother kinematic variations across adjacent energy bins. In contrast, at lower energies, slight variations in $E_\nu$ produce significant changes in the corresponding kinematic distributions, making training more challenging. The kinematic domains covered by the data for various neutrino energies are illustrated in Fig.~\ref{fig:nuwro}. At low $E_\nu$ energies, $350$ and $500$ MeV, mostly the quasi-elastic contribution is visible. In contrast, at higher energies, resonance contributions also play a significant role. Since higher-energy events dominate our training data, the models being trained first capture the structures associated with these higher energies.

\begin{figure}[b]\centering
    \includegraphics[width=0.49\textwidth]{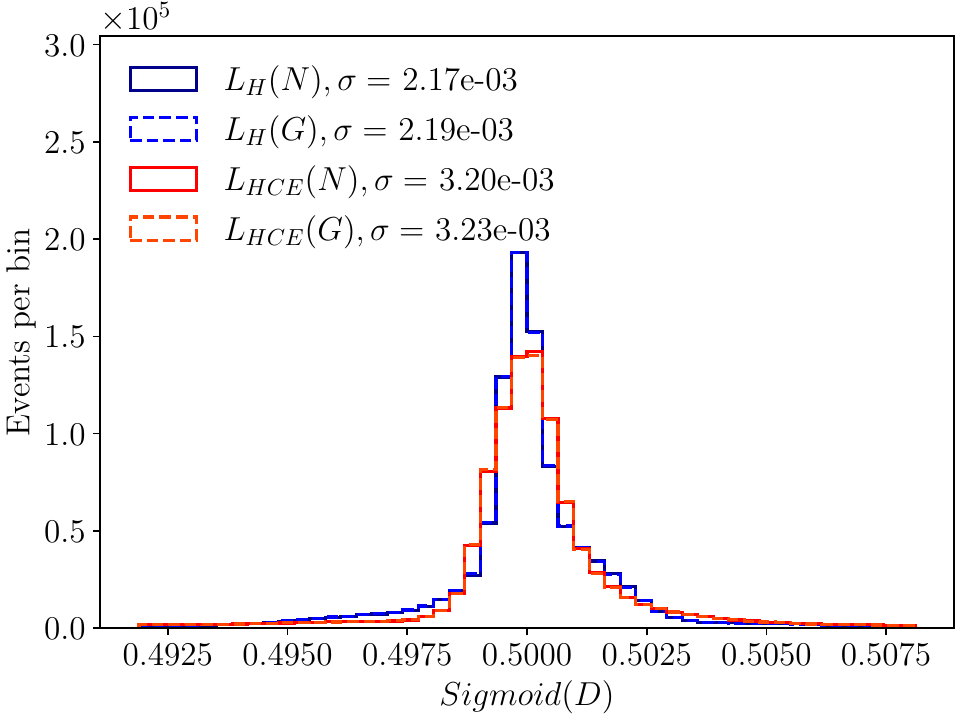}
    \includegraphics[width=0.49\textwidth]{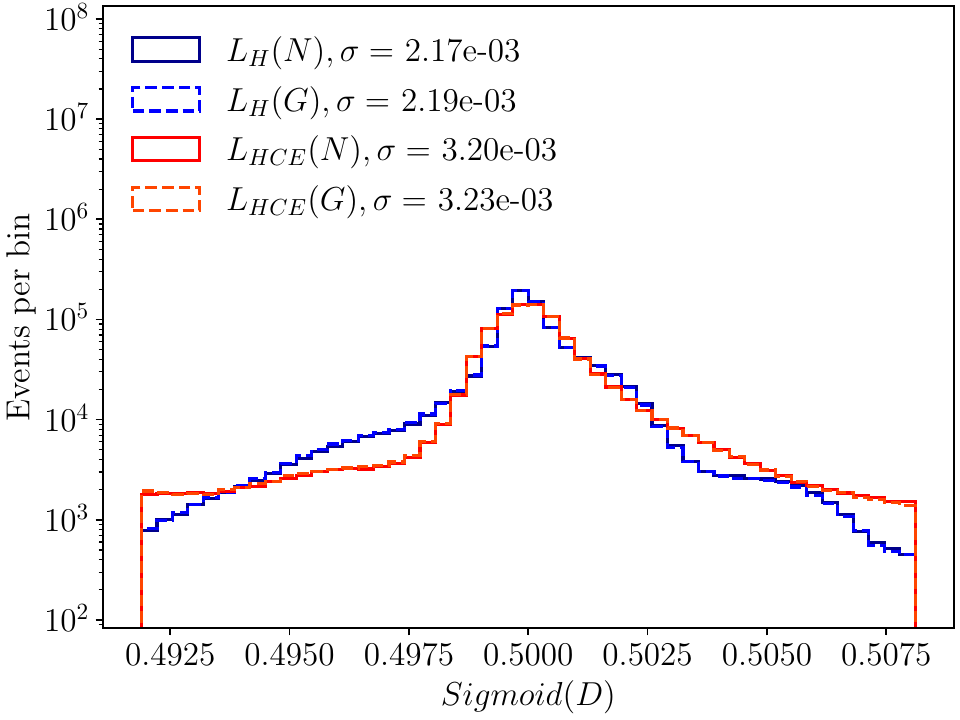}
    \caption{Same as Fig.~\ref{fig:350probs}, but for $E_\nu=5$ GeV.\label{fig:5000probs}}
\end{figure}

\begin{figure}[bthp]\centering
    \includegraphics[width=0.49\textwidth]{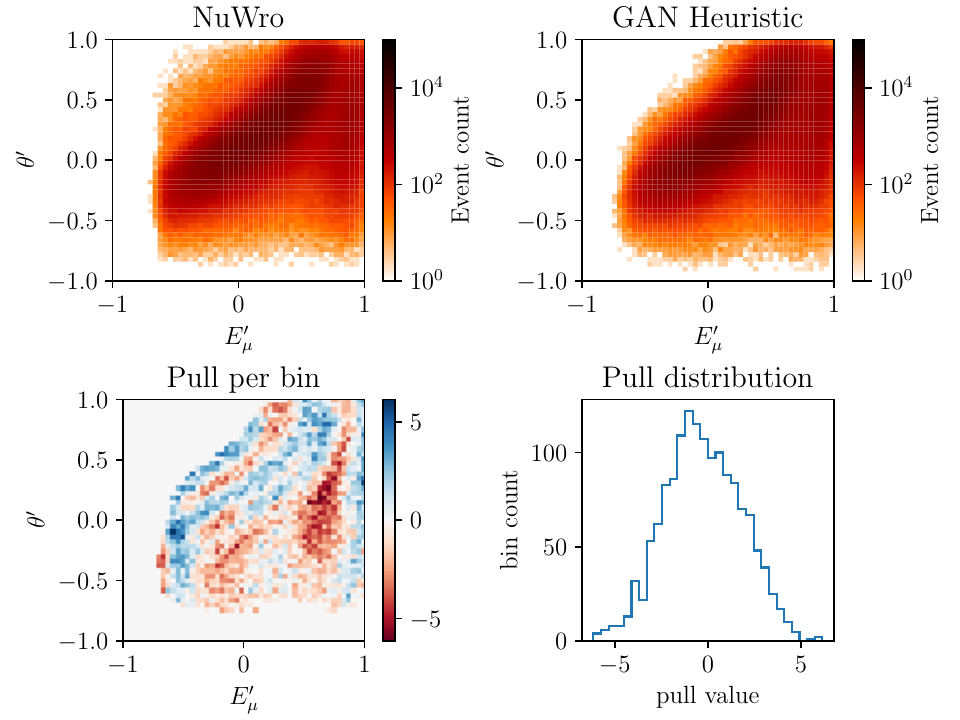}
    \includegraphics[width=0.49\textwidth]{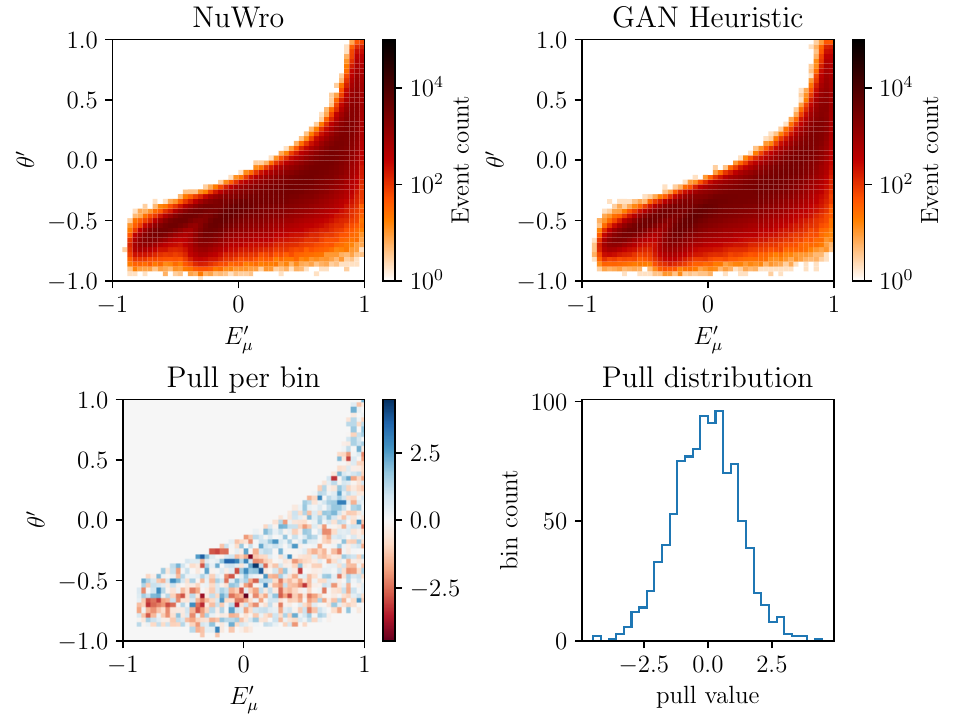}
    \caption{Kinematic distributions, pull per bin, and pull distributions for samples from \nuwro{} and GANs trained with the heuristic loss, for $E_\nu = 500$ MeV (first two rows) and $E_\nu = 3$ GeV (last two rows). \label{fig:thetaemuH}}
\end{figure}
\begin{figure}[bthp]\centering
    \includegraphics[width=0.49\textwidth]{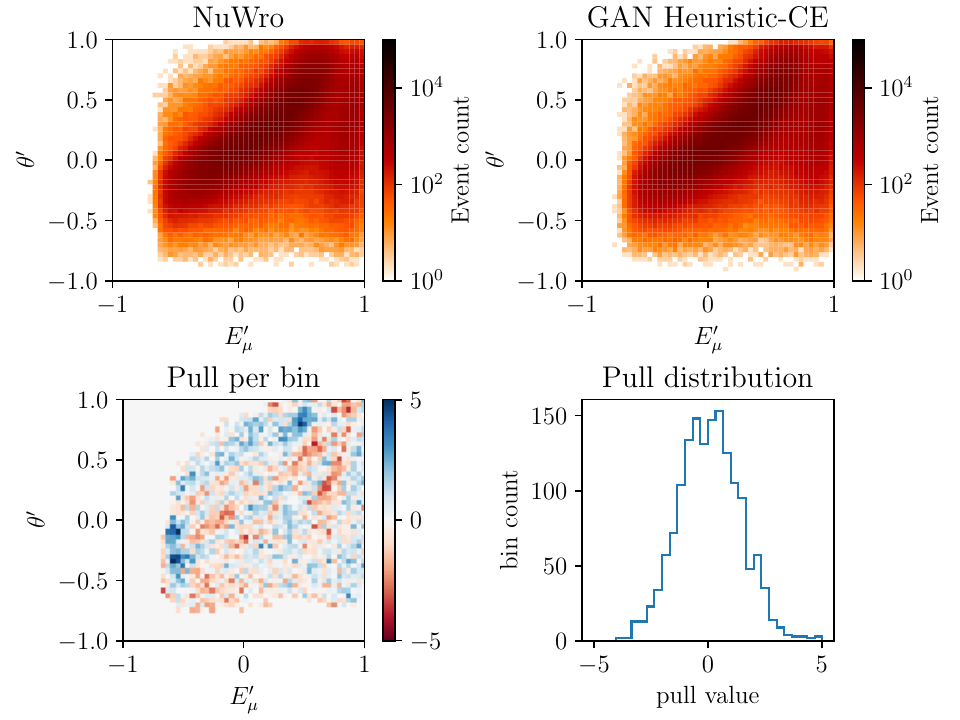}
    \includegraphics[width=0.49\textwidth]{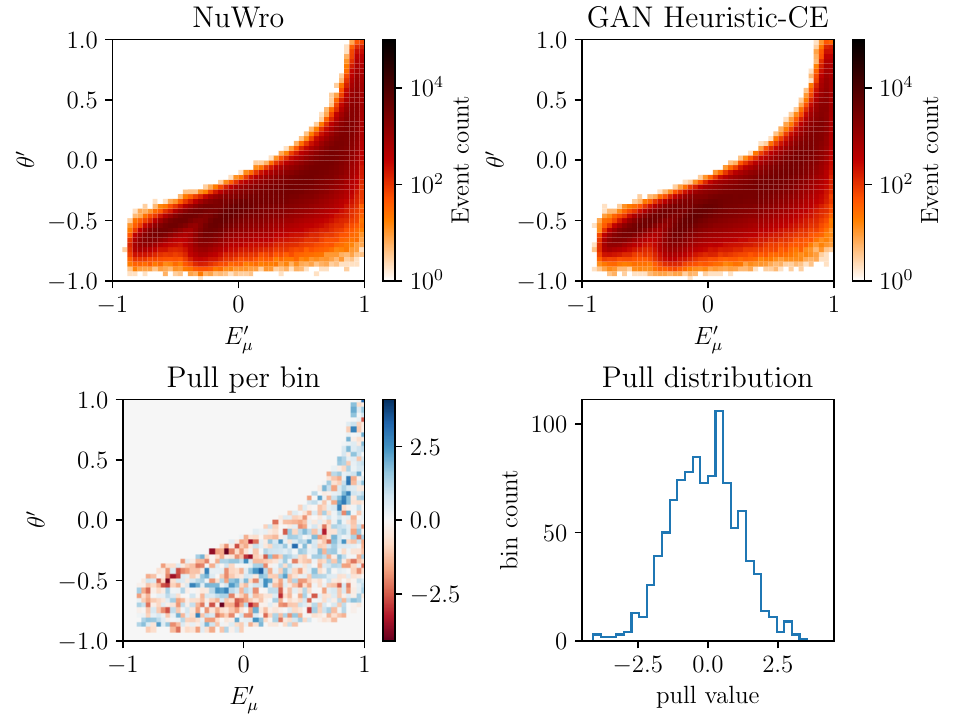}
    \caption{Same as in Fig.~\ref{fig:thetaemuH} but for model optimized with our loss (\ref{Eq:Glossce}). \label{fig:thetaemuHCE}}
\end{figure}

A special role in the optimization plays the discriminator, whose ability to distinguish between ``fake'' and ``true'' events is essential for obtaining a good generator. If the discriminator works either perfectly or imperfectly, then one can not get a good GAN model. 

For an optimal model of GAN, one can expect the response of the discriminator, which is a probability given by the sigmoid function $Sigmoid(D) = {1}/{1+\exp(D)}$, to be distributed around the value $0.5$. Such distributions are shown in 
Figs. \ref{fig:350probs} and \ref{fig:5000probs}, for low and higher neutrino energy events with \( E_\nu = 350 \) MeV and \( E_\nu = 5 \) GeV, respectively. These figures display the discriminator distributions for \nuwro{} (``true'') and generated by the GAN distribution of events. The GAN was trained with heuristic loss \( L_H \) defined in Eq. \ref{Eq:Glossheuristic}. We observe that the models trained with the heuristic loss exhibit greater dispersion, in low neutrino energy, in their predictions, whereas at higher energies, their distributions are more tightly constrained. Moreover, the lower energy distribution is peaked at a value lower than $0.5$ in contrast to the higher energy distribution. These deficiencies of the discriminator, in low neutrino energies, lead to inefficiency in the optimization of the GAN model.  

It  turns out that to optimize efficiently for both low and high neutrino energy cases, it is sufficient to consider the following loss function:
\begin{eqnarray}
L_{HCE}(G) & = & \frac{1}{B} \sum_{k=1}^B \left\{ \log\left[1 - D(G(\mathbf{x}_k, E_{\nu,k}), E_{\nu,k})\right]  - \log\left[D(G(\mathbf{x}_k, E_{\nu,k}), E_{\nu,k})\right] \right\},
\label{Eq:Glosshce}
\end{eqnarray}
which combines the original non-saturating heuristic loss (\ref{Eq:Glossheuristic}) with the maximization term from the conventional cross-entropy (\ref{Eq:Glossce}). 

The distributions of discriminator probabilities, obtained with  (\ref{Eq:Glosshce}), are also shown in Figs.~\ref{fig:350probs} and \ref{fig:5000probs}. The pathological shape of the discriminator distribution at low energies disappears, and it peaks at $0.5$.
\begin{figure*}[bthp]\centering
    \includegraphics[width=0.4\textwidth]{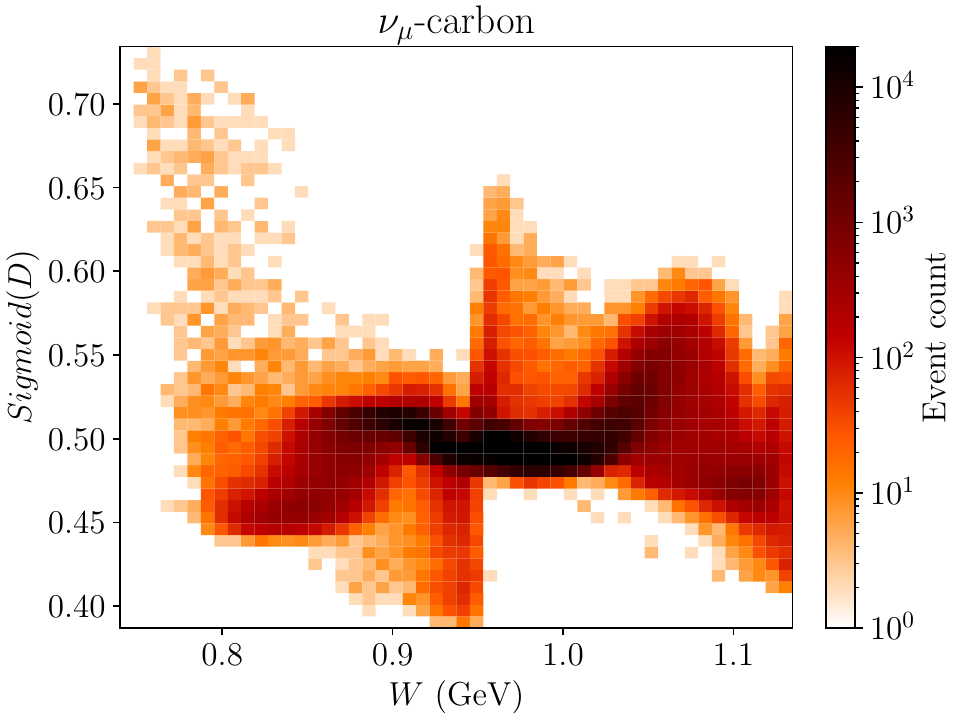}
    \includegraphics[width=0.4\textwidth]{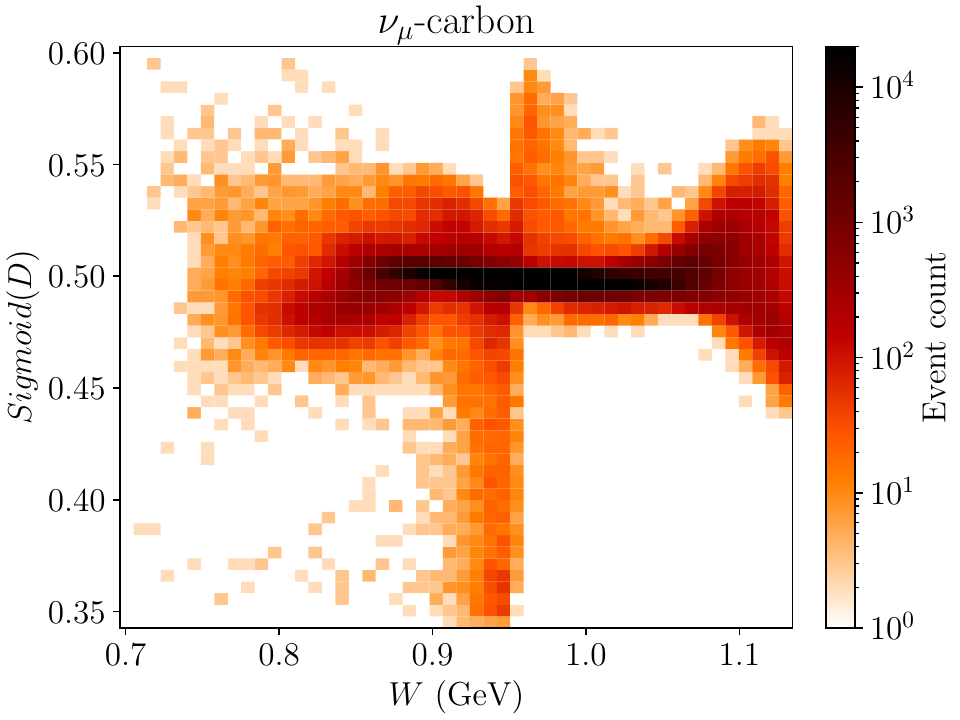}
    \caption{Dependence on $W$ of the discriminator-predicted probabilities for events generated by GANs trained with the loss functions in Eqs. (\ref{Eq:Glossheuristic}) (left) and (\ref{Eq:Glossce}) (right), for $E_\nu = 350$ MeV. \label{fig:wprobs350}}
\end{figure*}
\begin{figure*}[bthp]\centering
    \includegraphics[width=0.4\textwidth]{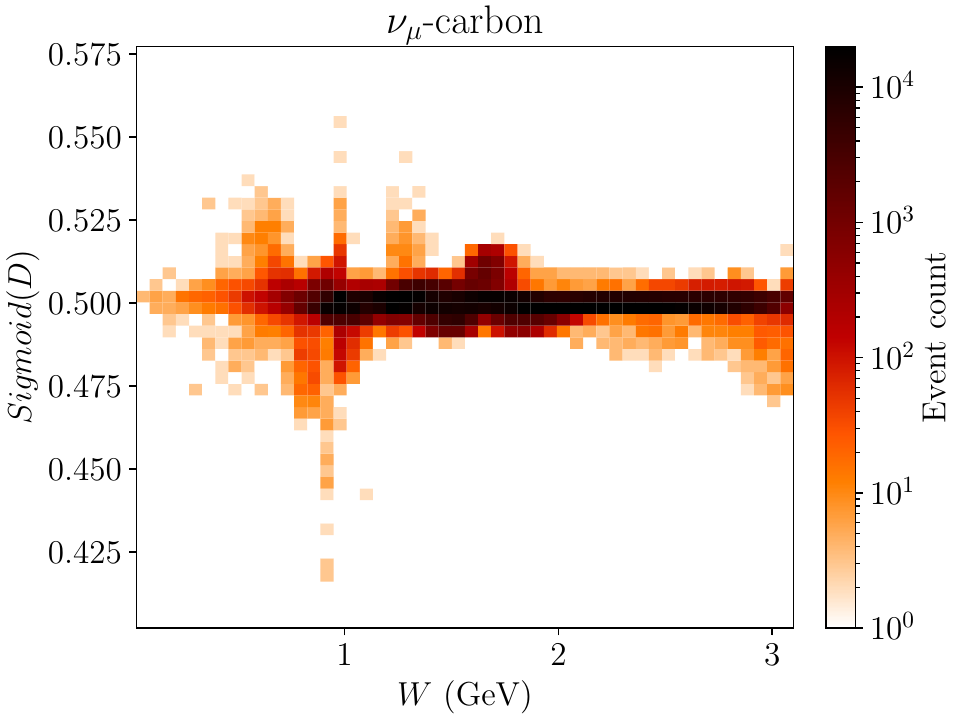}
    \includegraphics[width=0.4\textwidth]{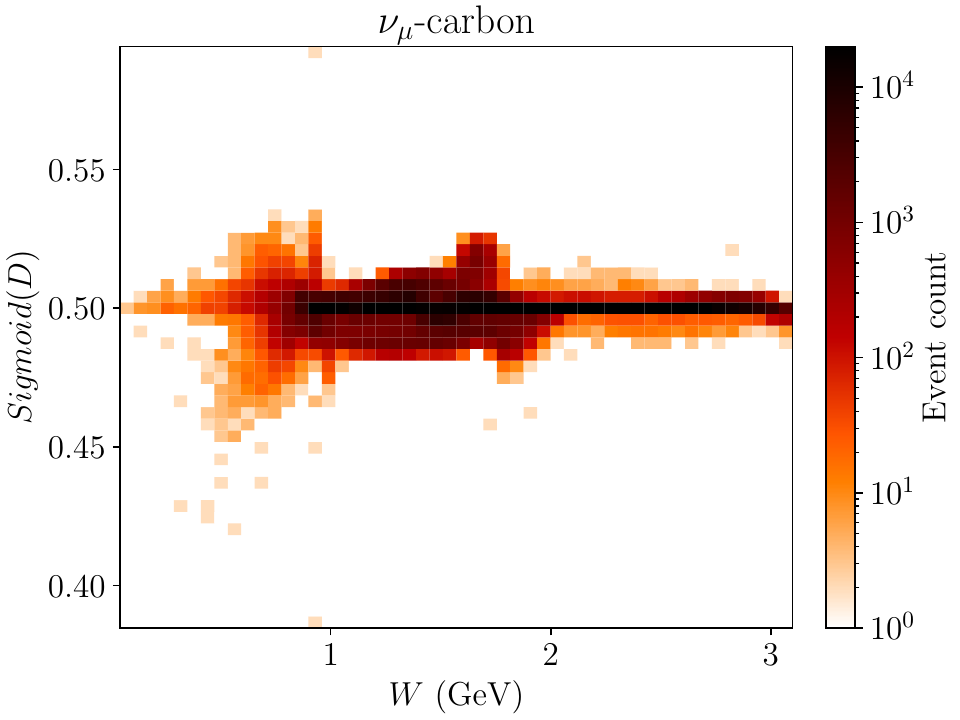}
    \caption{Same as in Fig.~\ref{fig:wprobs350} but for $E_\nu=5$ GeV. \label{fig:wprobs5000}}
\end{figure*}

Additionally, in Figs. \ref{fig:thetaemuH} and \ref{fig:thetaemuHCE} we show 2-dimensional distributions of proxy variables for low and higher neutrino energies (\( E_\nu = 0.5 \) and \( E_\nu = 3 \) GeV) generated by \nuwro{} and GAN models obtained with heuristic and proposed loss by us, respectively. We observe that at lower neutrino energies, the GAN model trained with heuristic loss fails to reproduce a fraction of events on the boundary of the allowed region.

Eventually, in Figs.~\ref{fig:wprobs350} and \ref{fig:wprobs5000}, we show a comparison between $W$-dependence of probability distributions predicted by discriminators for the events generated by the GANs obtained using  (\ref{Eq:Glossheuristic}) and (\ref{Eq:Glossce}) loss functions. 
We observe that the model optimized with the loss proposed in this paper exhibits a distribution concentrated around the probability $0.5$ for both presented neutrino energies. 

In contrast, for the model with heuristic loss for low neutrino energy, Fig.~\ref{fig:wprobs350}, besides the events concentrated around probability $0.5$, there are two groups distributed around probability values $0.45$ and $0.55$ for $W$ about $0.85$ and $1.05$~GeV, respectively. These two regions correspond to quasi-elastic and $\Delta$ resonance production contributions.  

For higher neutrino energies, Fig.~\ref{fig:wprobs5000}, the $W$-distributions for both models are very similar.

\bibliographystyle{apsrev4-2}
\bibliography{dnn,bibmoje,physics,bibdata}

\begin{thebibliography}{46}%
\makeatletter
\providecommand \@ifxundefined [1]{%
 \@ifx{#1\undefined}
}%
\providecommand \@ifnum [1]{%
 \ifnum #1\expandafter \@firstoftwo
 \else \expandafter \@secondoftwo
 \fi
}%
\providecommand \@ifx [1]{%
 \ifx #1\expandafter \@firstoftwo
 \else \expandafter \@secondoftwo
 \fi
}%
\providecommand \natexlab [1]{#1}%
\providecommand \enquote  [1]{``#1''}%
\providecommand \bibnamefont  [1]{#1}%
\providecommand \bibfnamefont [1]{#1}%
\providecommand \citenamefont [1]{#1}%
\providecommand \href@noop [0]{\@secondoftwo}%
\providecommand \href [0]{\begingroup \@sanitize@url \@href}%
\providecommand \@href[1]{\@@startlink{#1}\@@href}%
\providecommand \@@href[1]{\endgroup#1\@@endlink}%
\providecommand \@sanitize@url [0]{\catcode `\\12\catcode `\$12\catcode
  `\&12\catcode `\#12\catcode `\^12\catcode `\_12\catcode `\%12\relax}%
\providecommand \@@startlink[1]{}%
\providecommand \@@endlink[0]{}%
\providecommand \url  [0]{\begingroup\@sanitize@url \@url }%
\providecommand \@url [1]{\endgroup\@href {#1}{\urlprefix }}%
\providecommand \urlprefix  [0]{URL }%
\providecommand \Eprint [0]{\href }%
\providecommand \doibase [0]{https://doi.org/}%
\providecommand \selectlanguage [0]{\@gobble}%
\providecommand \bibinfo  [0]{\@secondoftwo}%
\providecommand \bibfield  [0]{\@secondoftwo}%
\providecommand \translation [1]{[#1]}%
\providecommand \BibitemOpen [0]{}%
\providecommand \bibitemStop [0]{}%
\providecommand \bibitemNoStop [0]{.\EOS\space}%
\providecommand \EOS [0]{\spacefactor3000\relax}%
\providecommand \BibitemShut  [1]{\csname bibitem#1\endcsname}%
\let\auto@bib@innerbib\@empty
\bibitem [{\citenamefont {Avanzini}\ \emph {et~al.}(2022)\citenamefont
  {Avanzini} \emph {et~al.}}]{Avanzini:2021qlx}%
  \BibitemOpen
  \bibfield  {author} {\bibinfo {author} {\bibfnamefont {M.~B.}\ \bibnamefont
  {Avanzini}} \emph {et~al.},\ }\href
  {https://doi.org/10.1103/PhysRevD.105.092004} {\bibfield  {journal} {\bibinfo
   {journal} {Phys. Rev. D}\ }\textbf {\bibinfo {volume} {105}},\ \bibinfo
  {pages} {092004} (\bibinfo {year} {2022})},\ \Eprint
  {https://arxiv.org/abs/2112.09194} {arXiv:2112.09194 [hep-ex]} \BibitemShut
  {NoStop}%
\bibitem [{\citenamefont {Alvarez-Ruso}\ \emph {et~al.}(2018)\citenamefont
  {Alvarez-Ruso} \emph {et~al.}}]{NuSTEC:2017hzk}%
  \BibitemOpen
  \bibfield  {author} {\bibinfo {author} {\bibfnamefont {L.}~\bibnamefont
  {Alvarez-Ruso}} \emph {et~al.} (\bibinfo {collaboration} {NuSTEC}),\ }\href
  {https://doi.org/10.1016/j.ppnp.2018.01.006} {\bibfield  {journal} {\bibinfo
  {journal} {Prog. Part. Nucl. Phys.}\ }\textbf {\bibinfo {volume} {100}},\
  \bibinfo {pages} {1} (\bibinfo {year} {2018})},\ \Eprint
  {https://arxiv.org/abs/1706.03621} {arXiv:1706.03621 [hep-ph]} \BibitemShut
  {NoStop}%
\bibitem [{\citenamefont {Abe}\ \emph {et~al.}(2015)\citenamefont {Abe} \emph
  {et~al.}}]{Hyper-KamiokandeProto-:2015xww}%
  \BibitemOpen
  \bibfield  {author} {\bibinfo {author} {\bibfnamefont {K.}~\bibnamefont
  {Abe}} \emph {et~al.} (\bibinfo {collaboration} {Hyper-Kamiokande
  Proto-Collaboration}),\ }\href {https://doi.org/10.1093/ptep/ptv061}
  {\bibfield  {journal} {\bibinfo  {journal} {PTEP}\ }\textbf {\bibinfo
  {volume} {2015}},\ \bibinfo {pages} {053C02} (\bibinfo {year}
  {2015})}\BibitemShut {NoStop}%
\bibitem [{\citenamefont {Abi}\ \emph {et~al.}(2020)\citenamefont {Abi} \emph
  {et~al.}}]{DUNE:2020lwj}%
  \BibitemOpen
  \bibfield  {author} {\bibinfo {author} {\bibfnamefont {B.}~\bibnamefont
  {Abi}} \emph {et~al.} (\bibinfo {collaboration} {DUNE Collaboration}),\
  }\href {https://doi.org/10.1088/1748-0221/15/08/T08008} {\bibfield  {journal}
  {\bibinfo  {journal} {JINST}\ }\textbf {\bibinfo {volume} {15}}\bibinfo
  {number} { (08)},\ \bibinfo {pages} {T08008}}\BibitemShut {NoStop}%
\bibitem [{\citenamefont {Campbell}\ \emph {et~al.}(2024)\citenamefont
  {Campbell} \emph {et~al.}}]{campbell2024eventgeneratorshighenergyphysics}%
  \BibitemOpen
\bibfield  {number} {  }\bibfield  {author} {\bibinfo {author} {\bibfnamefont
  {J.~M.}\ \bibnamefont {Campbell}} \emph {et~al.},\ }\href
  {https://doi.org/10.21468/SciPostPhys.16.5.130} {\bibfield  {journal}
  {\bibinfo  {journal} {SciPost Phys.}\ }\textbf {\bibinfo {volume} {16}},\
  \bibinfo {pages} {130} (\bibinfo {year} {2024})},\ \Eprint
  {https://arxiv.org/abs/2203.11110} {arXiv:2203.11110 [hep-ph]} \BibitemShut
  {NoStop}%
\bibitem [{\citenamefont {Hayato}\ and\ \citenamefont
  {Pickering}(2021)}]{Hayato:2021heg}%
  \BibitemOpen
  \bibfield  {author} {\bibinfo {author} {\bibfnamefont {Y.}~\bibnamefont
  {Hayato}}\ and\ \bibinfo {author} {\bibfnamefont {L.}~\bibnamefont
  {Pickering}},\ }\href {https://doi.org/10.1140/epjs/s11734-021-00287-7}
  {\bibfield  {journal} {\bibinfo  {journal} {Eur. Phys. J. ST}\ }\textbf
  {\bibinfo {volume} {230}},\ \bibinfo {pages} {4469} (\bibinfo {year}
  {2021})},\ \Eprint {https://arxiv.org/abs/2106.15809} {arXiv:2106.15809
  [hep-ph]} \BibitemShut {NoStop}%
\bibitem [{\citenamefont {Andreopoulos}\ \emph {et~al.}(2010)\citenamefont
  {Andreopoulos} \emph {et~al.}}]{Andreopoulos:2009rq}%
  \BibitemOpen
  \bibfield  {author} {\bibinfo {author} {\bibfnamefont {C.}~\bibnamefont
  {Andreopoulos}} \emph {et~al.},\ }\href
  {https://doi.org/10.1016/j.nima.2009.12.009} {\bibfield  {journal} {\bibinfo
  {journal} {Nucl. Instrum. Meth. A}\ }\textbf {\bibinfo {volume} {614}},\
  \bibinfo {pages} {87} (\bibinfo {year} {2010})}\BibitemShut {NoStop}%
\bibitem [{\citenamefont {Mosel}\ and\ \citenamefont
  {Gallmeister}(2019)}]{Mosel:2018qmv}%
  \BibitemOpen
  \bibfield  {author} {\bibinfo {author} {\bibfnamefont {U.}~\bibnamefont
  {Mosel}}\ and\ \bibinfo {author} {\bibfnamefont {K.}~\bibnamefont
  {Gallmeister}},\ }\href {https://doi.org/10.1103/PhysRevC.99.064605}
  {\bibfield  {journal} {\bibinfo  {journal} {Phys. Rev. C}\ }\textbf {\bibinfo
  {volume} {99}},\ \bibinfo {pages} {064605} (\bibinfo {year} {2019})},\
  \Eprint {https://arxiv.org/abs/1811.10637} {arXiv:1811.10637 [nucl-ex]}
  \BibitemShut {NoStop}%
\bibitem [{\citenamefont {Golan}\ \emph {et~al.}(2012)\citenamefont {Golan},
  \citenamefont {Juszczak},\ and\ \citenamefont {Sobczyk}}]{Golan:2012wx}%
  \BibitemOpen
  \bibfield  {author} {\bibinfo {author} {\bibfnamefont {T.}~\bibnamefont
  {Golan}}, \bibinfo {author} {\bibfnamefont {C.}~\bibnamefont {Juszczak}},\
  and\ \bibinfo {author} {\bibfnamefont {J.~T.}\ \bibnamefont {Sobczyk}},\
  }\href {https://doi.org/10.1103/PhysRevC.86.015505} {\bibfield  {journal}
  {\bibinfo  {journal} {Phys. Rev. C}\ }\textbf {\bibinfo {volume} {86}},\
  \bibinfo {pages} {015505} (\bibinfo {year} {2012})},\ \Eprint
  {https://arxiv.org/abs/1202.4197} {arXiv:1202.4197 [nucl-th]} \BibitemShut
  {NoStop}%
\bibitem [{\citenamefont {Isaacson}\ \emph {et~al.}(2023)\citenamefont
  {Isaacson}, \citenamefont {Jay}, \citenamefont {Lovato}, \citenamefont
  {Machado},\ and\ \citenamefont {Rocco}}]{Isaacson:2022cwh}%
  \BibitemOpen
  \bibfield  {author} {\bibinfo {author} {\bibfnamefont {J.}~\bibnamefont
  {Isaacson}}, \bibinfo {author} {\bibfnamefont {W.~I.}\ \bibnamefont {Jay}},
  \bibinfo {author} {\bibfnamefont {A.}~\bibnamefont {Lovato}}, \bibinfo
  {author} {\bibfnamefont {P.~A.~N.}\ \bibnamefont {Machado}},\ and\ \bibinfo
  {author} {\bibfnamefont {N.}~\bibnamefont {Rocco}},\ }\href
  {https://doi.org/10.1103/PhysRevD.107.033007} {\bibfield  {journal} {\bibinfo
   {journal} {Phys. Rev. D}\ }\textbf {\bibinfo {volume} {107}},\ \bibinfo
  {pages} {033007} (\bibinfo {year} {2023})},\ \Eprint
  {https://arxiv.org/abs/2205.06378} {arXiv:2205.06378 [hep-ph]} \BibitemShut
  {NoStop}%
\bibitem [{\citenamefont {Tena-Vidal}\ \emph {et~al.}(2021)\citenamefont
  {Tena-Vidal} \emph {et~al.}}]{GENIE:2021zuu}%
  \BibitemOpen
  \bibfield  {author} {\bibinfo {author} {\bibfnamefont {J.}~\bibnamefont
  {Tena-Vidal}} \emph {et~al.} (\bibinfo {collaboration} {GENIE}),\ }\href
  {https://doi.org/10.1103/PhysRevD.104.072009} {\bibfield  {journal} {\bibinfo
   {journal} {Phys. Rev. D}\ }\textbf {\bibinfo {volume} {104}},\ \bibinfo
  {pages} {072009} (\bibinfo {year} {2021})},\ \Eprint
  {https://arxiv.org/abs/2104.09179} {arXiv:2104.09179 [hep-ph]} \BibitemShut
  {NoStop}%
\bibitem [{\citenamefont {Radovic}\ \emph {et~al.}(2018)\citenamefont
  {Radovic}, \citenamefont {Williams}, \citenamefont {Rousseau}, \citenamefont
  {Kagan}, \citenamefont {Bonacorsi}, \citenamefont {Himmel}, \citenamefont
  {Aurisano}, \citenamefont {Terao},\ and\ \citenamefont
  {Wongjirad}}]{Radovic:2018dip}%
  \BibitemOpen
  \bibfield  {author} {\bibinfo {author} {\bibfnamefont {A.}~\bibnamefont
  {Radovic}}, \bibinfo {author} {\bibfnamefont {M.}~\bibnamefont {Williams}},
  \bibinfo {author} {\bibfnamefont {D.}~\bibnamefont {Rousseau}}, \bibinfo
  {author} {\bibfnamefont {M.}~\bibnamefont {Kagan}}, \bibinfo {author}
  {\bibfnamefont {D.}~\bibnamefont {Bonacorsi}}, \bibinfo {author}
  {\bibfnamefont {A.}~\bibnamefont {Himmel}}, \bibinfo {author} {\bibfnamefont
  {A.}~\bibnamefont {Aurisano}}, \bibinfo {author} {\bibfnamefont
  {K.}~\bibnamefont {Terao}},\ and\ \bibinfo {author} {\bibfnamefont
  {T.}~\bibnamefont {Wongjirad}},\ }\href
  {https://doi.org/10.1038/s41586-018-0361-2} {\bibfield  {journal} {\bibinfo
  {journal} {Nature}\ }\textbf {\bibinfo {volume} {560}},\ \bibinfo {pages}
  {41} (\bibinfo {year} {2018})}\BibitemShut {NoStop}%
\bibitem [{\citenamefont {Alanazi}\ \emph {et~al.}(2021)\citenamefont
  {Alanazi}, \citenamefont {Sato}, \citenamefont {Ambrozewicz}, \citenamefont
  {Hiller-Blin}, \citenamefont {Melnitchouk}, \citenamefont {Battaglieri},
  \citenamefont {Liu},\ and\ \citenamefont {Li}}]{Alanazi_2021}%
  \BibitemOpen
  \bibfield  {author} {\bibinfo {author} {\bibfnamefont {Y.}~\bibnamefont
  {Alanazi}}, \bibinfo {author} {\bibfnamefont {N.}~\bibnamefont {Sato}},
  \bibinfo {author} {\bibfnamefont {P.}~\bibnamefont {Ambrozewicz}}, \bibinfo
  {author} {\bibfnamefont {A.}~\bibnamefont {Hiller-Blin}}, \bibinfo {author}
  {\bibfnamefont {W.}~\bibnamefont {Melnitchouk}}, \bibinfo {author}
  {\bibfnamefont {M.}~\bibnamefont {Battaglieri}}, \bibinfo {author}
  {\bibfnamefont {T.}~\bibnamefont {Liu}},\ and\ \bibinfo {author}
  {\bibfnamefont {Y.}~\bibnamefont {Li}},\ }in\ \href
  {https://doi.org/10.24963/ijcai.2021/588} {\emph {\bibinfo {booktitle}
  {Proceedings of the Thirtieth International Joint Conference on Artificial
  Intelligence}}},\ \bibinfo {series and number} {IJCAI-2021}\ (\bibinfo
  {publisher} {International Joint Conferences on Artificial Intelligence
  Organization},\ \bibinfo {year} {2021})\ p.\ \bibinfo {pages}
  {4286–4293}\BibitemShut {NoStop}%
\bibitem [{\citenamefont {Butter}\ and\ \citenamefont
  {et~al.}(2022)}]{Butter2022}%
  \BibitemOpen
  \bibfield  {author} {\bibinfo {author} {\bibfnamefont {A.}~\bibnamefont
  {Butter}}\ and\ \bibinfo {author} {\bibnamefont {et~al.}},\ }\href@noop {}
  {\bibfield  {journal} {\bibinfo  {journal} {arXiv preprint arXiv:2203.07460}\
  } (\bibinfo {year} {2022})}\BibitemShut {NoStop}%
\bibitem [{\citenamefont {Goodfellow}\ \emph {et~al.}(2014)\citenamefont
  {Goodfellow}, \citenamefont {Pouget-Abadie}, \citenamefont {Mirza},
  \citenamefont {Xu}, \citenamefont {Warde-Farley}, \citenamefont {Ozair},
  \citenamefont {Courville},\ and\ \citenamefont
  {Bengio}}]{goodfellow2014generativeadversarialnetworks}%
  \BibitemOpen
  \bibfield  {author} {\bibinfo {author} {\bibfnamefont {I.~J.}\ \bibnamefont
  {Goodfellow}}, \bibinfo {author} {\bibfnamefont {J.}~\bibnamefont
  {Pouget-Abadie}}, \bibinfo {author} {\bibfnamefont {M.}~\bibnamefont
  {Mirza}}, \bibinfo {author} {\bibfnamefont {B.}~\bibnamefont {Xu}}, \bibinfo
  {author} {\bibfnamefont {D.}~\bibnamefont {Warde-Farley}}, \bibinfo {author}
  {\bibfnamefont {S.}~\bibnamefont {Ozair}}, \bibinfo {author} {\bibfnamefont
  {A.}~\bibnamefont {Courville}},\ and\ \bibinfo {author} {\bibfnamefont
  {Y.}~\bibnamefont {Bengio}},\ }\href {https://arxiv.org/abs/1406.2661}
  {\bibinfo {title} {Generative adversarial networks}} (\bibinfo {year}
  {2014}),\ \Eprint {https://arxiv.org/abs/1406.2661} {arXiv:1406.2661
  [stat.ML]} \BibitemShut {NoStop}%
\bibitem [{\citenamefont {de~Oliveira}\ \emph {et~al.}(2017)\citenamefont
  {de~Oliveira}, \citenamefont {Paganini},\ and\ \citenamefont
  {Nachman}}]{deOliveira:2017pjk}%
  \BibitemOpen
  \bibfield  {author} {\bibinfo {author} {\bibfnamefont {L.}~\bibnamefont
  {de~Oliveira}}, \bibinfo {author} {\bibfnamefont {M.}~\bibnamefont
  {Paganini}},\ and\ \bibinfo {author} {\bibfnamefont {B.}~\bibnamefont
  {Nachman}},\ }\href {https://doi.org/10.1007/s41781-017-0004-6} {\bibfield
  {journal} {\bibinfo  {journal} {Comput. Softw. Big Sci.}\ }\textbf {\bibinfo
  {volume} {1}},\ \bibinfo {pages} {4} (\bibinfo {year} {2017})},\ \Eprint
  {https://arxiv.org/abs/1701.05927} {arXiv:1701.05927 [stat.ML]} \BibitemShut
  {NoStop}%
\bibitem [{\citenamefont {Monk}(2018)}]{Monk:2018zsb}%
  \BibitemOpen
  \bibfield  {author} {\bibinfo {author} {\bibfnamefont {J.~W.}\ \bibnamefont
  {Monk}},\ }\href {https://doi.org/10.1007/JHEP12(2018)021} {\bibfield
  {journal} {\bibinfo  {journal} {JHEP}\ }\textbf {\bibinfo {volume} {12}},\
  \bibinfo {pages} {021}},\ \Eprint {https://arxiv.org/abs/1807.03685}
  {arXiv:1807.03685 [hep-ph]} \BibitemShut {NoStop}%
\bibitem [{\citenamefont {Alanazi}\ \emph {et~al.}(2022)\citenamefont {Alanazi}
  \emph {et~al.}}]{Alanazi:2020jod}%
  \BibitemOpen
  \bibfield  {author} {\bibinfo {author} {\bibfnamefont {Y.}~\bibnamefont
  {Alanazi}} \emph {et~al.},\ }\href
  {https://doi.org/10.1103/PhysRevD.106.096002} {\bibfield  {journal} {\bibinfo
   {journal} {Phys. Rev. D}\ }\textbf {\bibinfo {volume} {106}},\ \bibinfo
  {pages} {096002} (\bibinfo {year} {2022})},\ \Eprint
  {https://arxiv.org/abs/2008.03151} {arXiv:2008.03151 [hep-ph]} \BibitemShut
  {NoStop}%
\bibitem [{\citenamefont {Badger}\ \emph {et~al.}(2023)\citenamefont {Badger}
  \emph {et~al.}}]{Butter:2022rso}%
  \BibitemOpen
  \bibfield  {author} {\bibinfo {author} {\bibfnamefont {S.}~\bibnamefont
  {Badger}} \emph {et~al.},\ }\href
  {https://doi.org/10.21468/SciPostPhys.14.4.079} {\bibfield  {journal}
  {\bibinfo  {journal} {SciPost Phys.}\ }\textbf {\bibinfo {volume} {14}},\
  \bibinfo {pages} {079} (\bibinfo {year} {2023})},\ \Eprint
  {https://arxiv.org/abs/2203.07460} {arXiv:2203.07460 [hep-ph]} \BibitemShut
  {NoStop}%
\bibitem [{\citenamefont {Ghosh}\ \emph {et~al.}(2022)\citenamefont {Ghosh},
  \citenamefont {Ju}, \citenamefont {Nachman},\ and\ \citenamefont
  {Siodmok}}]{Ghosh:2022zdz}%
  \BibitemOpen
  \bibfield  {author} {\bibinfo {author} {\bibfnamefont {A.}~\bibnamefont
  {Ghosh}}, \bibinfo {author} {\bibfnamefont {X.}~\bibnamefont {Ju}}, \bibinfo
  {author} {\bibfnamefont {B.}~\bibnamefont {Nachman}},\ and\ \bibinfo {author}
  {\bibfnamefont {A.}~\bibnamefont {Siodmok}},\ }\href
  {https://doi.org/10.1103/PhysRevD.106.096020} {\bibfield  {journal} {\bibinfo
   {journal} {Phys. Rev. D}\ }\textbf {\bibinfo {volume} {106}},\ \bibinfo
  {pages} {096020} (\bibinfo {year} {2022})},\ \Eprint
  {https://arxiv.org/abs/2203.12660} {arXiv:2203.12660 [hep-ph]} \BibitemShut
  {NoStop}%
\bibitem [{\citenamefont {Ilten}\ \emph {et~al.}(2023)\citenamefont {Ilten},
  \citenamefont {Menzo}, \citenamefont {Youssef},\ and\ \citenamefont
  {Zupan}}]{Ilten:2022jfm}%
  \BibitemOpen
  \bibfield  {author} {\bibinfo {author} {\bibfnamefont {P.}~\bibnamefont
  {Ilten}}, \bibinfo {author} {\bibfnamefont {T.}~\bibnamefont {Menzo}},
  \bibinfo {author} {\bibfnamefont {A.}~\bibnamefont {Youssef}},\ and\ \bibinfo
  {author} {\bibfnamefont {J.}~\bibnamefont {Zupan}},\ }\href
  {https://doi.org/10.21468/SciPostPhys.14.3.027} {\bibfield  {journal}
  {\bibinfo  {journal} {SciPost Phys.}\ }\textbf {\bibinfo {volume} {14}},\
  \bibinfo {pages} {027} (\bibinfo {year} {2023})},\ \Eprint
  {https://arxiv.org/abs/2203.04983} {arXiv:2203.04983 [hep-ph]} \BibitemShut
  {NoStop}%
\bibitem [{\citenamefont {Chan}\ \emph {et~al.}(2023)\citenamefont {Chan},
  \citenamefont {Ju}, \citenamefont {Kania}, \citenamefont {Nachman},
  \citenamefont {Sangli},\ and\ \citenamefont {Siodmok}}]{Chan:2023ume}%
  \BibitemOpen
  \bibfield  {author} {\bibinfo {author} {\bibfnamefont {J.}~\bibnamefont
  {Chan}}, \bibinfo {author} {\bibfnamefont {X.}~\bibnamefont {Ju}}, \bibinfo
  {author} {\bibfnamefont {A.}~\bibnamefont {Kania}}, \bibinfo {author}
  {\bibfnamefont {B.}~\bibnamefont {Nachman}}, \bibinfo {author} {\bibfnamefont
  {V.}~\bibnamefont {Sangli}},\ and\ \bibinfo {author} {\bibfnamefont
  {A.}~\bibnamefont {Siodmok}},\ }\href
  {https://doi.org/10.1007/JHEP09(2023)084} {\bibfield  {journal} {\bibinfo
  {journal} {JHEP}\ }\textbf {\bibinfo {volume} {09}},\ \bibinfo {pages}
  {084}},\ \Eprint {https://arxiv.org/abs/2305.17169} {arXiv:2305.17169
  [hep-ph]} \BibitemShut {NoStop}%
\bibitem [{\citenamefont {Bonilla}\ \emph {et~al.}(2025)\citenamefont
  {Bonilla}, \citenamefont {Graczyk}, \citenamefont {Ankowski}, \citenamefont
  {Banerjee}, \citenamefont {Kowal}, \citenamefont {Prasad},\ and\
  \citenamefont {Sobczyk}}]{bonilla2025generativeadversarialneuralnetworks}%
  \BibitemOpen
  \bibfield  {author} {\bibinfo {author} {\bibfnamefont {J.~L.}\ \bibnamefont
  {Bonilla}}, \bibinfo {author} {\bibfnamefont {K.~M.}\ \bibnamefont
  {Graczyk}}, \bibinfo {author} {\bibfnamefont {A.~M.}\ \bibnamefont
  {Ankowski}}, \bibinfo {author} {\bibfnamefont {R.~D.}\ \bibnamefont
  {Banerjee}}, \bibinfo {author} {\bibfnamefont {B.~E.}\ \bibnamefont {Kowal}},
  \bibinfo {author} {\bibfnamefont {H.}~\bibnamefont {Prasad}},\ and\ \bibinfo
  {author} {\bibfnamefont {J.~T.}\ \bibnamefont {Sobczyk}},\ }\href
  {https://doi.org/10.1103/l6td-93sr} {\bibfield  {journal} {\bibinfo
  {journal} {Phys. Rev. D}\ }\textbf {\bibinfo {volume} {112}},\ \bibinfo
  {pages} {013007} (\bibinfo {year} {2025})}\BibitemShut {NoStop}%
\bibitem [{\citenamefont {El~Baz}\ and\ \citenamefont
  {S\'anchez}(2024)}]{ElBaz:2023ijr}%
  \BibitemOpen
  \bibfield  {author} {\bibinfo {author} {\bibfnamefont {M.}~\bibnamefont
  {El~Baz}}\ and\ \bibinfo {author} {\bibfnamefont {F.}~\bibnamefont
  {S\'anchez}},\ }\href {https://doi.org/10.1103/PhysRevD.109.032008}
  {\bibfield  {journal} {\bibinfo  {journal} {Phys. Rev. D}\ }\textbf {\bibinfo
  {volume} {109}},\ \bibinfo {pages} {032008} (\bibinfo {year}
  {2024})}\BibitemShut {NoStop}%
\bibitem [{\citenamefont {El~Baz}\ \emph {et~al.}(2025)\citenamefont {El~Baz},
  \citenamefont {S{\'a}nchez}, \citenamefont {Jachowicz}, \citenamefont
  {Niewczas}, \citenamefont {Jha},\ and\ \citenamefont
  {Nikolakopoulos}}]{ElBaz:2025qjp}%
  \BibitemOpen
  \bibfield  {author} {\bibinfo {author} {\bibfnamefont {M.}~\bibnamefont
  {El~Baz}}, \bibinfo {author} {\bibfnamefont {F.}~\bibnamefont {S{\'a}nchez}},
  \bibinfo {author} {\bibfnamefont {N.}~\bibnamefont {Jachowicz}}, \bibinfo
  {author} {\bibfnamefont {K.}~\bibnamefont {Niewczas}}, \bibinfo {author}
  {\bibfnamefont {A.~K.}\ \bibnamefont {Jha}},\ and\ \bibinfo {author}
  {\bibfnamefont {A.}~\bibnamefont {Nikolakopoulos}},\ }\href
  {https://doi.org/10.1103/v7sz-vn3b} {\bibfield  {journal} {\bibinfo
  {journal} {Phys. Rev. D}\ }\textbf {\bibinfo {volume} {111}},\ \bibinfo
  {pages} {113001} (\bibinfo {year} {2025})},\ \Eprint
  {https://arxiv.org/abs/2502.14452} {arXiv:2502.14452 [hep-ex]} \BibitemShut
  {NoStop}%
\bibitem [{\citenamefont {Graczyk}\ \emph {et~al.}(2025)\citenamefont
  {Graczyk}, \citenamefont {Kowal}, \citenamefont {Ankowski}, \citenamefont
  {Banerjee}, \citenamefont {Bonilla}, \citenamefont {Prasad},\ and\
  \citenamefont {Sobczyk}}]{graczyk2024electronnucleuscrosssectionstransfer}%
  \BibitemOpen
  \bibfield  {author} {\bibinfo {author} {\bibfnamefont {K.~M.}\ \bibnamefont
  {Graczyk}}, \bibinfo {author} {\bibfnamefont {B.~E.}\ \bibnamefont {Kowal}},
  \bibinfo {author} {\bibfnamefont {A.~M.}\ \bibnamefont {Ankowski}}, \bibinfo
  {author} {\bibfnamefont {R.~D.}\ \bibnamefont {Banerjee}}, \bibinfo {author}
  {\bibfnamefont {J.~L.}\ \bibnamefont {Bonilla}}, \bibinfo {author}
  {\bibfnamefont {H.}~\bibnamefont {Prasad}},\ and\ \bibinfo {author}
  {\bibfnamefont {J.~T.}\ \bibnamefont {Sobczyk}},\ }\href
  {https://doi.org/10.1103/zxv6-22tz} {\bibfield  {journal} {\bibinfo
  {journal} {Phys. Rev. Lett.}\ }\textbf {\bibinfo {volume} {135}},\ \bibinfo
  {pages} {052502} (\bibinfo {year} {2025})}\BibitemShut {NoStop}%
\bibitem [{\citenamefont {Chappell}\ and\ \citenamefont
  {Whitehead}(2022)}]{Chappell:2022yxd}%
  \BibitemOpen
  \bibfield  {author} {\bibinfo {author} {\bibfnamefont {A.}~\bibnamefont
  {Chappell}}\ and\ \bibinfo {author} {\bibfnamefont {L.~H.}\ \bibnamefont
  {Whitehead}},\ }\href {https://doi.org/10.1140/epjc/s10052-022-11066-6}
  {\bibfield  {journal} {\bibinfo  {journal} {Eur. Phys. J. C}\ }\textbf
  {\bibinfo {volume} {82}},\ \bibinfo {pages} {1099} (\bibinfo {year}
  {2022})},\ \Eprint {https://arxiv.org/abs/2207.03139} {arXiv:2207.03139
  [hep-ex]} \BibitemShut {NoStop}%
\bibitem [{\citenamefont {Steiner}(2001)}]{STEINER200115845}%
  \BibitemOpen
  \bibfield  {author} {\bibinfo {author} {\bibfnamefont {G.}~\bibnamefont
  {Steiner}},\ }in\ \href
  {https://doi.org/https://doi.org/10.1016/B0-08-043076-7/01481-9} {\emph
  {\bibinfo {booktitle} {International Encyclopedia of the Social \& Behavioral
  Sciences}}},\ \bibinfo {editor} {edited by\ \bibinfo {editor} {\bibfnamefont
  {N.~J.}\ \bibnamefont {Smelser}}\ and\ \bibinfo {editor} {\bibfnamefont
  {P.~B.}\ \bibnamefont {Baltes}}}\ (\bibinfo  {publisher} {Pergamon},\
  \bibinfo {address} {Oxford},\ \bibinfo {year} {2001})\ pp.\ \bibinfo {pages}
  {15845--15851}\BibitemShut {NoStop}%
\bibitem [{\citenamefont {Pan}\ and\ \citenamefont
  {Yang}(2010)}]{transfer_learning_survey}%
  \BibitemOpen
  \bibfield  {author} {\bibinfo {author} {\bibfnamefont {S.~J.}\ \bibnamefont
  {Pan}}\ and\ \bibinfo {author} {\bibfnamefont {Q.}~\bibnamefont {Yang}},\
  }\href {https://doi.org/10.1109/TKDE.2009.191} {\bibfield  {journal}
  {\bibinfo  {journal} {IEEE Transactions on Knowledge and Data Engineering}\
  }\textbf {\bibinfo {volume} {22}},\ \bibinfo {pages} {1345} (\bibinfo {year}
  {2010})}\BibitemShut {NoStop}%
\bibitem [{\citenamefont {Zhuang}\ \emph {et~al.}(2020)\citenamefont {Zhuang},
  \citenamefont {Qi}, \citenamefont {Duan}, \citenamefont {Xi}, \citenamefont
  {Zhu}, \citenamefont {Zhu}, \citenamefont {Xiong},\ and\ \citenamefont
  {He}}]{zhuang2020comprehensive}%
  \BibitemOpen
  \bibfield  {author} {\bibinfo {author} {\bibfnamefont {F.}~\bibnamefont
  {Zhuang}}, \bibinfo {author} {\bibfnamefont {Z.}~\bibnamefont {Qi}}, \bibinfo
  {author} {\bibfnamefont {K.}~\bibnamefont {Duan}}, \bibinfo {author}
  {\bibfnamefont {D.}~\bibnamefont {Xi}}, \bibinfo {author} {\bibfnamefont
  {Y.}~\bibnamefont {Zhu}}, \bibinfo {author} {\bibfnamefont {H.}~\bibnamefont
  {Zhu}}, \bibinfo {author} {\bibfnamefont {H.}~\bibnamefont {Xiong}},\ and\
  \bibinfo {author} {\bibfnamefont {Q.}~\bibnamefont {He}},\ }\href@noop {}
  {\bibinfo {title} {A comprehensive survey on transfer learning}} (\bibinfo
  {year} {2020}),\ \Eprint {https://arxiv.org/abs/1911.02685} {arXiv:1911.02685
  [cs.LG]} \BibitemShut {NoStop}%
\bibitem [{\citenamefont {Tan}\ \emph {et~al.}(2018)\citenamefont {Tan},
  \citenamefont {Sun}, \citenamefont {Kong}, \citenamefont {Zhang},
  \citenamefont {Yang},\ and\ \citenamefont {Liu}}]{tan2018survey}%
  \BibitemOpen
  \bibfield  {author} {\bibinfo {author} {\bibfnamefont {C.}~\bibnamefont
  {Tan}}, \bibinfo {author} {\bibfnamefont {F.}~\bibnamefont {Sun}}, \bibinfo
  {author} {\bibfnamefont {T.}~\bibnamefont {Kong}}, \bibinfo {author}
  {\bibfnamefont {W.}~\bibnamefont {Zhang}}, \bibinfo {author} {\bibfnamefont
  {C.}~\bibnamefont {Yang}},\ and\ \bibinfo {author} {\bibfnamefont
  {C.}~\bibnamefont {Liu}},\ }\href@noop {} {\bibinfo {title} {A survey on deep
  transfer learning}} (\bibinfo {year} {2018}),\ \Eprint
  {https://arxiv.org/abs/1808.01974} {arXiv:1808.01974 [cs.LG]} \BibitemShut
  {NoStop}%
\bibitem [{\citenamefont {Weiss}\ \emph {et~al.}(2016)\citenamefont {Weiss},
  \citenamefont {Khoshgoftaar},\ and\ \citenamefont {Wang}}]{Weiss2016}%
  \BibitemOpen
  \bibfield  {author} {\bibinfo {author} {\bibfnamefont {K.}~\bibnamefont
  {Weiss}}, \bibinfo {author} {\bibfnamefont {T.~M.}\ \bibnamefont
  {Khoshgoftaar}},\ and\ \bibinfo {author} {\bibfnamefont {D.}~\bibnamefont
  {Wang}},\ }\href {https://doi.org/10.1186/s40537-016-0043-6} {\bibfield
  {journal} {\bibinfo  {journal} {Journal of Big Data}\ }\textbf {\bibinfo
  {volume} {3}},\ \bibinfo {pages} {9} (\bibinfo {year} {2016})}\BibitemShut
  {NoStop}%
\bibitem [{\citenamefont {Benhar}\ \emph {et~al.}(1994)\citenamefont {Benhar},
  \citenamefont {Fabrocini}, \citenamefont {Fantoni},\ and\ \citenamefont
  {Sick}}]{Benhar:1994hw}%
  \BibitemOpen
  \bibfield  {author} {\bibinfo {author} {\bibfnamefont {O.}~\bibnamefont
  {Benhar}}, \bibinfo {author} {\bibfnamefont {A.}~\bibnamefont {Fabrocini}},
  \bibinfo {author} {\bibfnamefont {S.}~\bibnamefont {Fantoni}},\ and\ \bibinfo
  {author} {\bibfnamefont {I.}~\bibnamefont {Sick}},\ }\href
  {https://doi.org/10.1016/0375-9474(94)90920-2} {\bibfield  {journal}
  {\bibinfo  {journal} {Nucl. Phys. A}\ }\textbf {\bibinfo {volume} {579}},\
  \bibinfo {pages} {493} (\bibinfo {year} {1994})}\BibitemShut {NoStop}%
\bibitem [{\citenamefont {Sobczyk}\ \emph {et~al.}(2005)\citenamefont
  {Sobczyk}, \citenamefont {Nowak},\ and\ \citenamefont
  {Graczyk}}]{Sobczyk:2004va}%
  \BibitemOpen
  \bibfield  {author} {\bibinfo {author} {\bibfnamefont {J.~T.}\ \bibnamefont
  {Sobczyk}}, \bibinfo {author} {\bibfnamefont {J.~A.}\ \bibnamefont {Nowak}},\
  and\ \bibinfo {author} {\bibfnamefont {K.~M.}\ \bibnamefont {Graczyk}},\
  }\bibfield  {booktitle} {\emph {\bibinfo {booktitle} {{Proceedings, 3rd
  International Workshop on Neutrino-nucleus interactions in the few GeV region
  (NUINT 04): Assergi, Italy, March 17-21, 2004}}},\ }\href
  {https://doi.org/10.1016/j.nuclphysbps.2004.11.218} {\bibfield  {journal}
  {\bibinfo  {journal} {Nucl. Phys. Proc. Suppl.}\ }\textbf {\bibinfo {volume}
  {139}},\ \bibinfo {pages} {266} (\bibinfo {year} {2005})},\ \bibinfo {note}
  {[,266(2004)]},\ \Eprint {https://arxiv.org/abs/hep-ph/0407277}
  {arXiv:hep-ph/0407277 [hep-ph]} \BibitemShut {NoStop}%
\bibitem [{\citenamefont {Juszczak}\ \emph {et~al.}(2006)\citenamefont
  {Juszczak}, \citenamefont {Nowak},\ and\ \citenamefont
  {Sobczyk}}]{Juszczak:2005zs}%
  \BibitemOpen
  \bibfield  {author} {\bibinfo {author} {\bibfnamefont {C.}~\bibnamefont
  {Juszczak}}, \bibinfo {author} {\bibfnamefont {J.~A.}\ \bibnamefont
  {Nowak}},\ and\ \bibinfo {author} {\bibfnamefont {J.~T.}\ \bibnamefont
  {Sobczyk}},\ }\href {https://doi.org/10.1016/j.nuclphysbps.2006.08.069}
  {\bibfield  {journal} {\bibinfo  {journal} {Nucl. Phys. B Proc. Suppl.}\
  }\textbf {\bibinfo {volume} {159}},\ \bibinfo {pages} {211} (\bibinfo {year}
  {2006})},\ \Eprint {https://arxiv.org/abs/hep-ph/0512365}
  {arXiv:hep-ph/0512365} \BibitemShut {NoStop}%
\bibitem [{\citenamefont {Graczyk}\ \emph {et~al.}(2009)\citenamefont
  {Graczyk}, \citenamefont {Kielczewska}, \citenamefont {Przewlocki},\ and\
  \citenamefont {Sobczyk}}]{Graczyk:2009qm}%
  \BibitemOpen
  \bibfield  {author} {\bibinfo {author} {\bibfnamefont {K.~M.}\ \bibnamefont
  {Graczyk}}, \bibinfo {author} {\bibfnamefont {D.}~\bibnamefont
  {Kielczewska}}, \bibinfo {author} {\bibfnamefont {P.}~\bibnamefont
  {Przewlocki}},\ and\ \bibinfo {author} {\bibfnamefont {J.~T.}\ \bibnamefont
  {Sobczyk}},\ }\href {https://doi.org/10.1103/PhysRevD.80.093001} {\bibfield
  {journal} {\bibinfo  {journal} {Phys. Rev.}\ }\textbf {\bibinfo {volume}
  {D80}},\ \bibinfo {pages} {093001} (\bibinfo {year} {2009})},\ \Eprint
  {https://arxiv.org/abs/0908.2175} {arXiv:0908.2175 [hep-ph]} \BibitemShut
  {NoStop}%
\bibitem [{\citenamefont {Banerjee}\ \emph {et~al.}(2024)\citenamefont
  {Banerjee}, \citenamefont {Ankowski}, \citenamefont {Graczyk}, \citenamefont
  {Kowal}, \citenamefont {Prasad},\ and\ \citenamefont
  {Sobczyk}}]{Banerjee:2023hub}%
  \BibitemOpen
  \bibfield  {author} {\bibinfo {author} {\bibfnamefont {R.~D.}\ \bibnamefont
  {Banerjee}}, \bibinfo {author} {\bibfnamefont {A.~M.}\ \bibnamefont
  {Ankowski}}, \bibinfo {author} {\bibfnamefont {K.~M.}\ \bibnamefont
  {Graczyk}}, \bibinfo {author} {\bibfnamefont {B.~E.}\ \bibnamefont {Kowal}},
  \bibinfo {author} {\bibfnamefont {H.}~\bibnamefont {Prasad}},\ and\ \bibinfo
  {author} {\bibfnamefont {J.~T.}\ \bibnamefont {Sobczyk}},\ }\href
  {https://doi.org/10.1103/PhysRevD.109.073004} {\bibfield  {journal} {\bibinfo
   {journal} {Phys. Rev. D}\ }\textbf {\bibinfo {volume} {109}},\ \bibinfo
  {pages} {073004} (\bibinfo {year} {2024})}\BibitemShut {NoStop}%
\bibitem [{\citenamefont {Jiang}\ \emph {et~al.}(2022)\citenamefont {Jiang},
  \citenamefont {Ankowski}, \citenamefont {Abrams}, \citenamefont {Gu},
  \citenamefont {Aljawrneh}, \citenamefont {Alsalmi}, \citenamefont {Bane},
  \citenamefont {Batz}, \citenamefont {Barcus}, \citenamefont {Barroso},
  \citenamefont {Bellini}, \citenamefont {Benhar}, \citenamefont {Bericic},
  \citenamefont {Biswas}, \citenamefont {Camsonne}, \citenamefont
  {Castellanos}, \citenamefont {Chen}, \citenamefont {Christy}, \citenamefont
  {Craycraft}, \citenamefont {Cruz-Torres}, \citenamefont {Dai}, \citenamefont
  {Day}, \citenamefont {Dirican}, \citenamefont {Dusa}, \citenamefont {Fuchey},
  \citenamefont {Gautam}, \citenamefont {Giusti}, \citenamefont {Gomez},
  \citenamefont {Gu}, \citenamefont {Hague}, \citenamefont {Hansen},
  \citenamefont {Hauenstein}, \citenamefont {Higinbotham}, \citenamefont
  {Hyde}, \citenamefont {Jerzyk}, \citenamefont {Keppel}, \citenamefont {Li},
  \citenamefont {Lindgren}, \citenamefont {Liu}, \citenamefont {Mariani},
  \citenamefont {McClellan}, \citenamefont {Meekins}, \citenamefont {Michaels},
  \citenamefont {Mihovilovic}, \citenamefont {Murphy}, \citenamefont {Nguyen},
  \citenamefont {Nycz}, \citenamefont {Ou}, \citenamefont {Pandey},
  \citenamefont {Pandey}, \citenamefont {Park}, \citenamefont {Perera},
  \citenamefont {Puckett}, \citenamefont {Santiesteban}, \citenamefont
  {\ifmmode~\check{S}\else \v{S}\fi{}irca}, \citenamefont {Su}, \citenamefont
  {Tang}, \citenamefont {Tian}, \citenamefont {Ton}, \citenamefont
  {Wojtsekhowski}, \citenamefont {Wood}, \citenamefont {Ye},\ and\
  \citenamefont {Zhang}}]{PhysRevD.105.112002}%
  \BibitemOpen
  \bibfield  {author} {\bibinfo {author} {\bibfnamefont {L.}~\bibnamefont
  {Jiang}}, \bibinfo {author} {\bibfnamefont {A.~M.}\ \bibnamefont {Ankowski}},
  \bibinfo {author} {\bibfnamefont {D.}~\bibnamefont {Abrams}}, \bibinfo
  {author} {\bibfnamefont {L.}~\bibnamefont {Gu}}, \bibinfo {author}
  {\bibfnamefont {B.}~\bibnamefont {Aljawrneh}}, \bibinfo {author}
  {\bibfnamefont {S.}~\bibnamefont {Alsalmi}}, \bibinfo {author} {\bibfnamefont
  {J.}~\bibnamefont {Bane}}, \bibinfo {author} {\bibfnamefont {A.}~\bibnamefont
  {Batz}}, \bibinfo {author} {\bibfnamefont {S.}~\bibnamefont {Barcus}},
  \bibinfo {author} {\bibfnamefont {M.}~\bibnamefont {Barroso}}, \bibinfo
  {author} {\bibfnamefont {V.}~\bibnamefont {Bellini}}, \bibinfo {author}
  {\bibfnamefont {O.}~\bibnamefont {Benhar}}, \bibinfo {author} {\bibfnamefont
  {J.}~\bibnamefont {Bericic}}, \bibinfo {author} {\bibfnamefont
  {D.}~\bibnamefont {Biswas}}, \bibinfo {author} {\bibfnamefont
  {A.}~\bibnamefont {Camsonne}}, \bibinfo {author} {\bibfnamefont
  {J.}~\bibnamefont {Castellanos}}, \bibinfo {author} {\bibfnamefont {J.-P.}\
  \bibnamefont {Chen}}, \bibinfo {author} {\bibfnamefont {M.~E.}\ \bibnamefont
  {Christy}}, \bibinfo {author} {\bibfnamefont {K.}~\bibnamefont {Craycraft}},
  \bibinfo {author} {\bibfnamefont {R.}~\bibnamefont {Cruz-Torres}}, \bibinfo
  {author} {\bibfnamefont {H.}~\bibnamefont {Dai}}, \bibinfo {author}
  {\bibfnamefont {D.}~\bibnamefont {Day}}, \bibinfo {author} {\bibfnamefont
  {A.}~\bibnamefont {Dirican}}, \bibinfo {author} {\bibfnamefont {S.-C.}\
  \bibnamefont {Dusa}}, \bibinfo {author} {\bibfnamefont {E.}~\bibnamefont
  {Fuchey}}, \bibinfo {author} {\bibfnamefont {T.}~\bibnamefont {Gautam}},
  \bibinfo {author} {\bibfnamefont {C.}~\bibnamefont {Giusti}}, \bibinfo
  {author} {\bibfnamefont {J.}~\bibnamefont {Gomez}}, \bibinfo {author}
  {\bibfnamefont {C.}~\bibnamefont {Gu}}, \bibinfo {author} {\bibfnamefont
  {T.~J.}\ \bibnamefont {Hague}}, \bibinfo {author} {\bibfnamefont {J.-O.}\
  \bibnamefont {Hansen}}, \bibinfo {author} {\bibfnamefont {F.}~\bibnamefont
  {Hauenstein}}, \bibinfo {author} {\bibfnamefont {D.~W.}\ \bibnamefont
  {Higinbotham}}, \bibinfo {author} {\bibfnamefont {C.}~\bibnamefont {Hyde}},
  \bibinfo {author} {\bibfnamefont {Z.}~\bibnamefont {Jerzyk}}, \bibinfo
  {author} {\bibfnamefont {C.}~\bibnamefont {Keppel}}, \bibinfo {author}
  {\bibfnamefont {S.}~\bibnamefont {Li}}, \bibinfo {author} {\bibfnamefont
  {R.}~\bibnamefont {Lindgren}}, \bibinfo {author} {\bibfnamefont
  {H.}~\bibnamefont {Liu}}, \bibinfo {author} {\bibfnamefont {C.}~\bibnamefont
  {Mariani}}, \bibinfo {author} {\bibfnamefont {R.~E.}\ \bibnamefont
  {McClellan}}, \bibinfo {author} {\bibfnamefont {D.}~\bibnamefont {Meekins}},
  \bibinfo {author} {\bibfnamefont {R.}~\bibnamefont {Michaels}}, \bibinfo
  {author} {\bibfnamefont {M.}~\bibnamefont {Mihovilovic}}, \bibinfo {author}
  {\bibfnamefont {M.}~\bibnamefont {Murphy}}, \bibinfo {author} {\bibfnamefont
  {D.}~\bibnamefont {Nguyen}}, \bibinfo {author} {\bibfnamefont
  {M.}~\bibnamefont {Nycz}}, \bibinfo {author} {\bibfnamefont {L.}~\bibnamefont
  {Ou}}, \bibinfo {author} {\bibfnamefont {B.}~\bibnamefont {Pandey}}, \bibinfo
  {author} {\bibfnamefont {V.}~\bibnamefont {Pandey}}, \bibinfo {author}
  {\bibfnamefont {K.}~\bibnamefont {Park}}, \bibinfo {author} {\bibfnamefont
  {G.}~\bibnamefont {Perera}}, \bibinfo {author} {\bibfnamefont {A.~J.~R.}\
  \bibnamefont {Puckett}}, \bibinfo {author} {\bibfnamefont {S.~N.}\
  \bibnamefont {Santiesteban}}, \bibinfo {author} {\bibfnamefont
  {S.}~\bibnamefont {\ifmmode~\check{S}\else \v{S}\fi{}irca}}, \bibinfo
  {author} {\bibfnamefont {T.}~\bibnamefont {Su}}, \bibinfo {author}
  {\bibfnamefont {L.}~\bibnamefont {Tang}}, \bibinfo {author} {\bibfnamefont
  {Y.}~\bibnamefont {Tian}}, \bibinfo {author} {\bibfnamefont {N.}~\bibnamefont
  {Ton}}, \bibinfo {author} {\bibfnamefont {B.}~\bibnamefont {Wojtsekhowski}},
  \bibinfo {author} {\bibfnamefont {S.}~\bibnamefont {Wood}}, \bibinfo {author}
  {\bibfnamefont {Z.}~\bibnamefont {Ye}},\ and\ \bibinfo {author}
  {\bibfnamefont {J.}~\bibnamefont {Zhang}} (\bibinfo {collaboration}
  {Jefferson Lab Hall A Collaboration}),\ }\href
  {https://doi.org/10.1103/PhysRevD.105.112002} {\bibfield  {journal} {\bibinfo
   {journal} {Phys. Rev. D}\ }\textbf {\bibinfo {volume} {105}},\ \bibinfo
  {pages} {112002} (\bibinfo {year} {2022})}\BibitemShut {NoStop}%
\bibitem [{\citenamefont {Jiang}\ \emph {et~al.}(2023)\citenamefont {Jiang},
  \citenamefont {Ankowski}, \citenamefont {Abrams}, \citenamefont {Gu},
  \citenamefont {Aljawrneh}, \citenamefont {Alsalmi}, \citenamefont {Bane},
  \citenamefont {Batz}, \citenamefont {Barcus}, \citenamefont {Barroso},
  \citenamefont {Bellini}, \citenamefont {Benhar}, \citenamefont {Bericic},
  \citenamefont {Biswas}, \citenamefont {Camsonne}, \citenamefont
  {Castellanos}, \citenamefont {Chen}, \citenamefont {Christy}, \citenamefont
  {Craycraft}, \citenamefont {Cruz-Torres}, \citenamefont {Dai}, \citenamefont
  {Day}, \citenamefont {Dirican}, \citenamefont {Dusa}, \citenamefont {Fuchey},
  \citenamefont {Gautam}, \citenamefont {Giusti}, \citenamefont {Gomez},
  \citenamefont {Gu}, \citenamefont {Hague}, \citenamefont {Hansen},
  \citenamefont {Hauenstein}, \citenamefont {Higinbotham}, \citenamefont
  {Hyde}, \citenamefont {Jerzyk}, \citenamefont {Johnson}, \citenamefont
  {Keppel}, \citenamefont {Lanham}, \citenamefont {Li}, \citenamefont
  {Lindgren}, \citenamefont {Liu}, \citenamefont {Mariani}, \citenamefont
  {McClellan}, \citenamefont {Meekins}, \citenamefont {Michaels}, \citenamefont
  {Mihovilovic}, \citenamefont {Murphy}, \citenamefont {Nguyen}, \citenamefont
  {Nycz}, \citenamefont {Ou}, \citenamefont {Pandey}, \citenamefont {Pandey},
  \citenamefont {Park}, \citenamefont {Perera}, \citenamefont {Puckett},
  \citenamefont {Santiesteban}, \citenamefont {\ifmmode~\check{S}\else
  \v{S}\fi{}irca}, \citenamefont {Su}, \citenamefont {Tang}, \citenamefont
  {Tian}, \citenamefont {Ton}, \citenamefont {Wojtsekhowski}, \citenamefont
  {Wood}, \citenamefont {Ye},\ and\ \citenamefont
  {Zhang}}]{PhysRevD.107.012005}%
  \BibitemOpen
  \bibfield  {author} {\bibinfo {author} {\bibfnamefont {L.}~\bibnamefont
  {Jiang}}, \bibinfo {author} {\bibfnamefont {A.~M.}\ \bibnamefont {Ankowski}},
  \bibinfo {author} {\bibfnamefont {D.}~\bibnamefont {Abrams}}, \bibinfo
  {author} {\bibfnamefont {L.}~\bibnamefont {Gu}}, \bibinfo {author}
  {\bibfnamefont {B.}~\bibnamefont {Aljawrneh}}, \bibinfo {author}
  {\bibfnamefont {S.}~\bibnamefont {Alsalmi}}, \bibinfo {author} {\bibfnamefont
  {J.}~\bibnamefont {Bane}}, \bibinfo {author} {\bibfnamefont {A.}~\bibnamefont
  {Batz}}, \bibinfo {author} {\bibfnamefont {S.}~\bibnamefont {Barcus}},
  \bibinfo {author} {\bibfnamefont {M.}~\bibnamefont {Barroso}}, \bibinfo
  {author} {\bibfnamefont {V.}~\bibnamefont {Bellini}}, \bibinfo {author}
  {\bibfnamefont {O.}~\bibnamefont {Benhar}}, \bibinfo {author} {\bibfnamefont
  {J.}~\bibnamefont {Bericic}}, \bibinfo {author} {\bibfnamefont
  {D.}~\bibnamefont {Biswas}}, \bibinfo {author} {\bibfnamefont
  {A.}~\bibnamefont {Camsonne}}, \bibinfo {author} {\bibfnamefont
  {J.}~\bibnamefont {Castellanos}}, \bibinfo {author} {\bibfnamefont {J.-P.}\
  \bibnamefont {Chen}}, \bibinfo {author} {\bibfnamefont {M.~E.}\ \bibnamefont
  {Christy}}, \bibinfo {author} {\bibfnamefont {K.}~\bibnamefont {Craycraft}},
  \bibinfo {author} {\bibfnamefont {R.}~\bibnamefont {Cruz-Torres}}, \bibinfo
  {author} {\bibfnamefont {H.}~\bibnamefont {Dai}}, \bibinfo {author}
  {\bibfnamefont {D.}~\bibnamefont {Day}}, \bibinfo {author} {\bibfnamefont
  {A.}~\bibnamefont {Dirican}}, \bibinfo {author} {\bibfnamefont {S.-C.}\
  \bibnamefont {Dusa}}, \bibinfo {author} {\bibfnamefont {E.}~\bibnamefont
  {Fuchey}}, \bibinfo {author} {\bibfnamefont {T.}~\bibnamefont {Gautam}},
  \bibinfo {author} {\bibfnamefont {C.}~\bibnamefont {Giusti}}, \bibinfo
  {author} {\bibfnamefont {J.}~\bibnamefont {Gomez}}, \bibinfo {author}
  {\bibfnamefont {C.}~\bibnamefont {Gu}}, \bibinfo {author} {\bibfnamefont
  {T.~J.}\ \bibnamefont {Hague}}, \bibinfo {author} {\bibfnamefont {J.-O.}\
  \bibnamefont {Hansen}}, \bibinfo {author} {\bibfnamefont {F.}~\bibnamefont
  {Hauenstein}}, \bibinfo {author} {\bibfnamefont {D.~W.}\ \bibnamefont
  {Higinbotham}}, \bibinfo {author} {\bibfnamefont {C.}~\bibnamefont {Hyde}},
  \bibinfo {author} {\bibfnamefont {Z.}~\bibnamefont {Jerzyk}}, \bibinfo
  {author} {\bibfnamefont {A.~M.}\ \bibnamefont {Johnson}}, \bibinfo {author}
  {\bibfnamefont {C.}~\bibnamefont {Keppel}}, \bibinfo {author} {\bibfnamefont
  {C.}~\bibnamefont {Lanham}}, \bibinfo {author} {\bibfnamefont
  {S.}~\bibnamefont {Li}}, \bibinfo {author} {\bibfnamefont {R.}~\bibnamefont
  {Lindgren}}, \bibinfo {author} {\bibfnamefont {H.}~\bibnamefont {Liu}},
  \bibinfo {author} {\bibfnamefont {C.}~\bibnamefont {Mariani}}, \bibinfo
  {author} {\bibfnamefont {R.~E.}\ \bibnamefont {McClellan}}, \bibinfo {author}
  {\bibfnamefont {D.}~\bibnamefont {Meekins}}, \bibinfo {author} {\bibfnamefont
  {R.}~\bibnamefont {Michaels}}, \bibinfo {author} {\bibfnamefont
  {M.}~\bibnamefont {Mihovilovic}}, \bibinfo {author} {\bibfnamefont
  {M.}~\bibnamefont {Murphy}}, \bibinfo {author} {\bibfnamefont
  {D.}~\bibnamefont {Nguyen}}, \bibinfo {author} {\bibfnamefont
  {M.}~\bibnamefont {Nycz}}, \bibinfo {author} {\bibfnamefont {L.}~\bibnamefont
  {Ou}}, \bibinfo {author} {\bibfnamefont {B.}~\bibnamefont {Pandey}}, \bibinfo
  {author} {\bibfnamefont {V.}~\bibnamefont {Pandey}}, \bibinfo {author}
  {\bibfnamefont {K.}~\bibnamefont {Park}}, \bibinfo {author} {\bibfnamefont
  {G.}~\bibnamefont {Perera}}, \bibinfo {author} {\bibfnamefont {A.~J.~R.}\
  \bibnamefont {Puckett}}, \bibinfo {author} {\bibfnamefont {S.~N.}\
  \bibnamefont {Santiesteban}}, \bibinfo {author} {\bibfnamefont
  {S.}~\bibnamefont {\ifmmode~\check{S}\else \v{S}\fi{}irca}}, \bibinfo
  {author} {\bibfnamefont {T.}~\bibnamefont {Su}}, \bibinfo {author}
  {\bibfnamefont {L.}~\bibnamefont {Tang}}, \bibinfo {author} {\bibfnamefont
  {Y.}~\bibnamefont {Tian}}, \bibinfo {author} {\bibfnamefont {N.}~\bibnamefont
  {Ton}}, \bibinfo {author} {\bibfnamefont {B.}~\bibnamefont {Wojtsekhowski}},
  \bibinfo {author} {\bibfnamefont {S.}~\bibnamefont {Wood}}, \bibinfo {author}
  {\bibfnamefont {Z.}~\bibnamefont {Ye}},\ and\ \bibinfo {author}
  {\bibfnamefont {J.}~\bibnamefont {Zhang}} (\bibinfo {collaboration} {The
  Jefferson Lab Hall A Collaboration}),\ }\href
  {https://doi.org/10.1103/PhysRevD.107.012005} {\bibfield  {journal} {\bibinfo
   {journal} {Phys. Rev. D}\ }\textbf {\bibinfo {volume} {107}},\ \bibinfo
  {pages} {012005} (\bibinfo {year} {2023})}\BibitemShut {NoStop}%
\bibitem [{\citenamefont {Ba}\ \emph {et~al.}(2016)\citenamefont {Ba},
  \citenamefont {Kiros},\ and\ \citenamefont
  {Hinton}}]{ba2016layernormalization}%
  \BibitemOpen
  \bibfield  {author} {\bibinfo {author} {\bibfnamefont {J.~L.}\ \bibnamefont
  {Ba}}, \bibinfo {author} {\bibfnamefont {J.~R.}\ \bibnamefont {Kiros}},\ and\
  \bibinfo {author} {\bibfnamefont {G.~E.}\ \bibnamefont {Hinton}},\ }\href
  {https://arxiv.org/abs/1607.06450} {\bibinfo {title} {Layer normalization}}
  (\bibinfo {year} {2016}),\ \Eprint {https://arxiv.org/abs/1607.06450}
  {arXiv:1607.06450 [stat.ML]} \BibitemShut {NoStop}%
\bibitem [{\citenamefont {Glorot}\ \emph {et~al.}(2011)\citenamefont {Glorot},
  \citenamefont {Bordes},\ and\ \citenamefont {Bengio}}]{pmlr-v15-glorot11a}%
  \BibitemOpen
  \bibfield  {author} {\bibinfo {author} {\bibfnamefont {X.}~\bibnamefont
  {Glorot}}, \bibinfo {author} {\bibfnamefont {A.}~\bibnamefont {Bordes}},\
  and\ \bibinfo {author} {\bibfnamefont {Y.}~\bibnamefont {Bengio}},\ }in\
  \href {https://proceedings.mlr.press/v15/glorot11a.html} {\emph {\bibinfo
  {booktitle} {Proceedings of the Fourteenth International Conference on
  Artificial Intelligence and Statistics}}},\ \bibinfo {series} {Proceedings of
  Machine Learning Research}, Vol.~\bibinfo {volume} {15},\ \bibinfo {editor}
  {edited by\ \bibinfo {editor} {\bibfnamefont {G.}~\bibnamefont {Gordon}},
  \bibinfo {editor} {\bibfnamefont {D.}~\bibnamefont {Dunson}},\ and\ \bibinfo
  {editor} {\bibfnamefont {M.}~\bibnamefont {Dudík}}}\ (\bibinfo  {publisher}
  {PMLR},\ \bibinfo {address} {Fort Lauderdale, FL, USA},\ \bibinfo {year}
  {2011})\ pp.\ \bibinfo {pages} {315--323}\BibitemShut {NoStop}%
\bibitem [{\citenamefont {{He}}\ \emph {et~al.}(2015)\citenamefont {{He}},
  \citenamefont {{Zhang}}, \citenamefont {{Ren}},\ and\ \citenamefont
  {{Sun}}}]{2015arXiv150201852H}%
  \BibitemOpen
  \bibfield  {author} {\bibinfo {author} {\bibfnamefont {K.}~\bibnamefont
  {{He}}}, \bibinfo {author} {\bibfnamefont {X.}~\bibnamefont {{Zhang}}},
  \bibinfo {author} {\bibfnamefont {S.}~\bibnamefont {{Ren}}},\ and\ \bibinfo
  {author} {\bibfnamefont {J.}~\bibnamefont {{Sun}}},\ }\href
  {https://doi.org/10.48550/arXiv.1502.01852} {\bibfield  {journal} {\bibinfo
  {journal} {arXiv e-prints}\ ,\ \bibinfo {eid} {arXiv:1502.01852}} (\bibinfo
  {year} {2015})},\ \Eprint {https://arxiv.org/abs/1502.01852}
  {arXiv:1502.01852 [cs.CV]} \BibitemShut {NoStop}%
\bibitem [{\citenamefont {Loshchilov}\ and\ \citenamefont
  {Hutter}(2019)}]{loshchilov2018decoupled}%
  \BibitemOpen
  \bibfield  {author} {\bibinfo {author} {\bibfnamefont {I.}~\bibnamefont
  {Loshchilov}}\ and\ \bibinfo {author} {\bibfnamefont {F.}~\bibnamefont
  {Hutter}},\ }in\ \href {https://openreview.net/forum?id=Bkg6RiCqY7} {\emph
  {\bibinfo {booktitle} {International Conference on Learning
  Representations}}}\ (\bibinfo {year} {2019})\BibitemShut {NoStop}%
\bibitem [{\citenamefont {Reddi}\ \emph {et~al.}(2018)\citenamefont {Reddi},
  \citenamefont {Kale},\ and\ \citenamefont {Kumar}}]{j.2018on}%
  \BibitemOpen
  \bibfield  {author} {\bibinfo {author} {\bibfnamefont {S.~J.}\ \bibnamefont
  {Reddi}}, \bibinfo {author} {\bibfnamefont {S.}~\bibnamefont {Kale}},\ and\
  \bibinfo {author} {\bibfnamefont {S.}~\bibnamefont {Kumar}},\ }in\ \href
  {https://openreview.net/forum?id=ryQu7f-RZ} {\emph {\bibinfo {booktitle}
  {International Conference on Learning Representations}}}\ (\bibinfo {year}
  {2018})\BibitemShut {NoStop}%
\bibitem [{\citenamefont {Demortier}\ and\ \citenamefont
  {Lyons}(2008)}]{DemortierEverythingYA}%
  \BibitemOpen
  \bibfield  {author} {\bibinfo {author} {\bibfnamefont {L.}~\bibnamefont
  {Demortier}}\ and\ \bibinfo {author} {\bibfnamefont {L.}~\bibnamefont
  {Lyons}},\ }\href
  {https://lucdemortier.github.io/assets/papers/cdf5776_pulls.pdf} {\bibfield
  {journal} {\bibinfo  {journal} {e-Print}\ } (\bibinfo {year}
  {2008})}\BibitemShut {NoStop}%
\bibitem [{\citenamefont {Rubner}\ \emph {et~al.}(1998)\citenamefont {Rubner},
  \citenamefont {Tomasi},\ and\ \citenamefont {Guibas}}]{710701}%
  \BibitemOpen
  \bibfield  {author} {\bibinfo {author} {\bibfnamefont {Y.}~\bibnamefont
  {Rubner}}, \bibinfo {author} {\bibfnamefont {C.}~\bibnamefont {Tomasi}},\
  and\ \bibinfo {author} {\bibfnamefont {L.}~\bibnamefont {Guibas}},\ }in\
  \href {https://doi.org/10.1109/ICCV.1998.710701} {\emph {\bibinfo {booktitle}
  {Sixth International Conference on Computer Vision (IEEE Cat.
  No.98CH36271)}}}\ (\bibinfo {year} {1998})\ pp.\ \bibinfo {pages}
  {59--66}\BibitemShut {NoStop}%
\end{thebibliography}%

\end{document}